\documentclass[10pt,journal,compsoc]{IEEEtran}
\ifCLASSOPTIONcompsoc
  \usepackage[nocompress]{cite}
\else
  \usepackage{cite}
\fi
\usepackage{hyperref}
\usepackage{tabularx}

\usepackage{graphicx}
\usepackage{subcaption}
\ifCLASSINFOpdf
\else
\fi
\usepackage{amsmath}
\usepackage{mathtools}
\usepackage{xparse}
\usepackage{booktabs}

\usepackage{algorithm,algpseudocode}
\newcommand{\subsup}[4]{ 
\ifthenelse{\equal{#1}{1}}
{ 
	{{\mathbf{#2}^ {#3}_ {#4} }} 
}
{ 
	{{{#2}^ {#3} }_{\hspace{-.045in} {#4} }}
}
}

\newcommand{\Quote}[1]{``{#1}''}
\usepackage{amssymb}
\newcommand{\R}{\mathbb{R}}

\DeclarePairedDelimiter{\norm}{\lVert}{\rVert}
\makeatletter
\newcommand{\@supnormstar}[1]{\norm*{#1}_2}
\newcommand{\@supnormnostar}[2][]{\norm[#1]{#2}_2}
\newcommand{\supnorm}{\@ifstar\@supnormstar\@supnormnostar}
\makeatother

\algnewcommand{\Initialize}[1]{%
  \Statex \textbf{Initialize:}
  \Statex \hspace*{\algorithmicindent}\parbox[t]{.8\linewidth}{\raggedright #1}
}

\algnewcommand\algorithmicforeach{\textbf{for each}}
\algdef{S}[FOR]{ForEach}[1]{\algorithmicforeach\ #1\ \algorithmicdo}

\newcolumntype{M}[1]{>{\centering\arraybackslash}m{#1}}

\begin{document}
%
\title{Skeleton Extraction from 3D Point Clouds by Decomposing the Object into Parts}
%
%
%
%

\author{Vijai~Jayadevan,~\IEEEmembership{}
        Edward~Delp,~\IEEEmembership{}
        and~Zygmunt~Pizlo
\IEEEcompsocitemizethanks{\IEEEcompsocthanksitem V. Jayadevan is with the Department
of Electrical and Computer Engineering, Purdue University, West Lafayette,
IN, 47906.\protect\\
E-mail: vthottat@purdue.edu}}

%
%

\markboth{Journal of \LaTeX\ Class Files,~Vol.~00, No.~00}%
{Jayadevan \MakeLowercase{\textit{et al.}}: Skeleton Extraction from 3D Point Clouds by Decomposing the Object into Parts}

\IEEEtitleabstractindextext{%
\begin{abstract}
Decomposing a point cloud into its components and extracting curve skeletons from point clouds are two related problems. Decomposition of a shape into its components is often obtained as a byproduct of skeleton extraction. In this work, we propose to extract curve skeletons, from unorganized point clouds, by decomposing the object into its parts, identifying part skeletons and then linking these part skeletons together to obtain the complete skeleton. We believe it is the most natural way to extract skeletons in the sense that this would be the way a human would approach the problem. Our parts are generalized cylinders (GCs). Since, the axis of a GC is an integral part of its definition, the parts have natural skeletal representations. We use translational symmetry, the fundamental property of GCs, to extract parts from point clouds. We demonstrate how this method can handle a large variety of shapes. We compare our method with state of the art methods and show how a part based approach can deal with some of the limitations of other methods. We present an improved version of an existing point set registration algorithm and demonstrate its utility in extracting parts from point clouds. We also show how this method can be used to extract skeletons from and identify parts of noisy point clouds. A part based approach also provides a natural and intuitive interface for user interaction. We demonstrate the ease with which mistakes, if any, can be fixed with minimal user interaction with the help of a graphical user interface.
\end{abstract}

}

\maketitle

\IEEEdisplaynontitleabstractindextext

%
\IEEEpeerreviewmaketitle


\IEEEraisesectionheading{\section{Introduction}\label{sec:introduction}}

Decomposing a complex 3D shape into its components is an important problem that has applications in many fields including computer graphics and computer vision. Psychological studies have shown that human shape perception and recognition is based on decomposing complex shapes into simple primitives \cite{Biederman_1993_recognizing}. Therefore, characterization of a shape by its parts and the connections between them is only natural, especially for applications involving human interaction. A good example in this regard is the work by Funkhouser e. al \cite{funkhouser2004modeling}, which aims at constructing detailed 3D models by assembling parts extracted from existing models. Part based approaches also have applications in other area like character animation \cite{baran2007automatic} and surface reconstruction \cite{li2010arterial, chen2013_3sweep, yin2014morfit}. 

To decompose a shape into its parts, we first need a good definition for a part. A good definition of a part would be general enough to capture the wide variety of part types available in the real world and at the same time be able to decompose a shape into meaningful units. In this context, generalized cylinders (GCs) introduced by Binford \cite{binford1971visual} would serve as a very good definition of a part. A wide variety of natural as well as man made objects have parts that can be well modeled by GCs. GCs are formed by sweeping a planar cross-section along a 3D axis, with uniform size scaling applied to the cross-section as it moves along the axis. An illustration of this is shown in Fig. \ref{fig:gc_illustration}. Translational symmetry is the fundamental property that defines a GC \cite{Pizlo_2014_Making}. In this work, we think of parts of a 3D shape as GCs and employ translational symmetry to extract them. 

\begin{figure*}[htp]
\centering
\begin{subfigure}[b]{0.3\textwidth}
\centering
	\includegraphics[width=\textwidth]{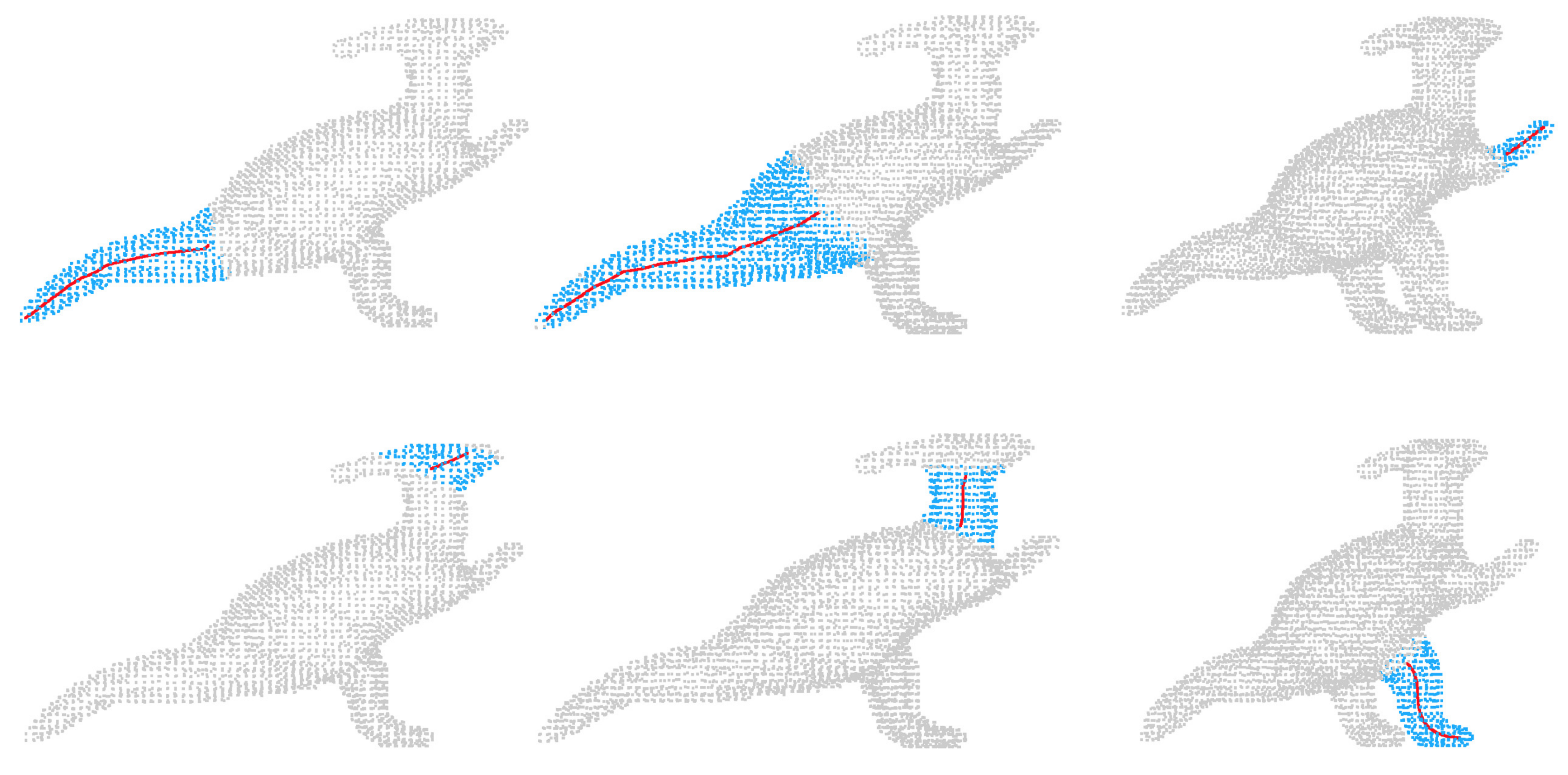}
	\caption{}
\end{subfigure}	
\begin{subfigure}[b]{0.3\textwidth}
\centering
	\includegraphics[width=\textwidth]{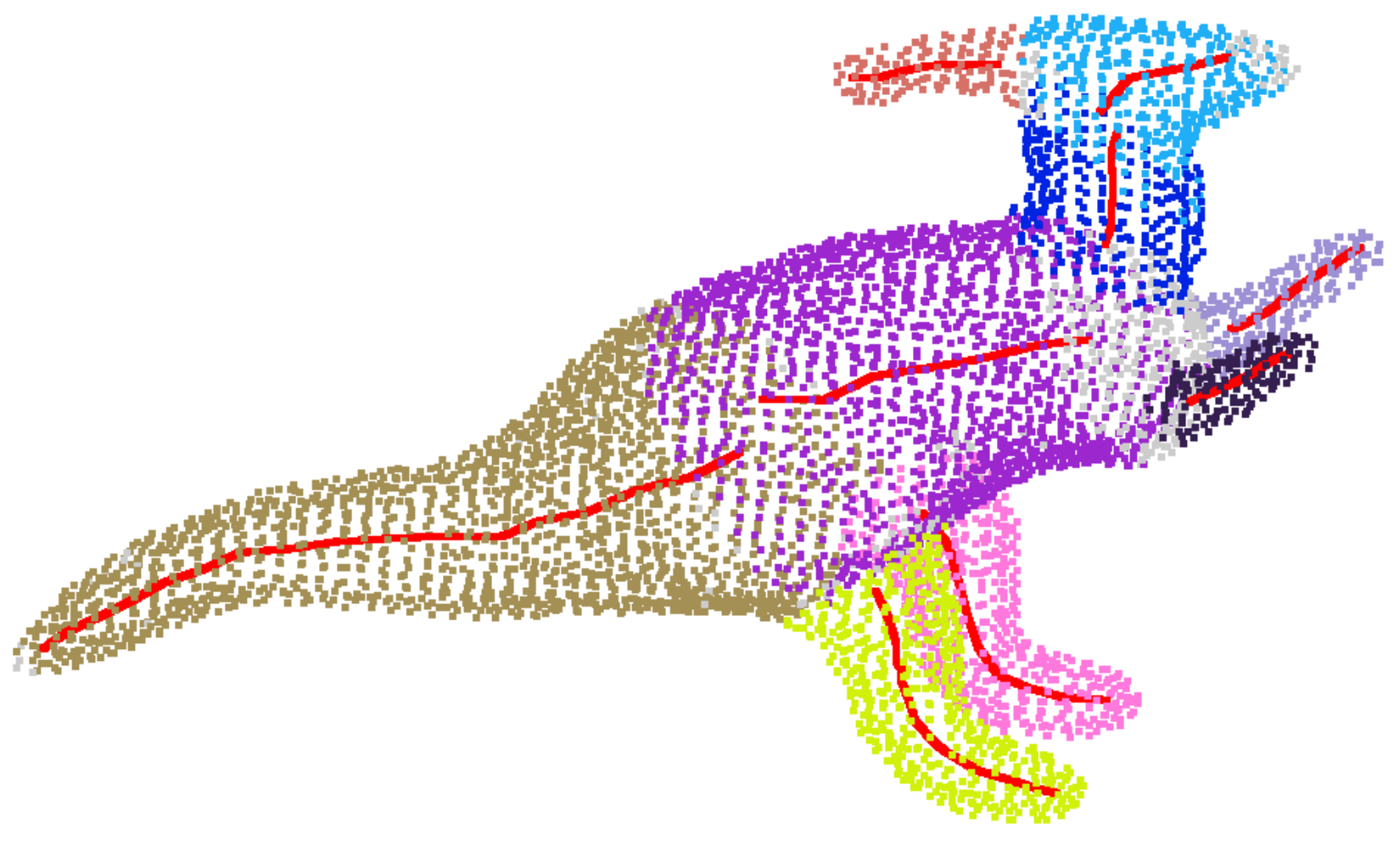}
	\caption{}
\end{subfigure}
\begin{subfigure}[b]{0.3\textwidth}
\centering
	\includegraphics[width=\textwidth]{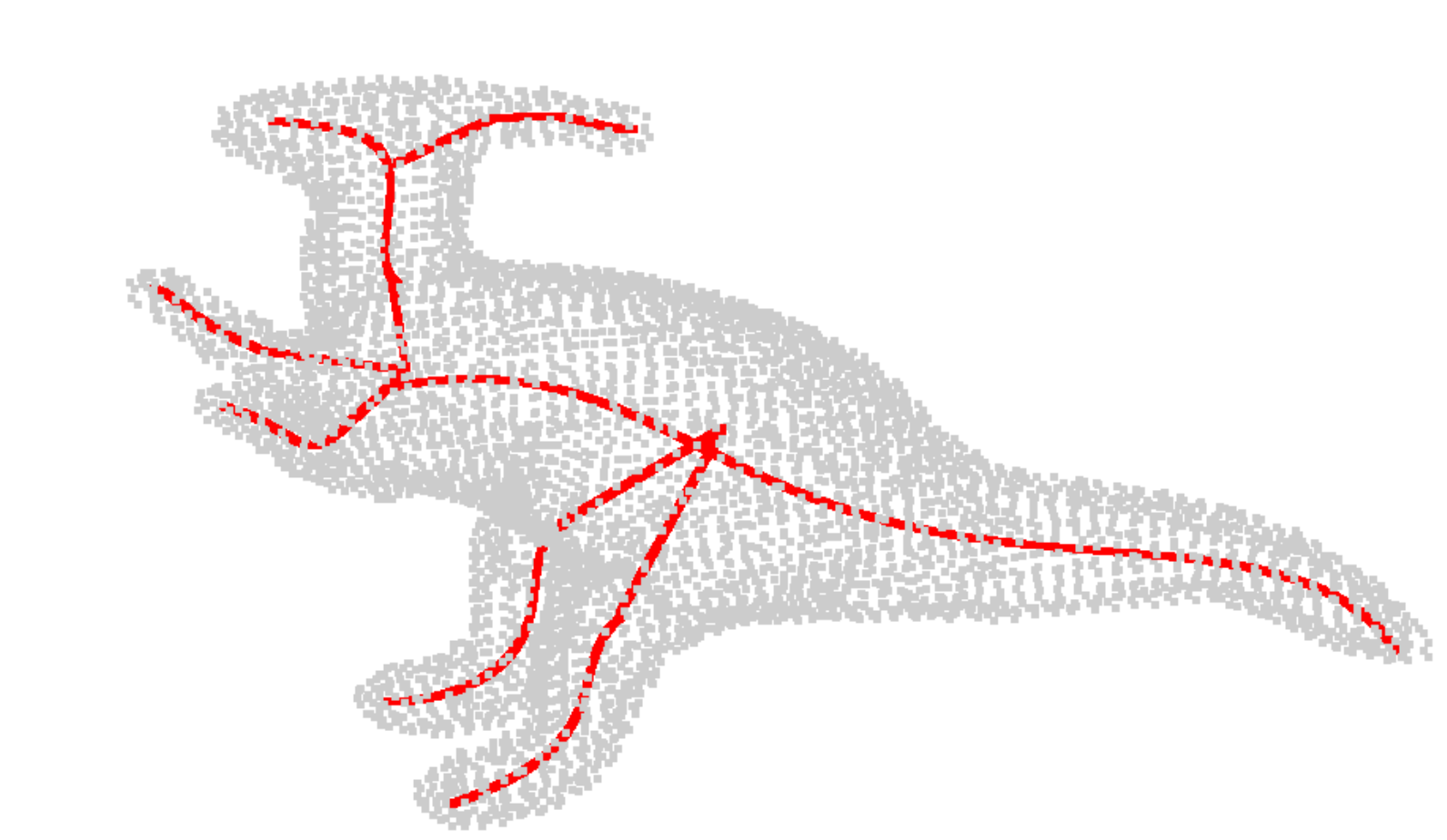}
	\caption{}
\end{subfigure}
\caption{An overview of the algorithm. (a) Candidate parts are first generated. The points representing the part are shown in blue and the skeletal representation of parts are shown in red. (b) An optimal subset of parts, that can represent the entire point cloud, are selected from the candidate parts. Different parts are shown in different colors. Also shown are the individual skeletons of each part. (c) Appropriate connections are made between skeletons of individual parts to form the final skeletal representation.}
\label{fig:overview}
\end{figure*}

Decomposing a shape into its components can be thought of as an inverse problem which is often times also ill-posed. It is an inverse problem because the forming of a complex 3D shape by interconnecting parts can be thought of as the forward problem and the ill-posedness comes from the fact that there is no unique solution to the problem. Solution to ill-posed inverse problems are obtained by imposing additional constraints (aka priors). In this context, translational symmetry can be thought of as the prior we use to identify parts which eventually leads to the decomposition of shapes.

Skeleton extraction methods often times lead to shape decomposition \cite{sharf2007fly,au2008skeleton,reniers2008computing}. But in our case it works in the opposite direction. The 3D-axis of a GC, which is integral to its definition, would serve as the skeleton of the part. Therefore, a GC based decomposition would naturally lead to skeletonization of the shape. There exist a method by Zhou et al. \cite{zhou2015generalized} which address the problem of decomposing shapes into GCs. As we point out later, though our method has similarities to their approach, there are significant differences stemming from the fact that our method works on unorganized point clouds and theirs on meshes.

As shown in Fig. \ref{fig:overview}, the algorithm has three main steps. In the first step, a few candidate parts are generated. Not all candidate parts are ideal to represent the actual parts of the object, and there is often overlap between different candidate parts. Therefore, an optimal subset of parts are then selected from these candidate parts in the next step. Note that, every part has an axis associated with it, which serves as its skeletal representation. Therefore, in the final step, the individual skeletons of the selected parts are linked appropriately to form the complete skeleton. 

In order to generate candidate parts, we first detect the cross-section of parts and then grow parts starting from this initial cross-section to obtain complete parts. An interesting fact to notice here is that growth explains why many things in nature, like limbs of animals and tree trunks, exhibit translational symmetry \cite{Pizlo_2014_Making}. We employ translational symmetry to grow parts. Specifically, we start with a initial cross-sectional cluster (a small cluster of points representing a thin cross-section) and grow this cluster by identifying similar clusters in its neighborhood via 3D point set registration. By combining ideas from \cite{Myronenko_2010_CPD} and \cite{Billings_2015_GIMLOP} we present a new registration algorithm that is tailor made for our purpose of growing parts via registration. The proposed algorithm makes use of the orientation information of the point clouds (i.e., the point normals) and this, as we show in the results section, helps it to achieve greater accuracy in the registration process. In the results section, we also compare our method to two state of the art methods for skeleton extraction and point out the advantages of using a part based method. We also present a graphical user interface (GUI) to demostrate the advantage of using a part based approach to skeletonization.  Specifically, we show the ease with which mistakes, if any, can be corrected with minimal user interaction using the GUI.

In the next section we review the past literature relevant to our work. In section \ref{sec:generating_candidate_parts}, we explain the procedure to generate candidate parts. Method for selecting optimal subset of parts and the method for linking selected parts are explained in Sections \ref{sec:optimal_part_selection} and \ref{sec:linking_parts} respectively. The GUI is described in section \ref{sec:GUI_Skel} and the results are presented in section \ref{sec:results_skel}. This is followed by sections on the implementation details and conclusion.


%
%
%
%


\begin{figure*}
\centering
\begin{subfigure}[b]{.63\textwidth}
\includegraphics[width=\textwidth]{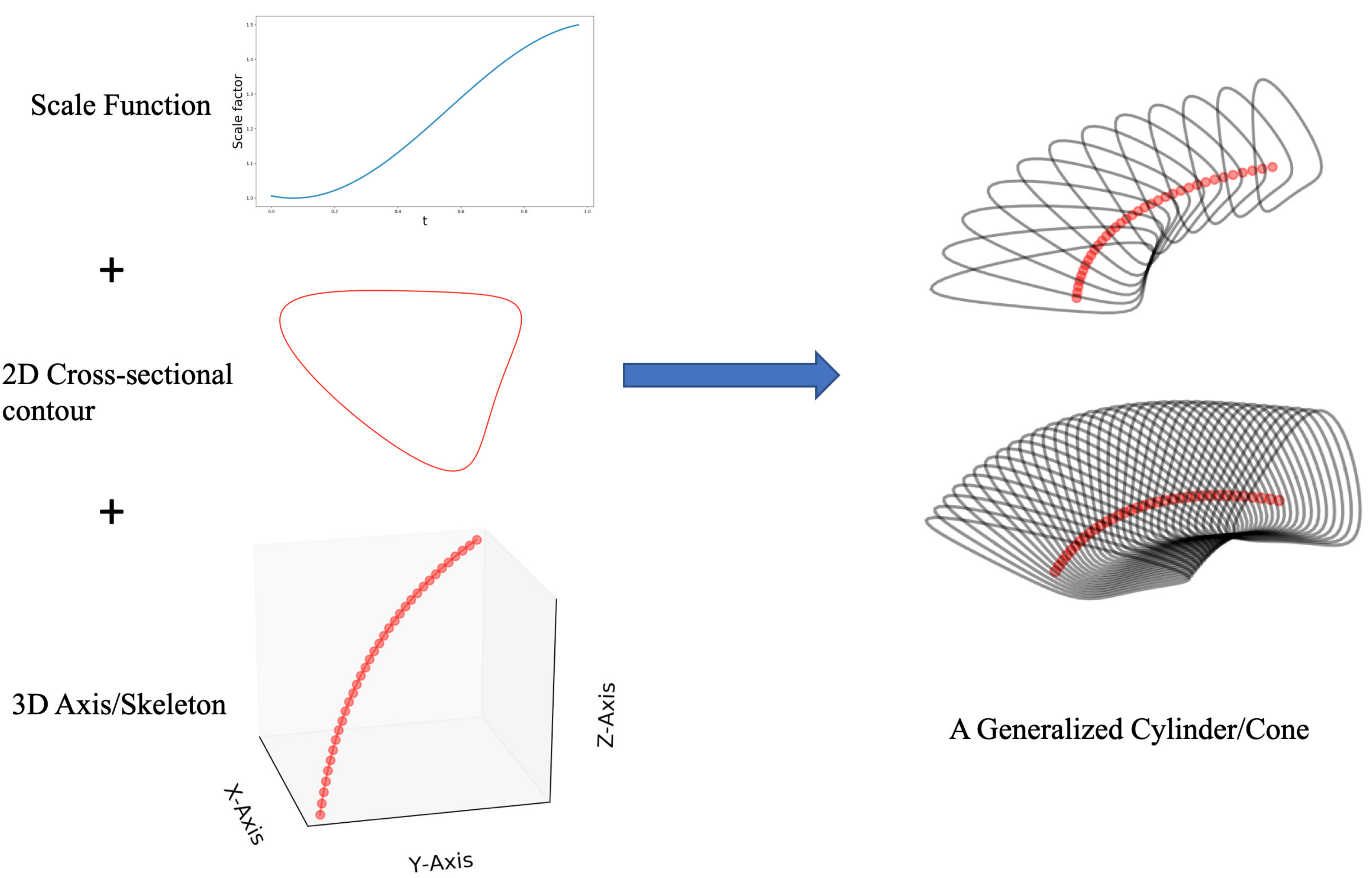}
\caption{}
\end{subfigure}\qquad
\begin{subfigure}[b]{.33\textwidth}
\hspace{7ex}
\includegraphics[width=0.5\textwidth]{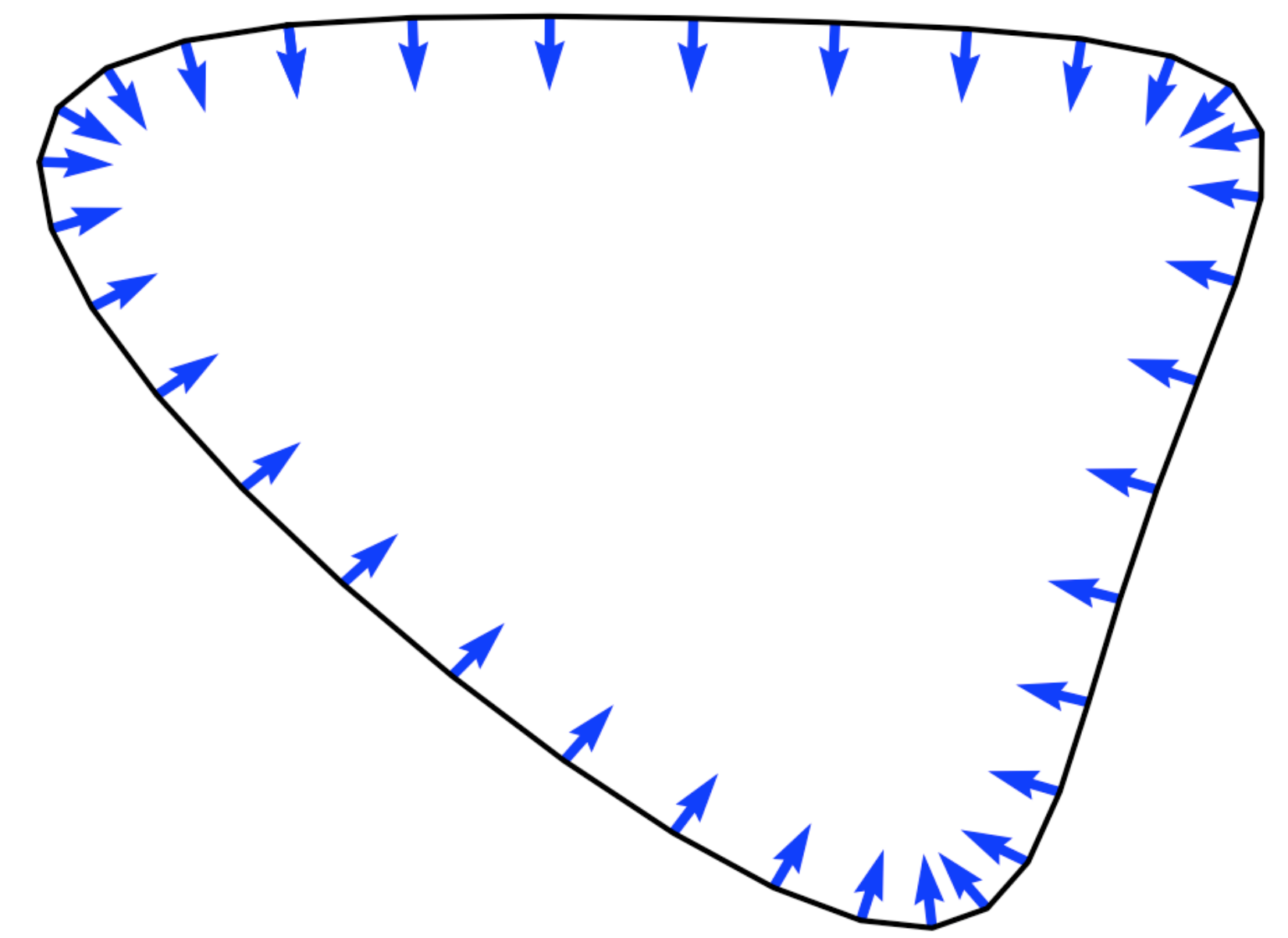}
\caption{}

\vspace{2ex}

\includegraphics[width=\textwidth]{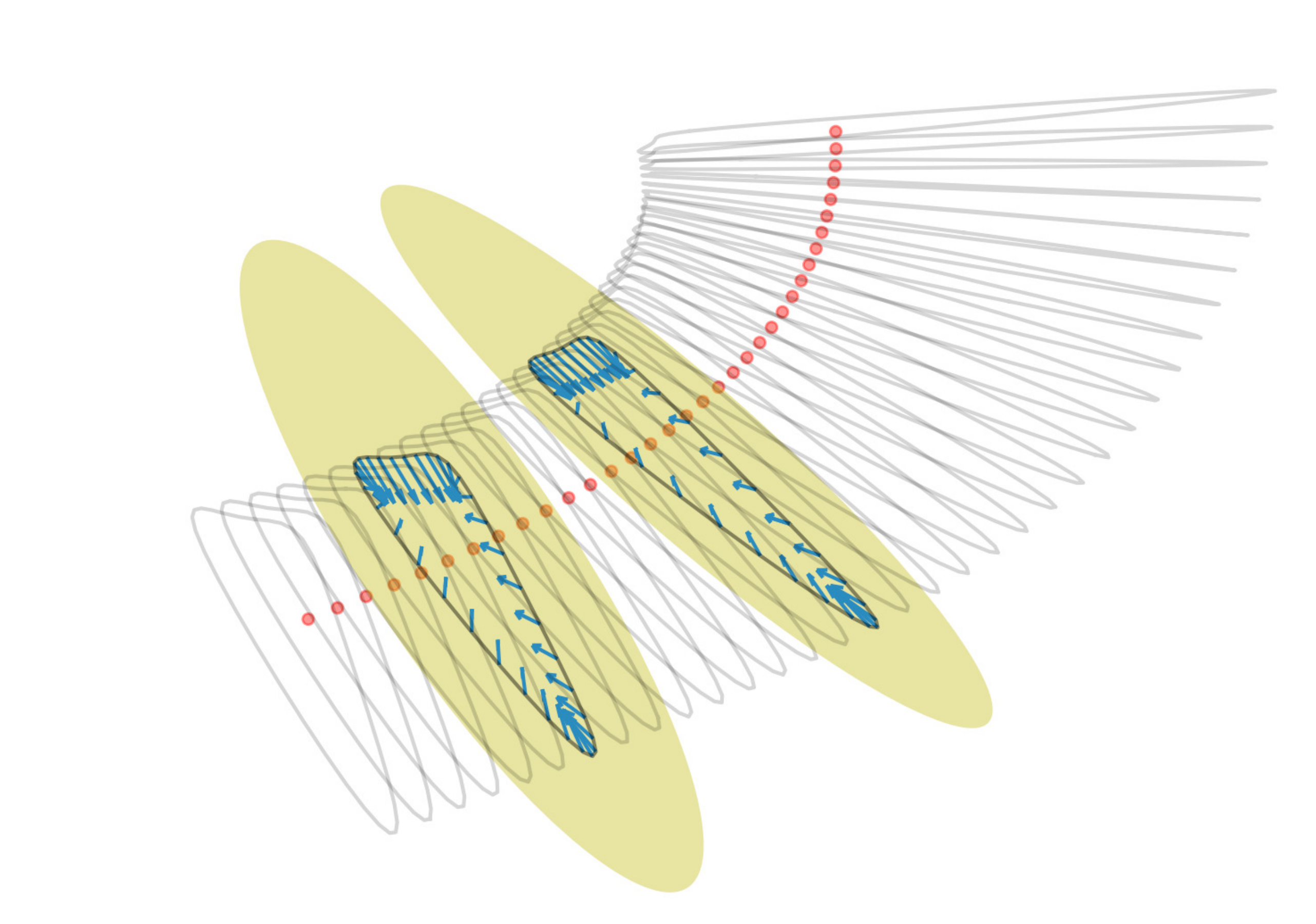}
\caption{}
\end{subfigure}
\caption{Our definition of a part is based on translational symmetry. (a) The three fundamental properties of 3D Axis, 2D cross-sectional contour and scale function define a part . Parts are formed by sweeping a planar cross-section through 3D space along a defined axis (a space curve) and simultaneously applying size scaling as the cross-section is swept along the axis. (b) The normals to the contour at the various points on the contour are shown. (c) At each point along the axis, the plane containing the corresponding 2D contour is perpendicular to the axis. Or in other words, the normal of the plane represents the tangent to the axis at that point. The normals of the 2D contour lie on the cross-sectional plane and hence they are perpendicular to the normal of the plane.}
\label{fig:gc_illustration}

\end{figure*}

\section{Related Work}
We organize the literature review into two sections. First, we discuss methods from the literature used for curve skeleton extraction. This is followed by a review of methods for shape decomposition. 

\subsubsection*{Skeleton Extraction} \label{sec:lit_review_skel}
There is a vast body of prior work addressing the problem of curve skeleton extraction. Methods exist to extract curve skeletons from complete surface models, like polygonal meshes, as well as unorganized point clouds. Methods belonging to the latter category are the most relevant to our work and hence we focus on these methods for the review. The reader may refer to \cite{tagliasacchi_2016_skel_survey} or \cite{cornea2007curve} for a comprehensive review of the various skeleton extraction methods that are available.

There exist a class of methods that rely on defining a field on the voxelized internal volume of shapes. In such methods, the ridges of the field are of particular interest when it comes to skeleton extraction. For example, in \cite{chuang2004potential}, the authors identify valleys, in the generalized potential field, which connect several seed points, to derive the axis of the GC representation of the shape. In \cite{siddiqi1999hamilton}, Siddiqi et al. extract skeletons by using the fact that the divergence of the euclidean distance transform would be zero everywhere except at the skeletal points. Song et al. \cite{song2018distance} improve upon the results obtained by Huang et al. \cite{huang2013L1} by using the distance field to guide the L1-median skeleton extraction. Note that the appropriate voxelization of unorganized point clouds is a non-trivial task. Apart from this, as we will demonstrate in the results section, these methods will not generalize well to certain shapes in comparison to a part-based approach.

By defining an appropriate real valued function on a compact manifold and then tracking the evolution of its level sets, Reeb graphs, which capture the topology of the manifold, can be created. Methods described in \cite{hilaga2001topology,natali2011graph,bucksch2010skeltre} employ Reeb graphs to extract skeletons. Authors in \cite{dey2006defining, ogniewicz1992voronoi} extract skeletons by employing Voronoi diagrams. Voronoi diagrams can be used to approximate medial surface of shapes and these can be further pruned to obtain curve skeletons. Tagliasacchi et al. \cite{tagliasacchi2009curve} proposed a method, based on the rotational symmetry axis, to extract skeletons from incomplete point clouds of generally cylindrical shapes. Sharf et al. \cite{sharf2007fly} proposed a method that is based on the evolution of a deformable model inside the point cloud. They obtain a first approximation to the skeleton by tracing the growing fronts and perform further filtering to obtain the final curve skeleton. Cao et al. \cite{cao2010point}  extend the mesh contraction method of Au et al. \cite{au2008skeleton} to point clouds and use Laplacian based contraction followed by topological thinning to obtain curve skeletons. In the results section, we compare our method with that of Cao et al. and point out the drawbacks of using a contraction based approach for skeleton extraction.

\subsubsection*{Shape Decomposition} \label{sec:lit_review_semantic}
A variety of methods address the problem of segmentation and semantic labelling of point clouds \cite{nguyen20133d}. Skeleton extraction methods can some time lead to decomposition of shapes \cite{sharf2007fly,au2008skeleton,reniers2008computing}. But model fitting methods, which addresses the problem of shape decomposition, are the most relevant to our work. Most primitive detection and fitting methods employ simple primitives for the purpose. For instance, Schnabel et al. \cite{schnabel2007efficient} describe an efficient RANSAC based method capable of fitting simple primitives like planes, spheres, cylinders, cones and tori. Attene et al. \cite{attene2006hierarchical} present a method to fit simple primitives to triangle meshes using a hierarchical face clustering approach. GCs, compared to these simple primitives, can represent a much broader range of shapes. 

Lit et al. \cite{li2001decomposing} present an early work on decomposing polygonal meshes into parts that resemble GCs. They extract approximate curve skeletons by mesh edge contraction and then identify the critical points (points where the topology or geometry changes) by sweeping a plane, perpendicular to the skeleton branches, over the mesh. The parts of the mesh between consecutive critical points are then extracted as components. Though it is not stated explicitly, this method uses the notion of translational symmetry to identify components/parts. One of the two methods to grow parts that we describe later has similarities with this method. However, unlike this method we operate on unorganized point clouds and hence cannot employ their method of skeleton extraction. Moreover, the second method to grow parts that we describe later is more sophisticated as it is based on the very definition of a GC (i.e., based on translational symmetry), and is more suitable for point clouds. In \cite{chuang2004potential}, the authors first derive a skeleton of the shape using a potential-based skeletonization approach and then generate the cross-sections with the help of seed points. But if the shape has to be decomposed into parts, each of which is a GC, the seed points have to be manually chosen. 

There are a number of automatic as well as interactive surface reconstruction methods that use a GC representation \cite{li2010arterial, chen2013_3sweep, yin2014morfit}. However, they do not explicitly aim at obtaining an optimal decomposition of the shape into GCs. With regards to obtaining a GC based decomposition, the work of Goyal et al. \cite{goyal2012towards} and Zhou et al. \cite{zhou2015generalized} are the most relevant to our work. Both these methods work on meshes unlike our method which takes unorganized point clouds as input. Goyal et al. obtain a direct parameterization of 3D meshes in terms of sets of locally prominent cross-sections (PCS). Each set of PCS represent a different sweep component. Zhou et al. propose a metric to measure cylindricity of a GC. They first extract an over-complete set of local cylinders using the method proposed by Tagliasacchi et al. \cite{tagliasacchi2009curve} and then combine these local cylinders to form longer candidate cylinders. The set of candidate cylinders is still over-complete and the partition of the shape into GCs is obtained by solving the exact cover problem. Our approach is similar to that of Zhou et al in the sense that we too first generate an overlapping set GCs and we choose the best subset of GCs that cover the shape. However, there are some significant differences between the two approaches. The most signifacnt difference being that our method operates on point clouds and not polygonal meshes. Our method to extract GCs is rooted in the very definition of a GC which is in turn based on translational symmetry. Their method of scoring the GCs is not applicable to us because we operate on point clouds. Therefore, we introduce metrics to score GCs based on factors like their self-similarity (i.e., based on how much one portion of the GC is similar to another portion of the same GC), length, straightness of axis etc. And finally, we frame the problem of choosing the final subset of GCs as a binary integer program. 

\section{Generating Candidate Parts} \label{sec:generating_candidate_parts}

As shown in Fig. \ref{fig:gc_illustration}, a GC is formed by sweeping a planar cross-section along a 3D axis, with size scaling applied to the cross-section, as it moves along the axis. Such objects were referred to as generalized cones by Binford \cite{binford1971visual}. As per this definition, every part has an axis. This axis serves as the skeletal representation of the part. According to this definition of a part, there is a cross-sectional plane associated with each point on the axis. And as shown in Fig. \ref{fig:gc_illustration} (c), the normal of the cross-sectional plane associated with a point on the axis, is the tangent to the axis at that point. This important observation will be employed later in detecting local cross-sectional planes. 

\begin{figure}[htp]
\centering
\includegraphics[width=0.5\textwidth]{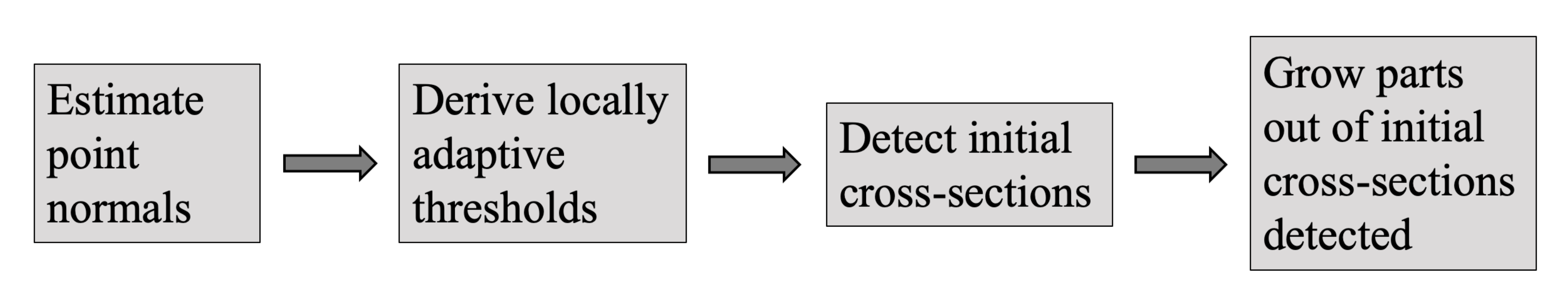}
\caption{Block diagram showing the different steps involved in generating candidate parts.}
\label{fig:bloc_diag_gen_parts}
\end{figure}

Fig. \ref{fig:bloc_diag_gen_parts}, shows the various steps involved in generating candidate parts. Given a point cloud, the first step in the algorithm is to estimate the point normals for each point in the cloud. The point normals are nothing but the local surface normals. At the end of this step, we would obtain an oriented set of points and the point normal orientations, as we show later, would play a very important role throughout the algorithm. In the next step, we derive local thresholds. The key idea here is that whenever thresholds are required to be applied (like estimating the closeness of a point in the cloud to some plane or deciding if two points in the point cloud can be considered as neighbors), locally adaptive thresholds  need to be employed. For instance, the sparsity/density of the point clouds need not be uniform through out the cloud. To deal with this issue, we assign a number, which represent the local distance threshold, to each point in the cloud. 

In the next step, we detect what we call \Quote{initial cross-sections}. An initial cross-section is a cluster of points in the cloud that are likely to represent one of the cross-sections of a part.  In Fig. \ref{fig:growing_stages}, cluster 0, represents such an initial cross-section. Such clusters serve as the starting points to \Quote{grow} parts. Keep in mind that all points in the point cloud are assumed to come from parts which are generalized cylinders. Therefore, each point in the cloud corresponds to one of the cross-sectional planes (like the yellow planes in Fig. \ref{fig:gc_illustration} (c)) of one of the parts. Therefore, in this step, given any point in the cloud, we would like to be able to estimate the orientation of the cross-sectional plane, passing through that point. Provided, we can find the orientation of the cross-sectional plane corresponding to a given point in the cloud, all points sufficiently close (the notion of being sufficiently close will be made more concrete later) to this cross-sectional plane would constitute what we call an initial cross-section.

Assume for a moment that given such an initial cross-section, we will be able to extract the part that the initial cross-section belongs to. For instance, in Fig. \ref{fig:growing_stages}, given cluster 0, let's assume we can grow the part and extract the torso of the dinosaur. We will explain how this can be done later. Since, we would like to extract all parts of an object we would need at least as many initial cross-sections as there are parts. In the example of the dinosaur (Fig. \ref{fig:growing_stages}), if we assume that the number of parts constituting the object is nine (i.e., four limbs, a tail, torso, neck, horn and face), we need at least nine initial cross-sections to extract all the parts. Hence, we cluster the point cloud into $M$ clusters and find the initial cross-section for each of these clusters. A large value is chosen for $M$ (close to hundred) to ensure that we detect initial cross-sections for all the parts. It is better to have some redundant parts than to miss out on extracting some parts.

As shown in Fig. \ref{fig:bloc_diag_gen_parts}, the next step is  \Quote{growing} parts out of these initial cross-sections. To understand this process consider  cluster 0, in Fig. \ref{fig:growing_stages}, which represents an initial cluster. Once we have such a cluster of points, we search in the neighborhood, of that cluster, for a similar group of points which could represent the neighboring cross-section of the initial cluster. Cluster 1 in Fig. \ref{fig:growing_stages} represents such a matching cluster. Note, we will make the notion of similarity/match concrete later. If we find a matching cluster, we repeat this process of searching for matching clusters. For instance, cluster 2 is found to be a match for cluster 1. Keep in mind that our parts are defined by translational symmetry. Therefore, for each cross-sectional cluster, we will find similar clusters in its neighborhood. Or in other words, the procedure just described for growing parts from initial cross-sections is inspired by the definition of parts as generalized cylinders. This process of growing a part is repeated until there are no more matches to be found. Note, the parts are to be grown, from the initial cluster, in both directions. In Fig. \ref{fig:growing_stages} , the clusters in the two directions are represented by positive and negative cluster numbers respectively. 

The three fundamental components of a part is its 2D cross-sectional contour, the scaling function and the 3D Axis. Since we are dealing with point clouds, the concept of a 2D contour lying on a plane does not make sense. Hence, the initial cross-section is a cluster of points and these thin clusters play the role of the 2D cross-sectional contour. In the growing stage we are attempting to determine matches for the thin cross-sections. Finding such matches would give us an estimate of the scaling function and also the estimate of the axis of the part. The procedure of finding matches and how such matches would lead to estimates of the scaling function and the axis would be explained in the sections below. But the key point to keep in mind is that the part detection algorithm is attempting to identify parts by using its very definition. Each of the steps in Fig. \ref{fig:bloc_diag_gen_parts} are explained in greater detail in the sections below.

\subsection{Estimating Point Normals} \label{sec:point_normal_estimation}
Before extracting the candidate parts, we estimate the local surface normal of the points in the 3D point cloud. We use the software CloudCompare to do this \cite{cloud_compare_2019}. A quadratic surface is fit on to the local neighborhood to estimate the normal direction. But the orientation of these normals (i.e., whether it points towards the inside or outside of the object) is still ambiguous. To ensure that the normals are oriented in a consistent fashion, the normals are re-oriented by propagating the normal orientation starting from a random point with the help of a Minimum Spanning Tree (MST).  We use $\mathbf{X} = \{\mathbf{x}^1,\mathbf{x}^2, \cdots, \mathbf{x}^N \}$ to represent the point cloud. Note, $\mathbf{x}^i = (\subsup{1}{x}{i}{p},\subsup{1}{x}{i}{n})$, where $\subsup{1}{x}{i}{p} \in \Re^3$ represents the position of the point and $\subsup{1}{x}{i}{n} \in \Re^3$ is a unit normal representing the orientation of the normal at the point.

\subsection{Deriving Locally-adaptive Thresholds} \label{sec:locally_adaptive_threshold}

We use minimum spanning tree (MST) to derive local distance thresholds. The distance to neighboring points in an MST, constructed on the 3D point cloud, is a good indication of the local sparsity of the point cloud. Therefore, to derive a locally adaptive distance threshold for a point in the cloud, the distances of the point to its neighbors, as per the MST constructed on the cloud, can be used. To compute the MST, we must have an initial graph on the point cloud. Instead of considering a fully connected graph as input to MST, we construct a graph where every point is connected to its hundred nearest neighbors. Though a higher number of nearest neighbors could be considered, in practice, we found it be sufficient to just consider the nearest hundred neighbors. The reason for not considering a fully connected graph is to reduce the time for MST computation. The time complexity for MST is $\mathcal{O}(E\log{V})$, where $V$ is the number of nodes/vertices in the graph and $E$ is the number of edges. A fully connected graph would have a time complexity of $\mathcal{O}(V^2\log{V})$, whereas considering a fixed number of nearest neighbors, say hundred in our case, will only have a time complexity of $\mathcal{O}(V\log{V})$. We found that using a fixed, albeit large enough, neighborhood to create the input graph to MST produced considerable savings in computation time in practice. We use the k-d tree data structure to speed up neighborhood queries. Specifically , we use the k-d tree implementation from the Scikit-learn package \cite{scikit_learn}. Kruskal's algorithm implemented in the SciPy package \cite{scipy_package} is used to compute MST.

As mentioned before, the locally adaptive threshold for points in the cloud is employed, for various purposes, throughout the algorithm. One purpose of the locally adaptive threshold is to create a connectivity matrix on the point cloud. Or in other words, the locally adaptive threshold derived for a point in the cloud is used to determine its neighbours. Therefore, once the MST is computed, we assign a distance threshold for each individual point in the cloud. I.e., for each point $\mathbf{x}^i$ in the point cloud, we set a distance threshold $\subsup{0}{\delta}{i}{cnct} =1.5 \subsup{0}{d}{i}{max}$, where $\subsup{0}{d}{i}{max}$ is the maximum among the distances to all the neighbors the point $\mathbf{x}^i$ is connected to, in the MST. Two points $\mathbf{x}^i$ and $\mathbf{x}^j$ in the point cloud are considered to be connected if $d_{eucl}(\mathbf{x}^i,\mathbf{x}^j) \leq $ maximum$(\subsup{0}{\delta}{i}{cnct},\subsup{0}{\delta}{j}{cnct})$, where $d_{eucl}(\mathbf{x}^i,\mathbf{x}^j)$ is the euclidean distance between the points. Based on this criterion a connectivity graph, $\mathbf{G}_{cnct}$, is constructed on the point cloud. 

\begin{figure}[htp]
\centering
\includegraphics[width=0.5\textwidth]{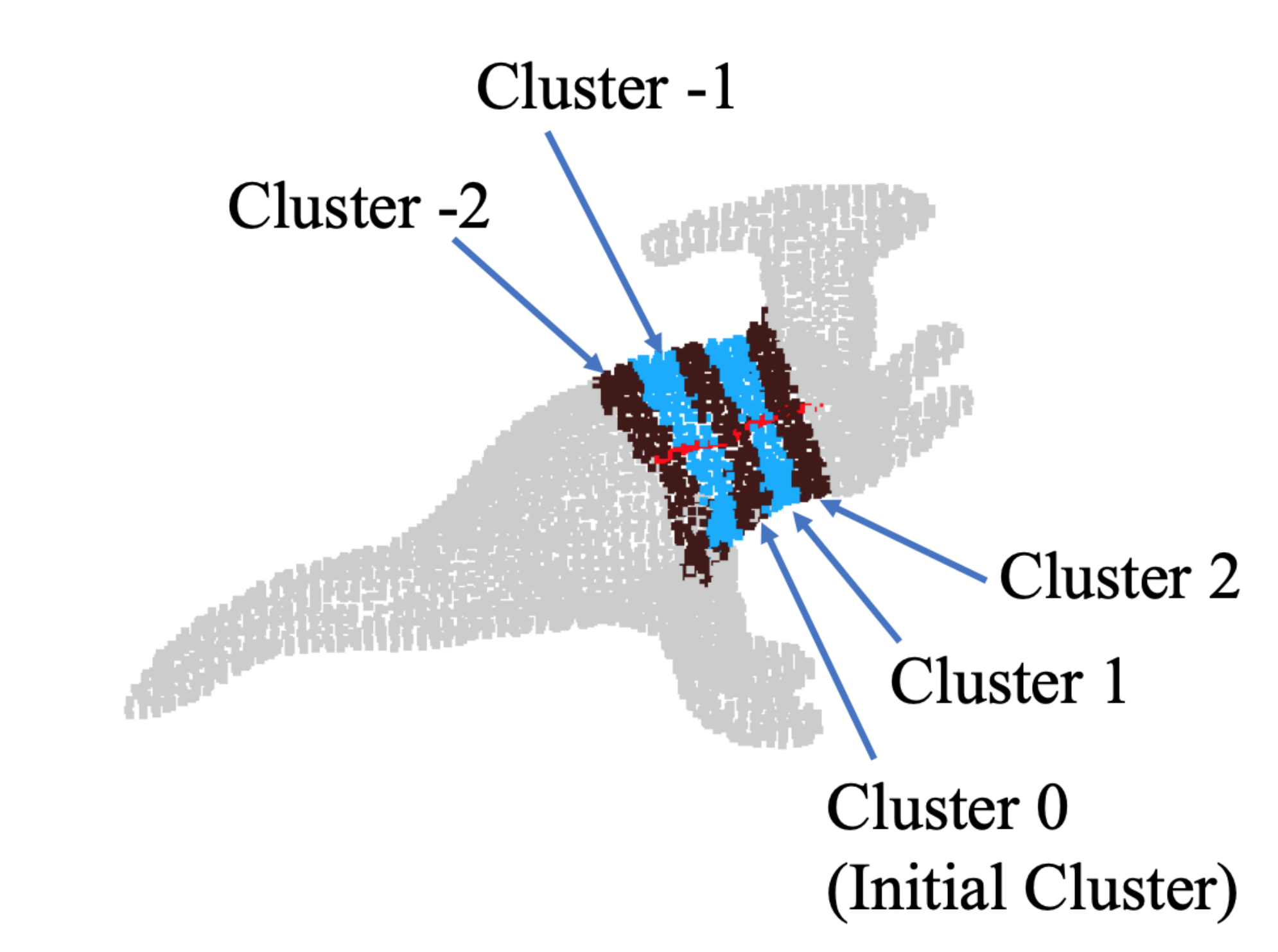}
\caption{Depiction of how a part is grown from an initial cross-sectional cluster (Cluster 0). Neighboring cross-sectional clusters are shown in alternating colors. The red curve represents the axis of the part.}
\label{fig:growing_stages}
\end{figure}

\begin{figure}[htp]
\centering
\begin{subfigure}[b]{0.4\textwidth}
	\includegraphics[width=\textwidth]{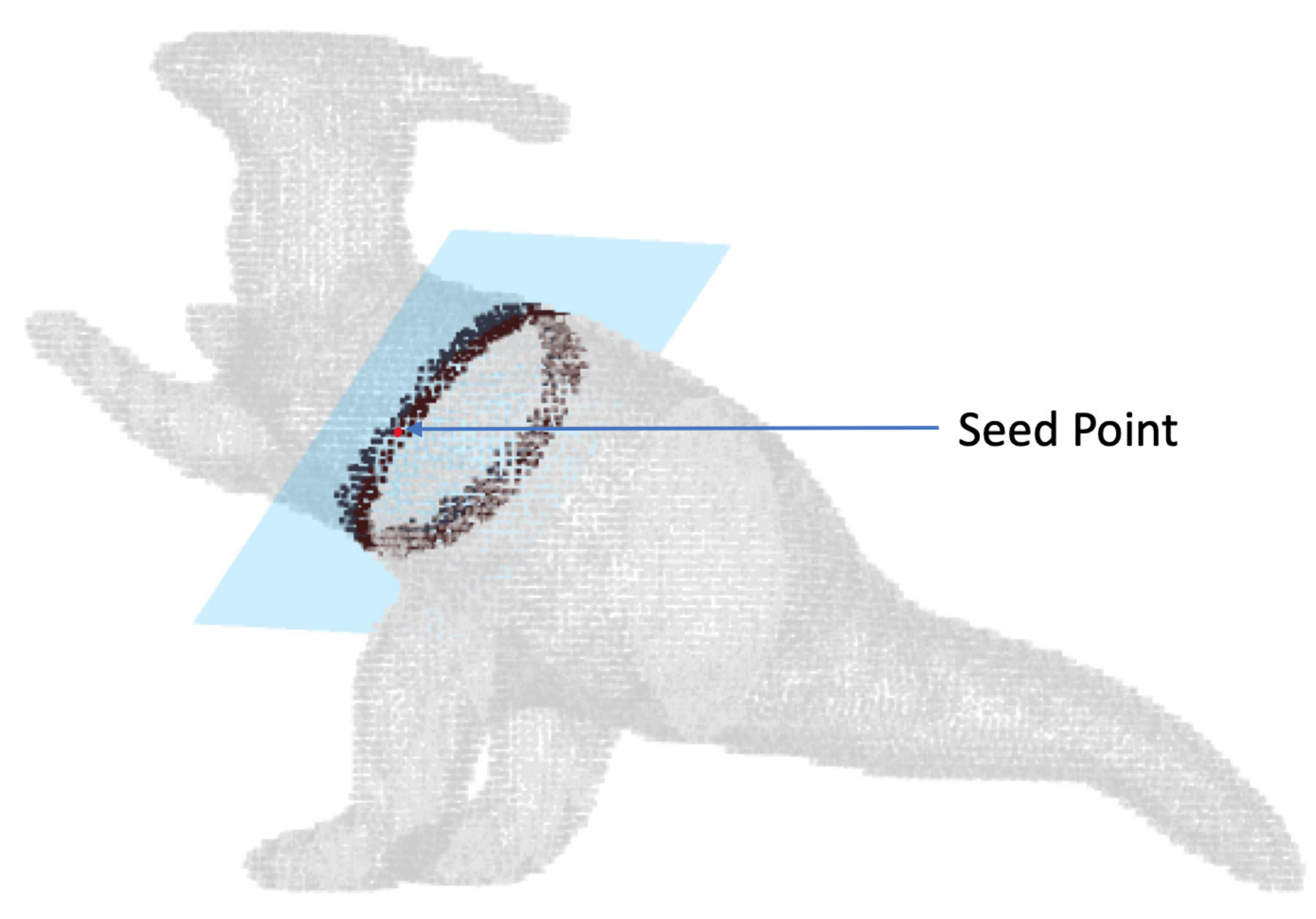}
	\caption{}
\end{subfigure}	
\begin{subfigure}[b]{0.44\textwidth}
\centering
	\includegraphics[width=\textwidth]{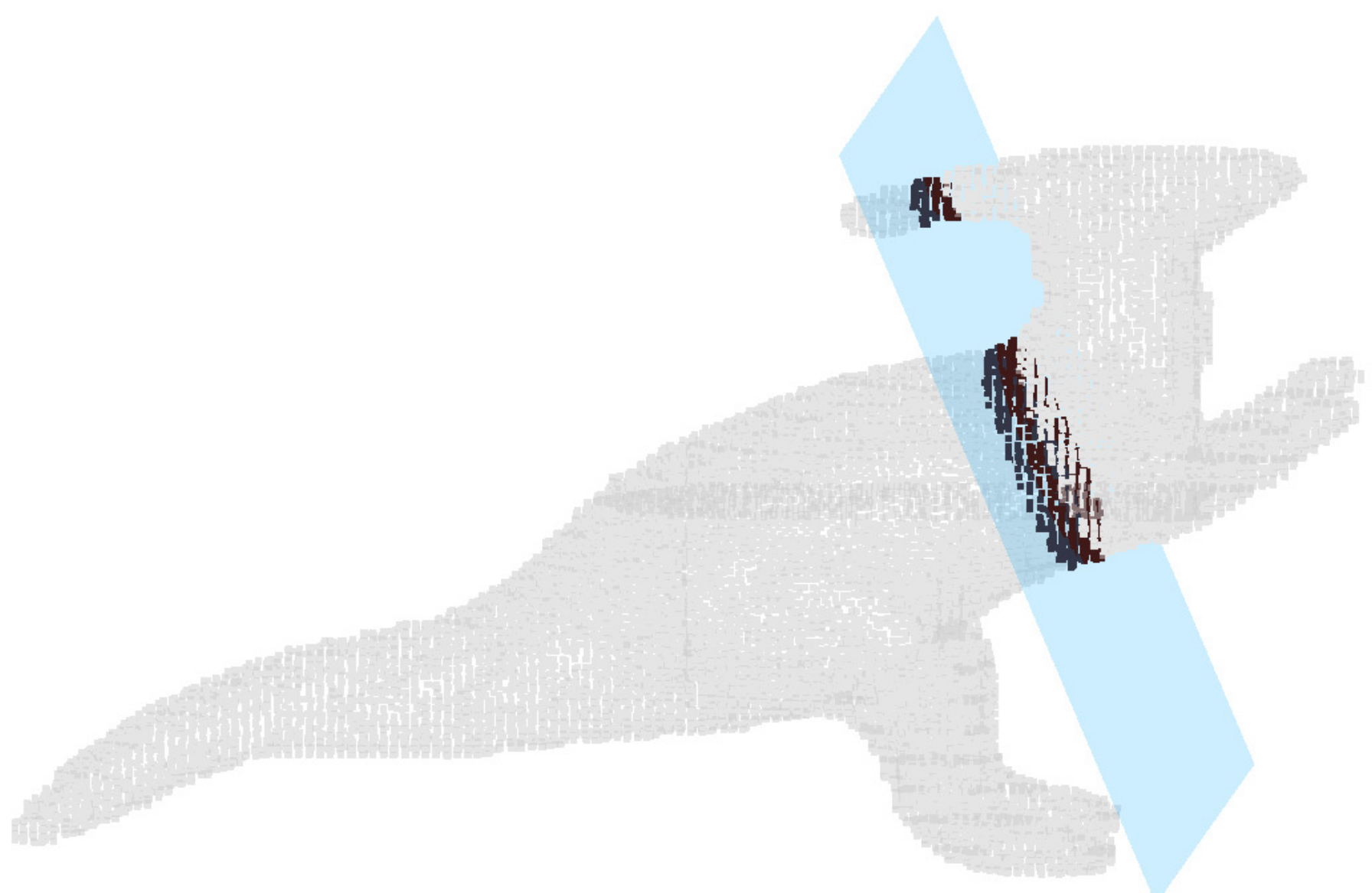}
	\caption{}
\end{subfigure}
\caption{(a) The cross-sectional plane is shown in blue, the thin cross-section associated with the plane is shown in brown and seed point is shown in red. (b) All points close to the cross-sectional plane, but not necessarily connected to the seed point.}
\label{fig:cross_section_demo}
\end{figure}

\subsection{Detecting Initial Cross-sections} \label{sec:init_cross-sections}
As stated earlier, parts are formed by scaled versions of a 2D cross-sectional contour translated along an axis. And since thin cross-sectional clusters serve as an approximation to the 2D contour of a part in a point cloud, we would like to detect them and then use them to grow parts. Given any point in the cloud, we need an algorithm to detect a thin cross-sectional cluster corresponding to that point. For instance, in Fig. \ref{fig:cross_section_demo} (a), given the red point, we need the algorithm to return the brown points. The red point is referred to as the seed point as it serves as the starting point for growing parts. As shown in  Fig. \ref{fig:cross_section_demo} (a), detecting the orientation of the cross-sectional plane (shown in blue) passing through the seed point would give us the thin cross-sectional cluster we are looking for. Because, once we estimate the orientation of the cross-sectional plane passing through the seed point, all points lying sufficiently close to that plane would represent the thin cross-sectional cluster. We need to consider only points lying close to the plane and at the same time connected to the seed point. As shown in Fig. \ref{fig:cross_section_demo} (b), considering all points near the plane that are not necessarily connected to the seed point could lead to the inclusion of some wrong points as part of the cross-section. Note that the approach described above is similar to that of Tagliasacchi et al. \cite{tagliasacchi2009curve}.

\begin{figure*}[htp]
\centering
\begin{subfigure}[b]{0.36\textwidth}
	\includegraphics[width=\textwidth]{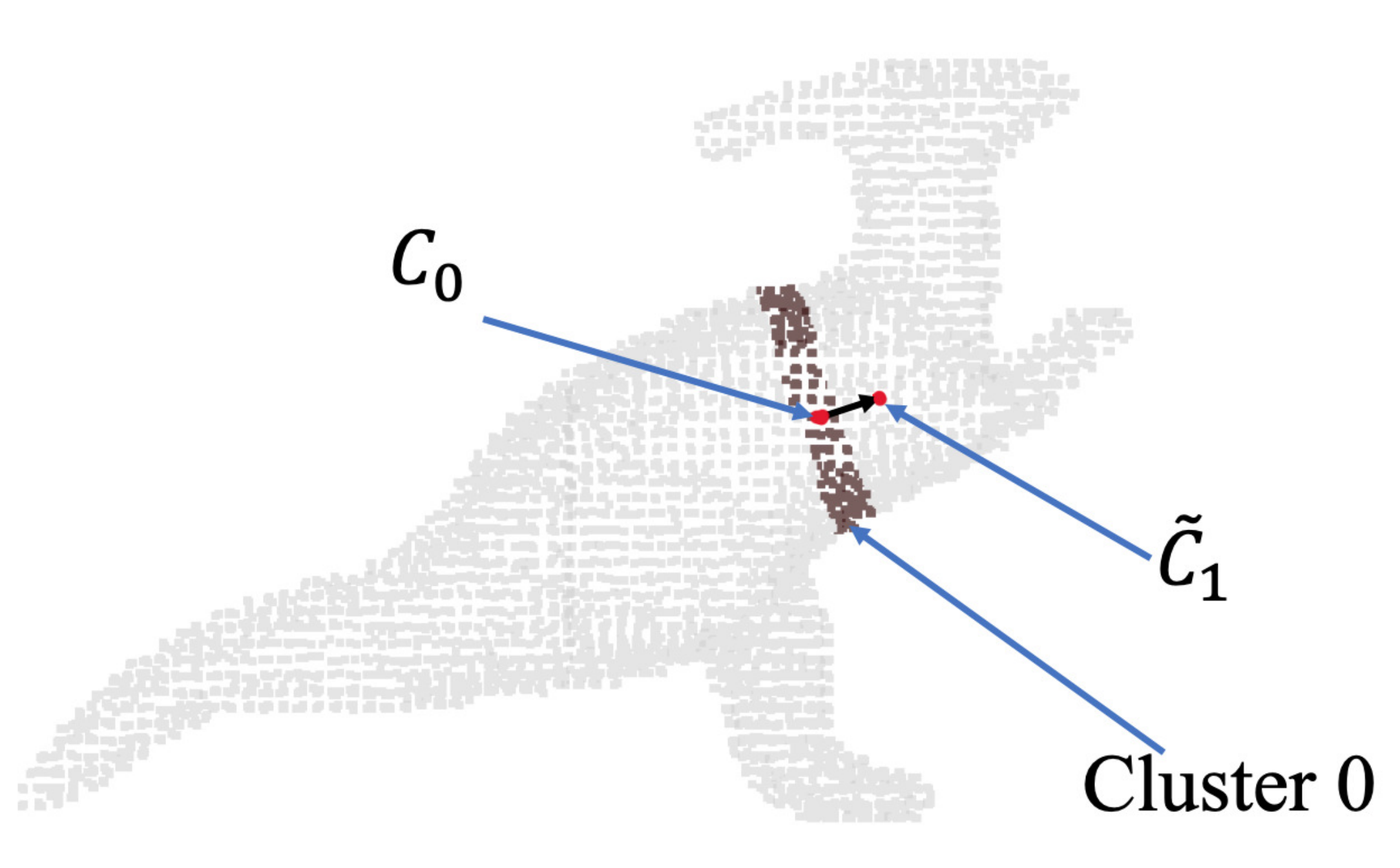}
	\caption{}
\end{subfigure}	
\begin{subfigure}[b]{0.42\textwidth}
\centering
	\includegraphics[width=\textwidth]{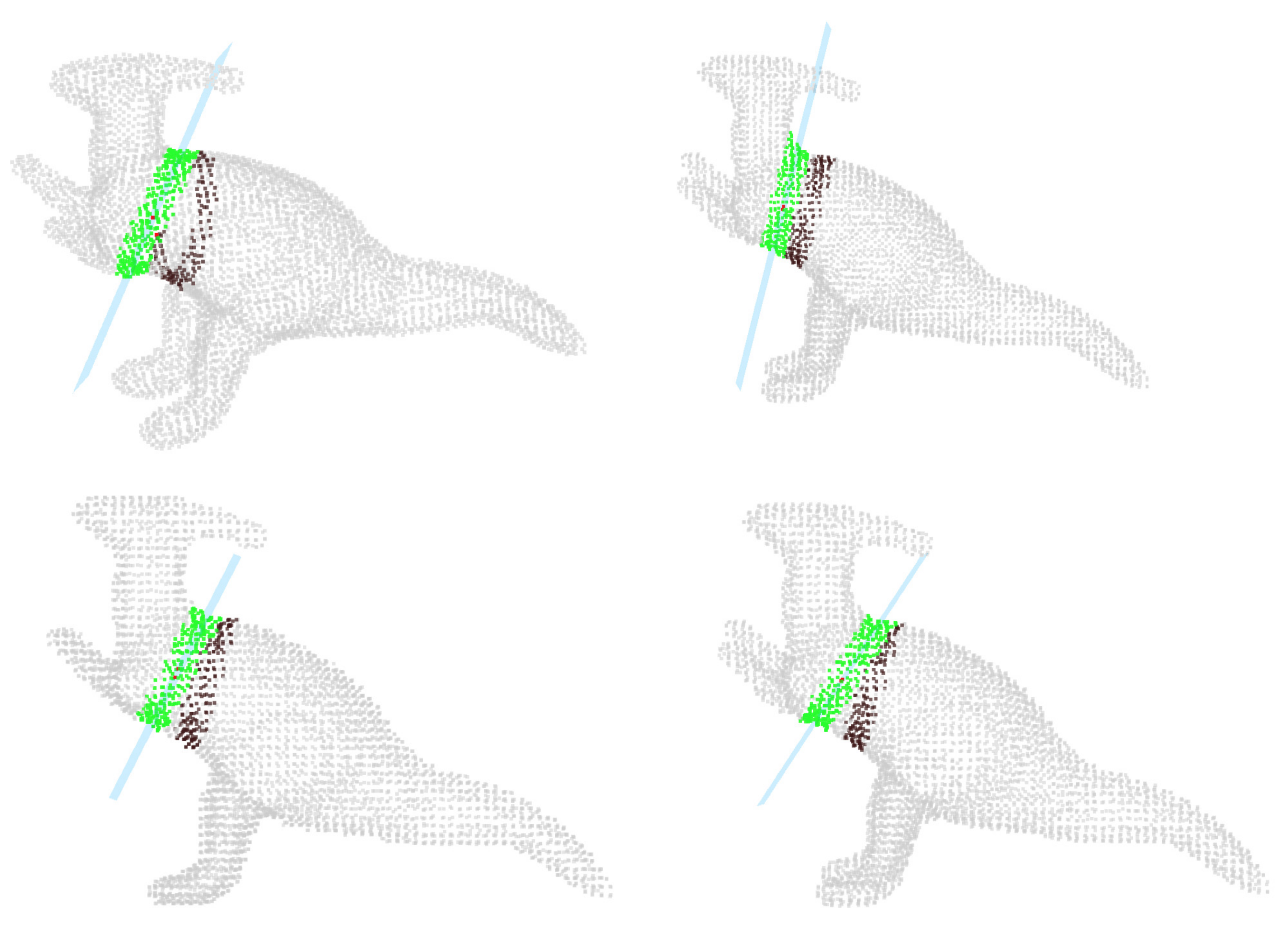}
	\caption{}
\end{subfigure}
\begin{subfigure}[b]{0.21\textwidth}
\centering
	\includegraphics[width=\textwidth]{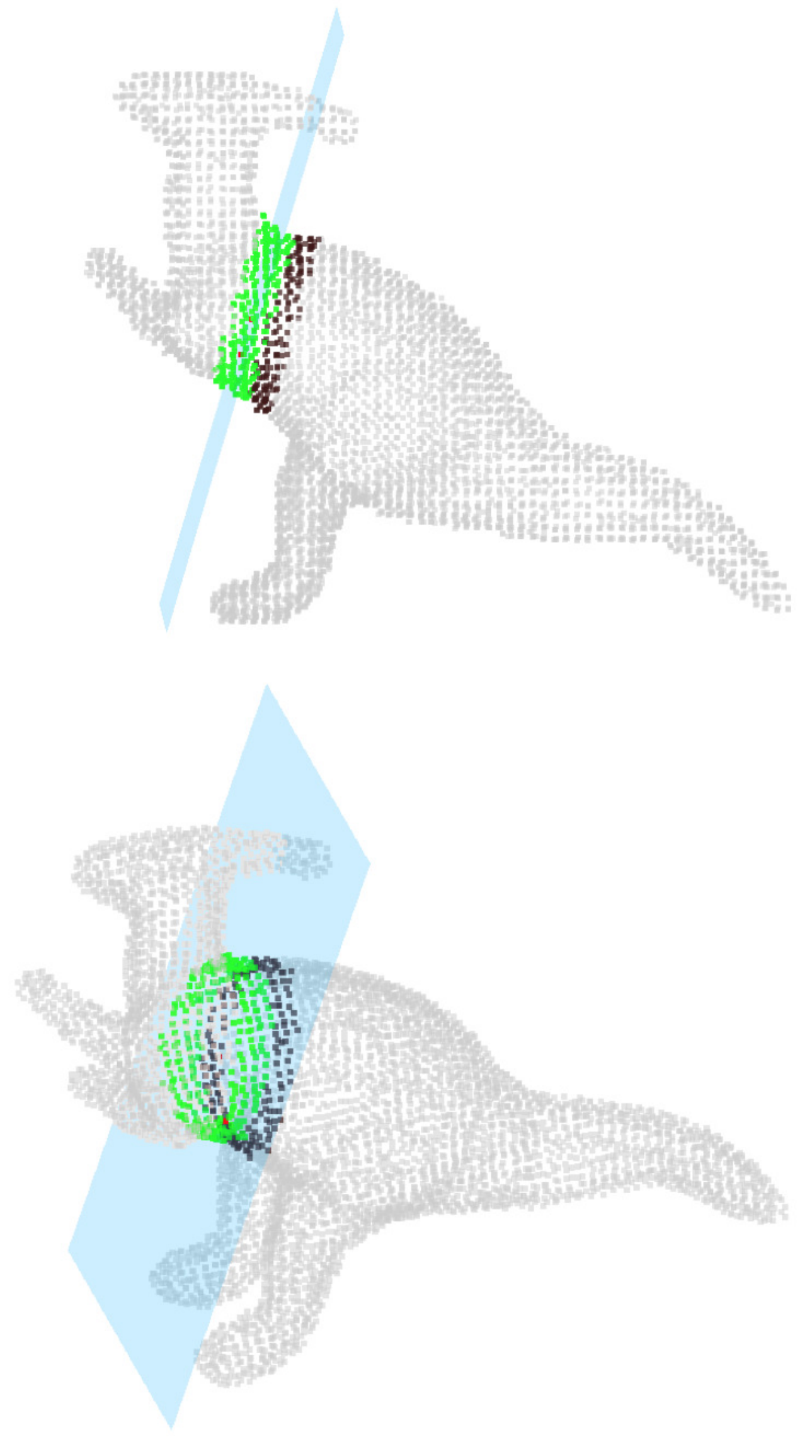}
	\caption{}
\end{subfigure}
\caption{Growing parts by method 1. (a) Step 1: take a small step along the normal of cross-sectional plane of Cluster 0  to obtain an estimate of the neighboring axis point $\widetilde{\mathbf{C}}_1$. (b) Step 2: Consider a set of planes (planes in $\mathbf{A}_{\theta} \bigtimes \mathbf{A}_{\phi}$) whose orientation is close to the orientation of the cross-sectional plane of Cluster 0. Assign a cost to each of these planes using the same cost function as the one in algorithm \ref{alg:cross-sectional_plane}. Only four planes (blue color) in the set $\mathbf{A}_{\theta} \bigtimes \mathbf{A}_{\phi}$ are shown. The green points represent inliers of the corresponding plane. (c) Step 3: choose the plane from the set  $\mathbf{A}_{\theta} \bigtimes \mathbf{A}_{\phi}$ which minimizes the cost computed. This plane represents the cross-sectional plane of the adjacent cross-section and the inlier set of this plane represents the adjacent cross-sectional cluster (Cluster 1 in our example). Top and bottom shows two views of the chosen plane.}
\label{fig:method1}
\end{figure*}

We need to detect initial cross-sectional clusters at multiple locations on the point cloud to be able to extract all the parts. As mentioned before, in the example of the dinosaur in Fig. \ref{fig:growing_stages}, we need at least nine initial cross-sections. Therefore, the point cloud is clustered into $M$ clusters using k-means clustering \cite{scikit_learn}. These clusters would determine the locations at which we would detect thin cross-sectional clusters from which candidate parts are grown. The idea is to pick one seed point per cluster and then find the thin cross-section to which the seed point belongs and finally grow parts out of this thin cross-section. For each cluster, we pick the point in the cluster which is closest to the k-means cluster center as the seed point. To find the cross-sectional plane that passes through a seed point, $\mathbf{s}^i$, we do an exhaustive search. Algorithm \ref{alg:cross-sectional_plane} shows the steps involved in searching for a cross-sectional plane that passes through a seed point. Note, the algorithm takes as input the connectivity matrix defined on the clusters, $\mathbf{G}_{clust}$. If any member of one cluster is connected to any member of the other cluster, according to $\mathbf{G}_{cnct}$, then two clusters are considered to be connected. Spherical coordinates are used to sample the unit sphere at regular angular intervals to obtain the normal directions for the planes. An angular interval of $\pi/6$ is used to sample both the zenith angle $\phi$ and the azimuthal angle $\theta$. Note, all the planes pass through the seed point, so picking a normal direction for the plane fully defines it. For each plane an inlier set,  $\mathbf{B}$, is  first computed. This inlier set, as shown in Fig. \ref{fig:cross_section_demo} (a),  contains all points that lie close to the plane and are connected to the seed point. A cluster dependent distance threshold, $\subsup{0}{\delta}{i}{pd} = median \big( \{ \subsup{0}{d}{j}{max} \mid \mathbf{x}^j \in \ $cluster$\ i \} \big)$, is used to determine if a point is considered close enough to a plane. We use the adaptive threshold derived earlier in section \ref{sec:locally_adaptive_threshold} here to decide which points lie close to the plane. 

A cost is then assigned to each plane. The cost, represented by $c(\theta,\phi)$, is based on the defenition of the part and measures the average length of projection of the point normals of the points in $\mathbf{B}$ on to the normal of the plane. Ideally, the point normals will be perpendicular to the normal of the cross-sectional plane and therefore will have a zero length projection on the normal of the cross-sectional plane. As shown in Fig. \ref{fig:gc_illustration} (c), the cross-sectional plane (shown in yellow) contains the contour. The normals to the contour lie on the plane and hence are parallel to the plane and have projection length of zero on to the normal of the plane. This implies that the plane that minimizes this cost, for a given seed point, would be the best estimate for the cross-sectional plane for that seed point. After the costs associated with each plane is computed the plane that minimizes this cost is selected as the cross-sectional plane. The inlier set of that plane defines the thin cross-section.

 \begin{algorithm}
 \caption{Algorithm for finding the cross-sectional plane}
 \label{alg:cross-sectional_plane}
 \begin{algorithmic}[1]
 \renewcommand{\algorithmicrequire}{\textbf{Input:}}
 \renewcommand{\algorithmicensure}{\textbf{Output:}}
 \Require
\Statex Point cloud : $\mathbf{X}$
\Statex Connectivity graph on $\mathbf{X}$: $\mathbf{G}_{cnct}$
\Statex Connectivity graph on the clusters: $\mathbf{G}_{clust}$
\Statex Distance Threshold: $\subsup{0}{\delta}{i}{pd}$
\Statex Seed point of $i^{th}$ cluster: $\mathbf{s}^i = (\subsup{1}{s}{i}{p},\subsup{1}{s}{i}{n})$
 \Ensure  $\theta,\phi$

 \Function{get\_inliers}{$\mathbf{X}, \mathbf{G}_{cnct}, \protect\subsup{0}{\delta}{i}{pd},  \mathbf{s}, \mathbf{\hat{n}}$}
\State $d \gets \mathbf{\hat{n}} \boldsymbol{\cdot} \subsup{1}{s}{i}{p} $ \Comment{ \Quote{$\boldsymbol{\cdot}$} represents dot product}
\Statex \hspace{1.25em} $\mathbf{A}$ is the set of points close to the plane
\Statex  \hspace{1.25em}  abs(): absolute value
\State $\mathbf{A} \gets \{ \mathbf{x}^i \ \mid$ abs(\ $\mathbf{\hat{n}} \boldsymbol{\cdot} \subsup{1}{x}{i}{p}  - d$\ ) $\leq \subsup{0}{\delta}{i}{pd} \} $
\State $\mathbf{B} \gets \{ \mathbf{x}^i \in \mathbf{A}  \mid \mathbf{x}^i $ is connected to $\mathbf{s}$ as per $\mathbf{G}_{cnct} \}$
\State \Return{$\mathbf{B}$}
 \EndFunction
 
 \Statex

\State $\mathbf{H} \gets \{ \mathbf{s}^j \mid$ cluster $j$ neighbor of cluster $i$ as per $ \mathbf{G}_{clust} \}$
\For {$\theta  \in \{0,\pi/6,2\pi/6, \cdots, 11\pi/6 \} $}
\For {$\phi \in \{0,\pi/6,\pi/3, \pi/2 \} $}
\State $\mathbf{\hat{n}} \gets [cos(\theta) sin(\phi), sin(\theta)sin(\phi), cos(\phi)]$
 \State $\mathbf{B} \gets$ \Call{get\_inliers}{$\mathbf{X}, \mathbf{G}_{cnct}, \protect\subsup{0}{\delta}{i}{pd},  \mathbf{s}^i, \mathbf{\hat{n}}$}\
 \State $c(\theta,\phi) \gets $mean$ \big( \{ $abs($ \mathbf{\hat{n}} \boldsymbol{\cdot} \subsup{1}{x}{i}{n} $)$ \ \mid \mathbf{x}^i \in \mathbf{B} \} \big)$
\EndFor
\EndFor
\Statex
\Return $\underset{\theta,\phi}{\arg\min} \ c(\theta,\phi) $
\end{algorithmic} 
\end{algorithm}

%
%
%

\begin{figure*}[htp]
\centering
\begin{subfigure}[b]{0.36\textwidth}
	\includegraphics[width=\textwidth]{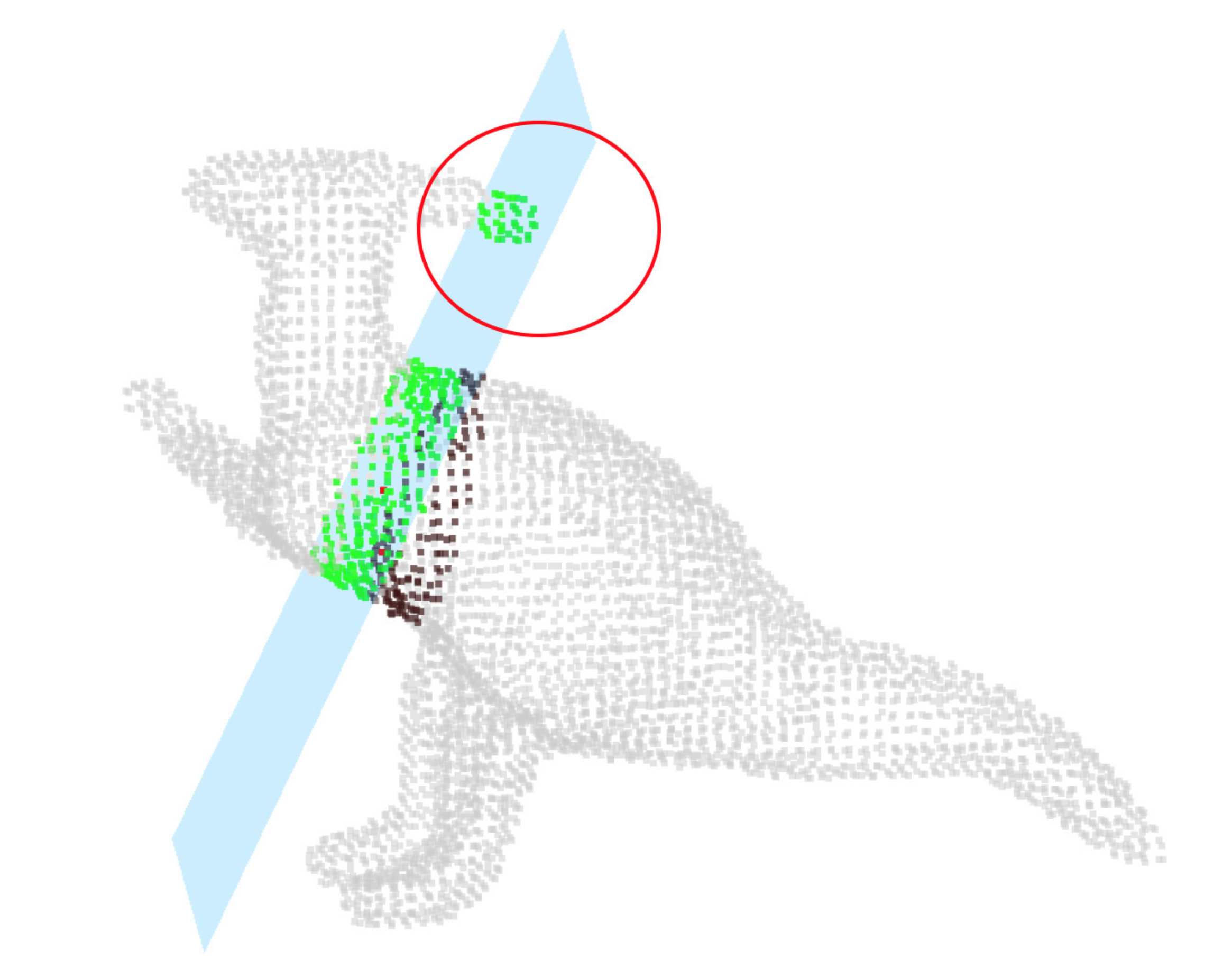}
	\caption{}
\end{subfigure}	
\begin{subfigure}[b]{0.34\textwidth}
\centering
	\includegraphics[width=\textwidth]{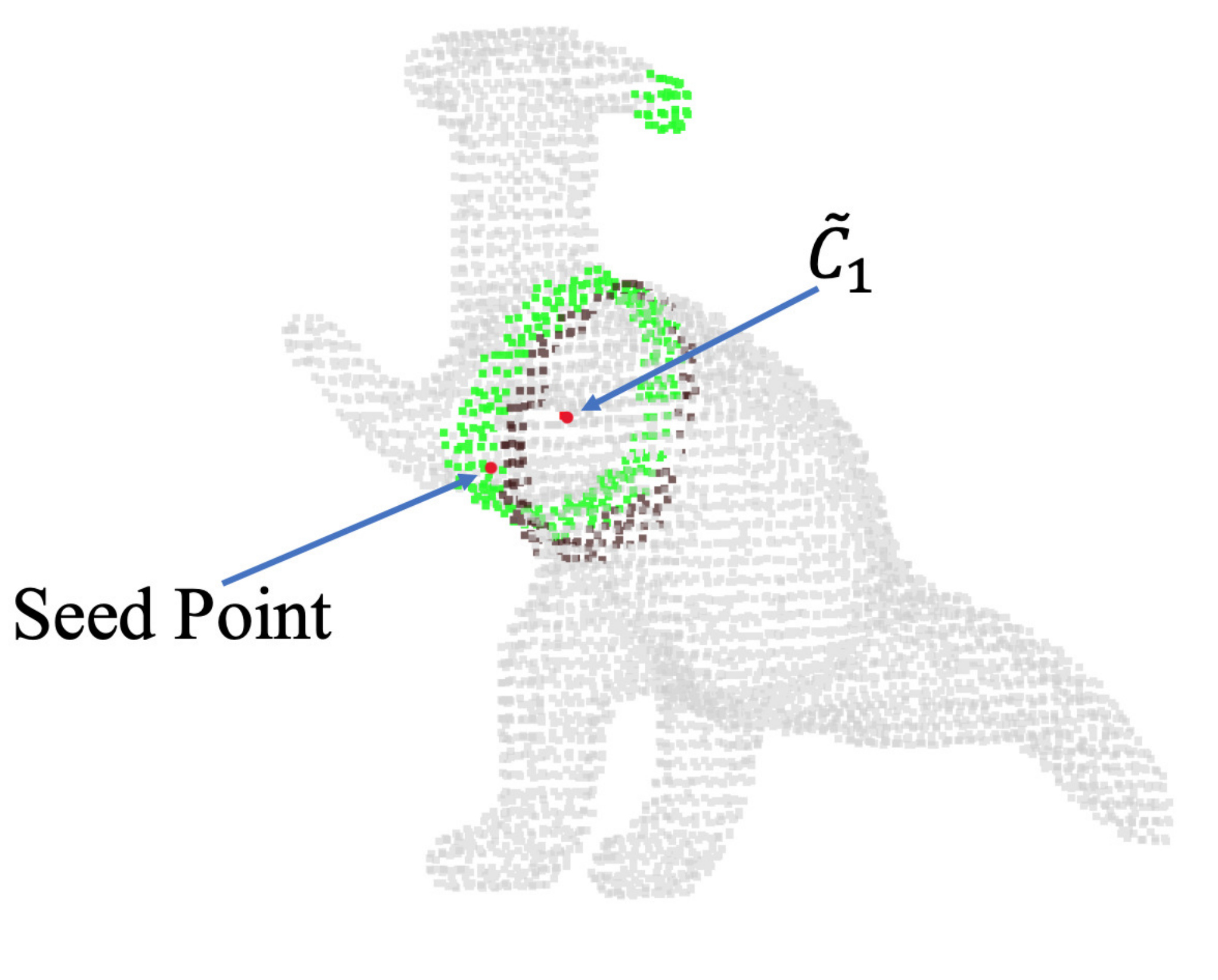}
	\caption{}
\end{subfigure}
\begin{subfigure}[b]{0.28\textwidth}
\centering
	\includegraphics[width=\textwidth]{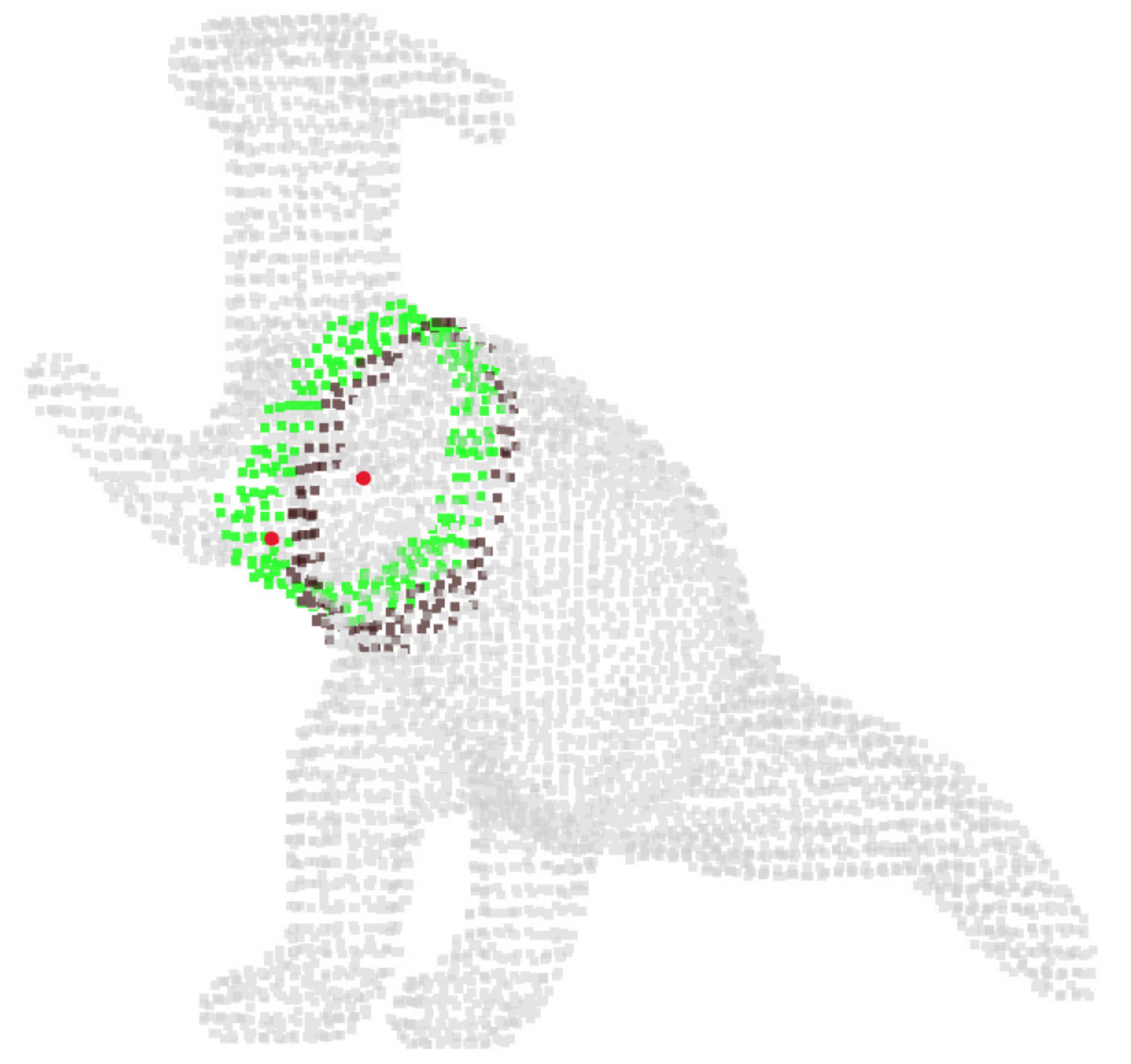}
	\caption{}
\end{subfigure}
\caption{Identifying a seed point. (a) All points lying close to the plane are shown in green. The brown points represent members of cluster 0. The points within the red circle are unwanted points, not part of the true cross-sectional cluster. (b) A seed point is identified as the point closest to $\widetilde{\mathbf{C}}_1$, from among the points lying close to the plane. The seed point and $\widetilde{\mathbf{C}}_1$ are shown in red. (c) Considering only points connected to the seed point removes the unwanted points and gives us the right cross-sectional cluster.}
\label{fig:nxt_seed_point}
\end{figure*}

\subsection{Growing Parts} \label{sec:growing_parts}
In the previous step, we identified $M$ initial cross-sectional clusters. Starting from these initial cross-sectional clusters, parts need to be grown. Two methods are used to grow parts from initial cross-sections. Both methods are again based on the definition of a part as a GC. We would refer to these methods as method 1 and method 2 for clarity. While method 1 is very simple, method 2 is much more sophisticated.  Before we describe the methods, we would describe three assumptions that these methods make. The first assumption is that the scaling function of the GC is smooth and the second assumption is that the 3D axis of the GC is smooth. Note that we are looking for matching cross-sectional cluster for the initial cross-sectional cluster in its neighborhood to grow a part. In this context, the first assumption implies that there would not be a sudden jump in the scale of a  cross-sectional cluster compared to its neighboring cross-sectional cluster. And the second assumption would mean that the orientation of the cross-sectional planes of neighboring cross-sectional clusters would be similar. As an example, in Fig. \ref{fig:gc_illustration} (c), the two yellow planes represent cross-sectional planes of two close-by cross-sections and therefore have similar orientation. Also, note that the scale function in Fig. \ref{fig:gc_illustration} (a) is smooth. The third assumption that both methods make is that the mean of all the points in a cross-sectional cluster of a part represents a point on the axis of the part. For instance, in Fig. \ref{fig:cross_section_demo} (a), if we take the centre of mass of all brown points, we would obtain a point in the interior of the torso of the dinosaur. The torso of the dinosaur is a GC and the centre of mass of the points in the cross-sectional cluster (shown in brown) is assumed to lie on the axis of this GC. We refer to the centre of mass of points in the cluster as the cluster center. The axis of the torso, shown in red in Fig. \ref{fig:growing_stages}, is obtained by joining the cluster centers of the blue and brown clusters. While this assumption is not generally true, we make such a simplifying assumption to estimate the axis points, given a cross-sectional cluster. We will demonstrate later that this assumption is a reasonable assumption for many real-world parts. The following sections will explain the two methods in detail.

\subsubsection{Method 1}
%



As stated above, this is a simple method for growing parts from initial cross-sectional clusters. Keep in mind that, given an initial cross-sectional cluster, we are looking for the neighboring cross-sectional cluster to grow the part. For instance, in Fig. \ref{fig:growing_stages}, we are looking for cluster 1, given the initial cluster, cluster 0. Fig. \ref{fig:method1}, shows the main steps in this method. As depicted in the figure, the first step is to obtain a rough estimate of the cluster center of the neighboring cluster, cluster 1. Let $\mathbf{C}_0$ denote the centre of cluster 0 and the normal of the cross-sectional plane corresponding to cluster 0 be denoted by $\hat{\mathbf{n}}_0$. By definition, the cross-sectional plane of cluster 0 is perpendicular to the axis of the GC (the torso of the dinosaur being the GC here) at $\mathbf{C}_0$. Or in other words, $\hat{\mathbf{n}}_0$ is the tangent to the axis of the GC at $\mathbf{C}_0$. Therefore, taking a small step along $\hat{\mathbf{n}}_0$ would give us a rough estimate of the axis point corresponding to cluster 1. Keep in mind that the tangent of a curve provides a linear approximation to the curve locally. Let this estimate of the neighboring axis point be denoted by $\widetilde{\mathbf{C}}_1$. I.e., $\widetilde{\mathbf{C}}_1 = \mathbf{C}_0 + \delta_{step} \hat{\mathbf{n}}_0$, where, $\delta_{step}$ represents a small step size. Fig. \ref{fig:method1} (a), depicts this process.

In the next step we estimate the orientation of the cross-sectional plane through $\widetilde{\mathbf{C}}_1$. An exhaustive, but local search is used to estimate the orientation of the cross-sectional plane. The exhaustive search is identical to the one described in algorithm \ref{alg:cross-sectional_plane}, but here we only consider planes whose normal orientation is close to $\hat{\mathbf{n}}_0$. This is because, according to the second assumption, the orientation of the cross-sectional plane of cluster 1 is close to the orientation of the cross-sectional plane of cluster 0 (because the 3D axis is smooth). To obtain planes whose orientation is close to $\hat{\mathbf{n}}_0$, an angle range, $\Delta_{ang}$, and a step number, $k_{step}$, is first chosen. Let $\theta_0$ and $\phi_0$ represent the azimuth and the zenith angles corresponding to $\hat{\mathbf{n}}_0$. Let $\mathbf{A}_{\theta}$ represent  $k_{step}$ equally spaced angles in the range $(\theta_0 - \Delta_{ang}, \theta_0 + \Delta_{ang} )$ and $\mathbf{A}_{\phi}$ represent $ k_{step}$ equally spaced angles in the range $(\phi_0 - \Delta_{ang}, \phi_0 + \Delta_{ang} )$. All planes represented by the cartesian product, $\mathbf{A}_{\theta} \bigtimes \mathbf{A}_{\phi}$, are considered. We use $\Delta_{ang} = 12.5$ and $k_{step} = 3$.  This ensures that all the planes considered have normals which are similar in orientation to $\hat{\mathbf{n}}_0$. The best plane is chosen by using the same criterion as the one described in algorithm \ref{alg:cross-sectional_plane}. I.e., for each plane in the set $\mathbf{A}_{\theta} \bigtimes \mathbf{A}_{\phi}$, we compute its inlier set (see algorithm \ref{alg:cross-sectional_plane}) and then choose the plane which minimizes the average length of projection of the point normals of the points in its inlier set on to the normal of the plane.

Considering all points lying close to a plane could lead to problems like the one shown in Fig. \ref{fig:nxt_seed_point} (a). To avoid this issue, we need to only consider points lying close to the plane and connected to a seed point. As shown in Fig. \ref{fig:nxt_seed_point} (b), the point closest to $\widetilde{\mathbf{C}}_1$, from amongst the points close to the plane (the green points), is chosen as the seed point. Choosing only those (green) points that are connected to the seed point would fix the issue of unwanted points being included in the cross-sectional cluster (Fig. \ref{fig:nxt_seed_point} (c)).

The steps described above can be used to find the neighboring cross-sectional cluster, cluster 1, given the initial cluster. This process can be repeated to obtain the neighboring cross-sectional cluster, cluster 2 in Fig. \ref{fig:growing_stages}, of cluster 1. There needs to be a stopping criterion for this process. Or in other words, there needs to be some criterion that can be checked to see if we have reached the end cluster of a part. Here is where the first assumption, mentioned earlier, comes into play. Since, the scale function of a part is smooth, the two neighboring cross-sections would have similar scale factors. A sudden jump in the scale factor of a neighboring cross-section would indicate that we have reached a junction. For example, in Fig. \ref{fig:growing_stages}, the search for the neighboring cross-sectional cluster for cluster 2 would fail as all the potential clusters considered would have a substantially different scale factor. 

A very simple metric is employed to measure the scale factor of a cross-sectional cluster. A local coordinate system is constructed, using PCA, for the cluster. Let the axes of this coordinate system be represented by $\mathbf{a}_1,\mathbf{a}_2$ and $\mathbf{a}_3$, with $\mathbf{a}_1$ representing the maximum variance direction and $\mathbf{a}_3$ representing the direction with the least variance. The axis $\mathbf{a}_3$, would have an orientation that is very close to the normal of the cross-sectional plane of the cross-sectional cluster because the cross-sectional cluster is thin and close to planar. Let the eigenvalues corresponding to axes $\mathbf{a}_1$ and $\mathbf{a}_2$ be $e_1$ and $e_2$ respectively. These eigenvalues serve as a good representation for the scale of the cluster. I.e., bigger cross-sectional clusters would have larger $e_1$ and $e_2$ values and vice-versa. Let $i$ and $j$ represent two neighboring cross-sections. We treat the combination of the two eigenvalues, $[e_1, e_2]$, as a vector. I.e., for cross-section $i$, we have a vector $[e^i_1, e^i_2]$ representing its scale. Similarly the vector $[e^j_1, e^j_2]$, represents the scale of cross-section $j$. We stop growing parts, if $ abs\big(1 - \frac{d_{eucl}( [e^i_1, e^i_2], [e^j_1, e^j_2])}{\supnorm*{[e^i_1, e^i_2]} } \big) > \Delta_{eg} $, where $d_{eucl}( [e^i_1, e^i_2], [e^j_1, e^j_2])$ represents the euclidean distance between the vectors, $abs()$ represents the absolute value and $\Delta_{eg}$ represents an acceptable threshold for scale difference between neighboring clusters. The fraction is measuring the percentage change in the scale between the cross-sectional clusters. A value of one for the fraction would mean that the size of the cross-section did not change (according to our metric). If the percentage change in scale of the next cross-sectional cluster compared to the current cross-sectional cluster is greater than a threshold, it implies that we have reached a junction and the growing process stops. The growing process also terminates if there are no further points to be added to the part. For instance, the growing process of the tail of the dinosaur halts in one of the directions due to this reason. The growth of the tail stops in the other direction because the junction where the legs meet the tail is detected. 

As mentioned earlier, this method is a very simple one in which we depend on the smoothness of the scale function to grow parts. The termination criterion for the growth process as well as the metric to compute the scale of a cross-sectional cluster are also simple. In the next section, we describe a sophisticated method, which we refer to as method 2, to grow parts which directly takes inspiration from the process in which GCs are formed from its components (scale function, cross-sectional contour and 3D axis). Though we use method 2 as the primary method to grow parts, there are scenarios in which method 1 would be more suitable. Such scenarios will also be pointed out in the following sections.

\subsubsection{Method 2} \label{sec:method_2}
Given a scale function, cross-sectional contour and 3D axis, a GC is formed by a similarity transformation (i.e., rigid transformation and uniform scaling) of the contour. I.e., the action referred to as sweeping the 2D contour along an axis, consists of three transformations (translation, rotation and uniform scaling) applied to the 2D contour. This implies that the relationship between adjacent cross-sections is defined by a similarity transformation. In method 2, like in method 1, parts are grown, starting from an initial cross-sectional cluster (like cluster 0 in Fig. \ref{fig:growing_stages}), by finding its neighboring cross-sectional cluster. Since, the relationship between neighboring clusters is defined by a similarity transformation, one way to search for a similar cluster of points is to do registration between the initial cross-section and the points in its neighborhood. I.e., through registration, we can estimate the similarity transform that relates the two neighboring cross-sections. In Fig. \ref{fig:growing_stages}, registration between cluster 0 and the points in its neighborhood would lead to the identification of cluster 1 as its neighbor. Similarly registration between cluster 1 and its neighborhood points would identify cluster 2 and so on.

\begin{figure*}[htp]
\centering
	\includegraphics[width=\textwidth]{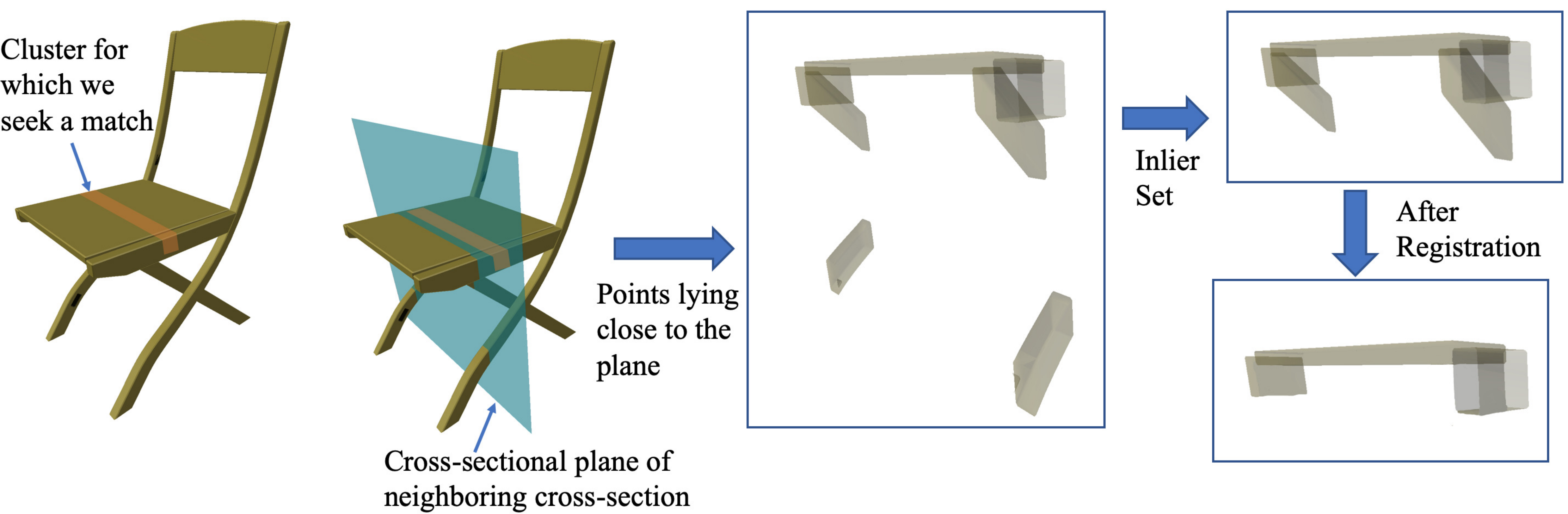}
\caption{Registration can help in removing some unwanted points from the neighboring cross-section.}
\label{fig:why_register}
\end{figure*}

Fig. \ref{fig:why_register}, demonstrates why method 2, which relies on registration to grow parts, is necessary to discover meaningful parts in the point cloud. The saffron colored portion represents the cross-section for which we seek to find a match. The plane shown in blue represents the actual cross-sectional plane of the neighboring cross-section.The points lying close to the cross-sectional plane, include many unwanted points as shown. Deriving a seed point (like in Fig. \ref{fig:nxt_seed_point} (b)) and looking for the inlier set (alll points connected to the seed point) will help remove some unwanted points like the portions of the chair's front legs. But portions of the chair's hind legs still remain in the cluster. Performing a registration would tell us that these points belonging to hind legs have no match in the cluster for which we seek a match. Such points for which we fail to find a match can be removed to exactly identify points belonging to the neighboring cross-section.


 \begin{algorithm}
 \caption{Algorithm to estimate parameters $\mathbf{R},s$, $\mathbf{t}$,$\alpha$ and $\sigma$}
 \label{alg:estimate_rst}
 \begin{algorithmic}[1]
 \renewcommand{\algorithmicrequire}{\textbf{Input:}}
 \renewcommand{\algorithmicensure}{\textbf{Output:}}
 \Require
\Statex Point set 1: $\mathbf{X}$
\Statex Point set 2: $\mathbf{Y}$

 \Ensure  $\mathbf{R},s$, $\mathbf{t}$, $\alpha$, $\sigma$
 
  \Statex
 
 \State Set {$\mathbf{R} = \mathbf{I}_3, s=1$ and $\mathbf{t} = (0,0,0)^T$ }
 
 \Statex
 
\While  {convergence not achieved}

 \Statex
\Statex \hspace{1.0 em} Compute $\mathbf{P}$:
 
 \Statex
 
\State $  p_{ji} = \frac{exp\big( - \frac{1}{2\sigma^2}  \supnorm*{\mathbf{x}_p^i -  s \mathbf{R} \ \subsup{1}{y}{j}{p} - \mathbf{t}}^2 + \alpha ({\mathbf{x}_n^i})^T \mathbf{R} \subsup{1}{y}{j}{n} \big)}{\sum\limits_{m=1}^M exp\big( - \frac{1}{2\sigma^2}  \supnorm*{\mathbf{x}_p^i -  s \mathbf{R} \ \subsup{1}{y}{m}{p} - \mathbf{t}}^2 + \alpha ({\mathbf{x}_n^i})^T \mathbf{R} \subsup{1}{y}{m}{n} \big)} $

 \Statex
 \Statex

\Statex \hspace{1.0 em}  Estimate the parameter values, $\theta$, using the above value of $\mathbf{P}$:

 \Statex

\State ${\hat{\mathbf{X}} = \mathbf{X}_p -  \frac{1}{N} \mathbf{1} \big( {\mathbf{X}_p^T\mathbf{P}\mathbf{1}} \big)^T}$  

\State ${\hat{\mathbf{Y}} = \mathbf{Y}_p -  \frac{1}{N} \mathbf{1} \big( s\mathbf{R} {\mathbf{Y}_p^T\mathbf{P}\mathbf{1}} \big)^T}$

\State $\mathbf{A} =  \hat{\mathbf{X}}^T \mathbf{P}^T \hat{\mathbf{Y}}$ 

\State $\mathbf{B} =  \mathbf{X}_n^T \mathbf{P}^T \mathbf{Y}_n $

\State ${K_1 = tr \big ( \hat{\mathbf{Y}}^T d \big( \mathbf{P} \mathbf{1}\big) \hat{\mathbf{Y}} \big) }$ 
\State ${K_2 = tr \big ( \hat{\mathbf{X}}^T d \big( \mathbf{P} \mathbf{1}\big) \hat{\mathbf{X}} \big) }$ 
\State Use the BFGS algorithm with the gradient equations defined in equation \ref{eq:final_gradients}, to compute optimal values for $\mathbf{R},s$, $\alpha$ and $\sigma$.

\Statex

\State $\mathbf{t} = \frac{1}{N} \big ( \mathbf{X}_p^T\mathbf{P}\mathbf{1} - s \mathbf{R} \mathbf{Y}_p^T\mathbf{P}\mathbf{1} \big )$

\Statex

\EndWhile \\
\Return $\mathbf{R},s$, $\mathbf{t}$, $\alpha$, $\sigma$
\end{algorithmic} 
\end{algorithm}


In \cite{Myronenko_2010_CPD}, Myronenko and Song present a solution to the registration problem by treating it as a probability density estimation problem. Note, though the authors call their method rigid registration, their method actually estimates the similarity transform that relates the two point clouds being registered. Myronenko and Song, however, only consider the location of the points being registered. In our case though, we have an oriented point set with each point having a position and an orientation (representing the local surface normal). As we demonstrate later, considering the orientation of the points is important in our case. Therefore, for the registration of two sets of oriented points, we modify the method in  \cite{Myronenko_2010_CPD} by drawing in ideas from \cite{Billings_2015_GIMLOP}. We first introduce some additional notation and then explain the registration process between two oriented sets of points in the next section. Later sections explain how this registration process can be employed to grow parts.

\subsubsection*{ Registartion of Two Oriented Point Sets} \label{sec:generic_registration}
Let $\mathbf{X} = \{\mathbf{x}^1,\mathbf{x}^2, \cdots, \mathbf{x}^N \}$ and $\mathbf{Y} = \{\mathbf{y}^1,\mathbf{y}^2, \cdots, \mathbf{y}^M \}$ represent the two point sets to be registered. As mentioned before, $\mathbf{x}^i = (\subsup{1}{x}{i}{p},\subsup{1}{x}{i}{n})$, where  $\subsup{1}{x}{i}{p}$ represents the location and $\subsup{1}{x}{i}{n}$ represent the point normal of $\mathbf{x}^i $. Similarly, $\mathbf{y}^i = (\subsup{1}{y}{i}{p},\subsup{1}{y}{i}{n})$. Let $\mathbf{X}_p = (\subsup{1}{x}{1}{p},\subsup{1}{x}{2}{p}, \cdots, \subsup{1}{x}{N}{p} )^T$ be a $N \times 3$ matrix representing the positions and $\mathbf{X}_n = (\subsup{1}{x}{1}{n},\subsup{1}{x}{2}{n}, \cdots, \subsup{1}{x}{M}{n} )^T$ be a $M \times 3$ matrix representing the point normals of $\mathbf{X}$. Similarly, let $\mathbf{Y}_p$ and $\mathbf{Y}_n$ represent the positions and point normals of points in $\mathbf{Y}$ respectively. The diagonal matrix formed from a vector $\mathbf{v}$ is represented by $diag(\mathbf{v})$. A column vector of all ones is represented by $\mathbf{1}$ and a $P \times P$  identity matrix is represented by $\mathbf{I}_P$. The trace of a matrix $\mathbf{A}$ is represented by ${tr(\mathbf{A})}$.

\begin{figure*}[htp]
\centering
\begin{subfigure}[b]{0.33\textwidth}
	\includegraphics[width=\textwidth]{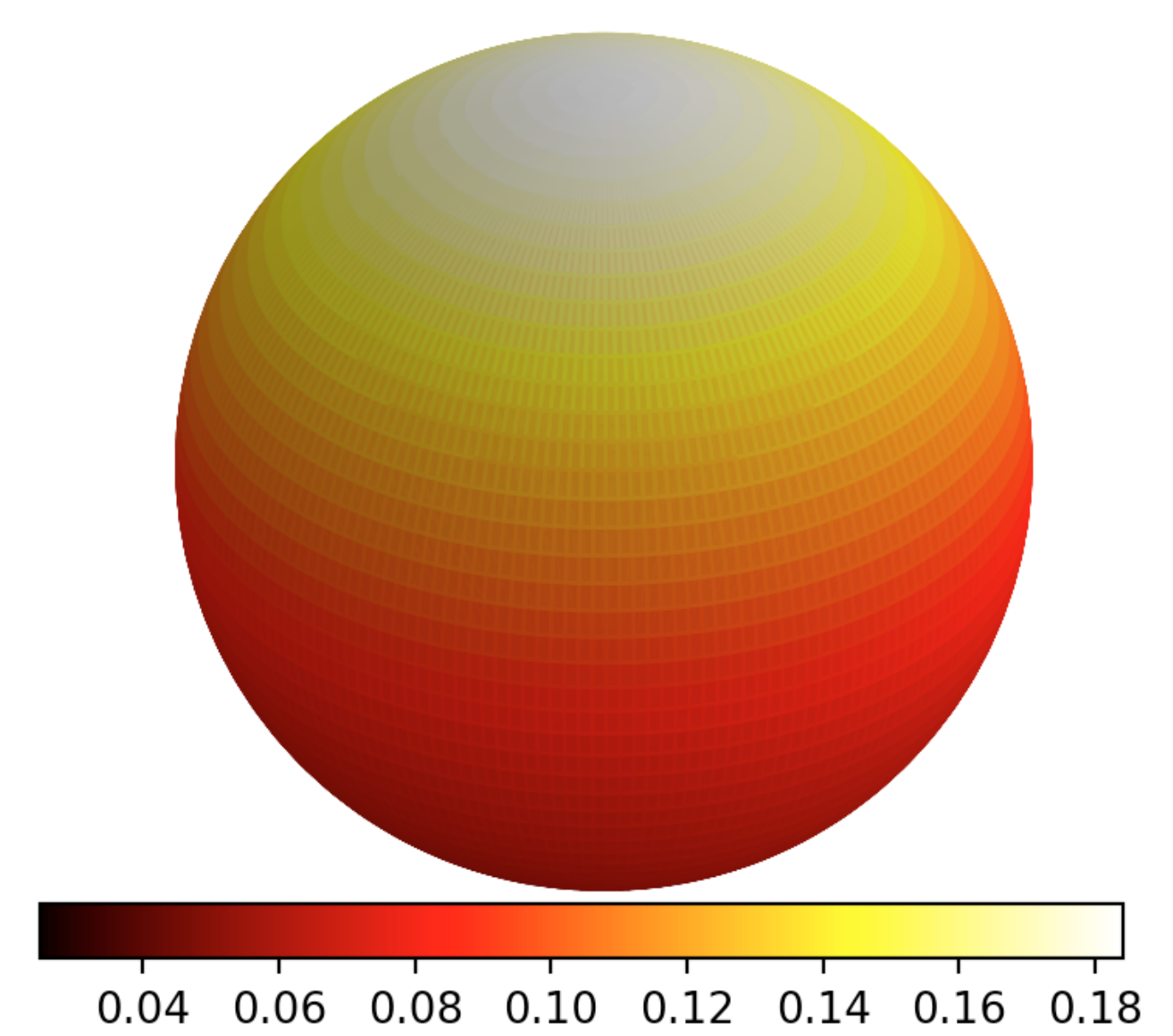}
	\caption{$\alpha$ = 1}
\end{subfigure}	
\begin{subfigure}[b]{0.33\textwidth}
\centering
	\includegraphics[width=\textwidth]{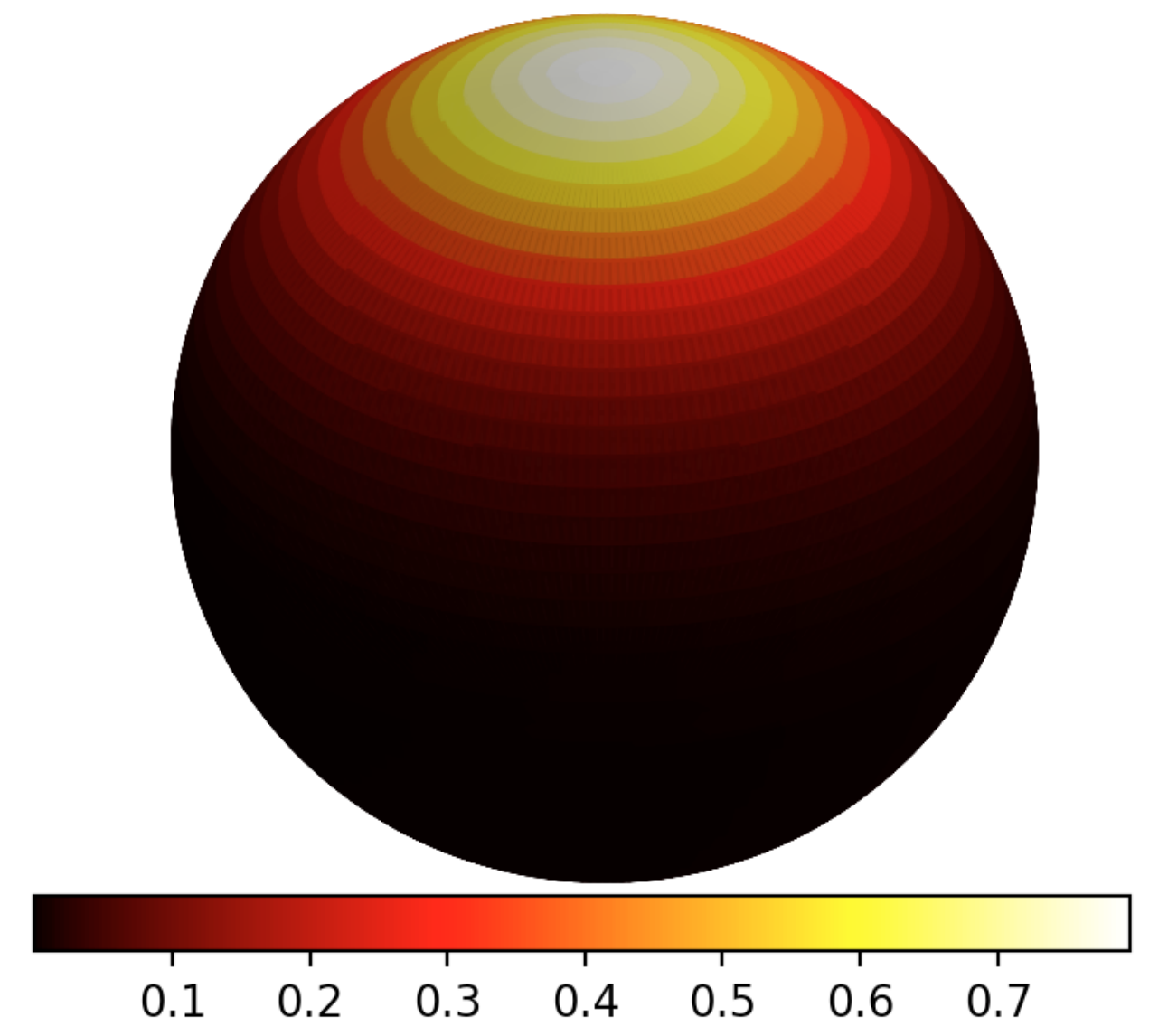}
	\caption{$\alpha$ = 5}
\end{subfigure}
\begin{subfigure}[b]{0.33\textwidth}
\centering
	\includegraphics[width=\textwidth]{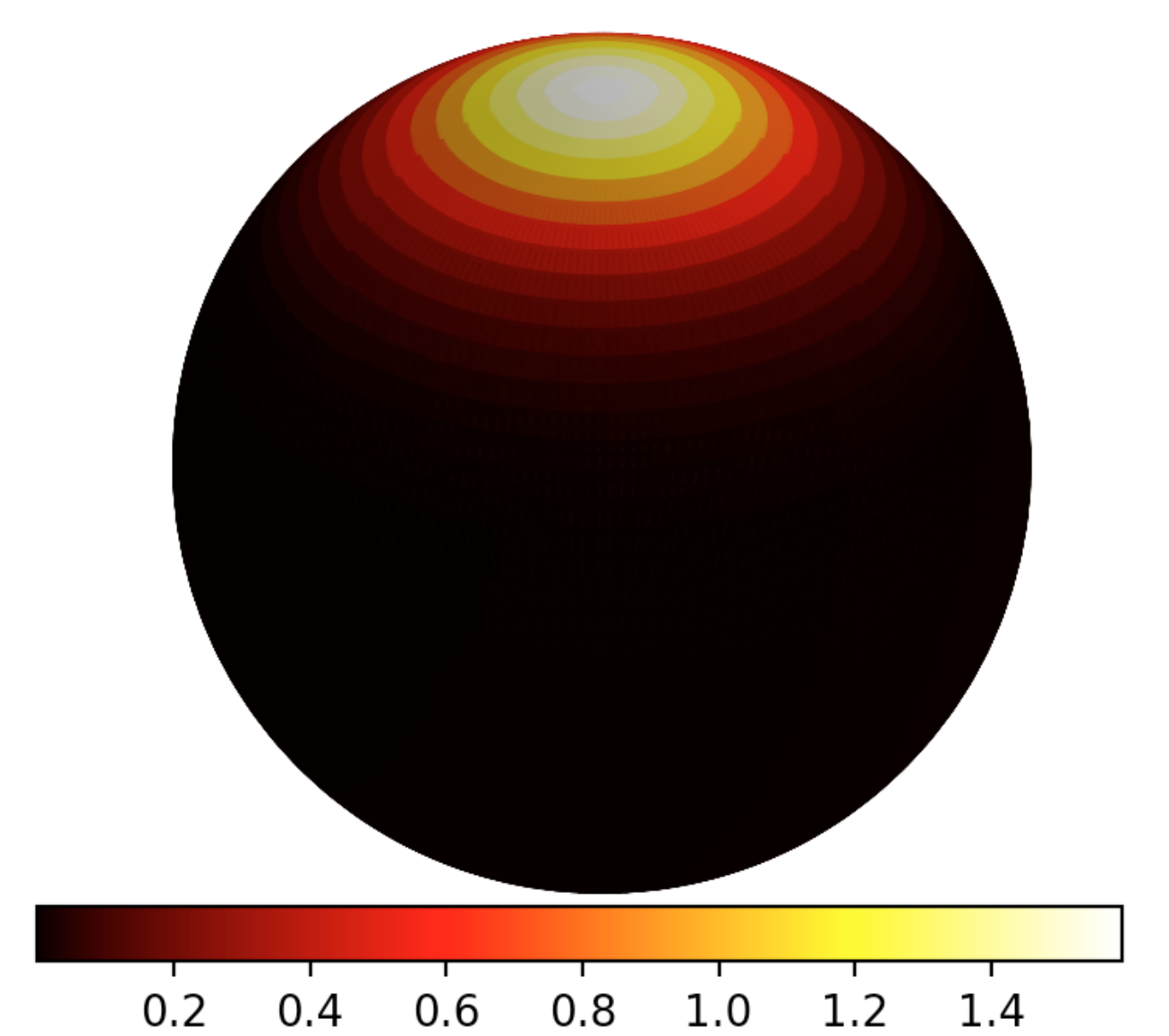}
	\caption{$\alpha$ = 10}
\end{subfigure}
\caption{The effect of the value of $\alpha$ on the Von Mises-Fisher distribution. The location of the north pole on the sphere represents the \Quote{mean direction}. The values on the color-bar represent probability density values. The greater the value of $\alpha$, the greater is the concentration of the distribution around the mean direction.}
\label{fig:alpha}
\end{figure*}

In \cite{Myronenko_2010_CPD}, the authors consider $\mathbf{Y}$ as the centroids of a Gaussian mixture model (GMM) and $\mathbf{X}$ as the data points generated by the GMM. The probability density function for the GMM is then given by:

 \begin{equation} 
 \label{eq:prob_x}
\begin{split}
p(\mathbf{x}) = \sum\limits_{j = 1}^{M} P(j)p(\mathbf{x} \mid j)
\end{split}
\end{equation}

where $p(\mathbf{x} \mid j) = \frac{1}{{(2 \pi \sigma^2}) ^ {3/2}} exp\big(- \frac{1}{2\sigma^2} \supnorm*{\mathbf{x}_p -  \subsup{1}{y}{j}{p}}^2 \ \big)$ and $P(j) = \frac{1}{M}$. In our case, we need the density to also depend on the point normal orientations. Therefore, we modify $p(\mathbf{x} \mid j) $ with an additional term that depends on the orientation as shown below:

 \begin{equation} 
 \label{eq:xconditionaly}
\begin{split}
p(\mathbf{x} \mid j) = \frac{\alpha\ exp\big(\alpha \subsup{1}{x}{T}{n} \subsup{1}{y}{j}{n} \big)}{2 \pi (exp(\alpha) - exp(-\alpha))}  \frac{exp\big(-  \frac{1}{2\sigma^2} \supnorm*{\mathbf{x}_p -  \subsup{1}{y}{j}{p}}^2 \ \big)}{( {(2 \pi \sigma^2}) ^ {3/2} )} 
\end{split}
\end{equation}
The above distribution is obtained by combining the original Gaussian distribution with another distribution, referred to as the Von Mises-Fisher distribution in directional statistics. The idea of combining these two distributions was first introduced by Billings and Taylor in \cite{Billings_2015_GIMLOP}. However, they do not involve a scaling parameter in their transformation. Since, our definition of a part involves uniform scaling, we cannot leave out the scale factor parameter from the formulation. Unlike \cite{Myronenko_2010_CPD}, we do not include a uniform distribution to account for noise. Fig. \ref{fig:alpha}, shows the effect of the value of $\alpha$ on the distribution. We place an upper limit of ten on the value of $\alpha$ to maintain a good balance between the importance of position and orientation of points.
 
 The central idea behind this approach to registration is to parameterize the distribution defined by $\mathbf{Y}$ with a rotation matrix $\mathbf{R}$, uniform scale factor $s$ and a translation vector $\mathbf{t}$, and then estimate these parameters using maximum likelihood estimation (MLE) by treating the point set $\mathbf{X}$ as the observed data. After the parameterization, $p(\mathbf{x} \mid j) $ is given by:
 
 \begin{equation} 
 \label{eq:reparam_xconditionaly}
\begin{split}
 p(\mathbf{x} \mid j) &= \frac{\alpha\ exp\big( - \frac{1}{2\sigma^2}  \supnorm*{\mathbf{x}_p -  s \mathbf{R} \ \subsup{1}{y}{j}{p} - \mathbf{t}}^2 + \alpha \subsup{1}{x}{T}{n} \mathbf{R} \subsup{1}{y}{j}{n} \big)}{2 \pi (exp(\alpha) - exp(-\alpha)) {(2 \pi \sigma^2}) ^ {3/2}}
\end{split}
\end{equation}

Like in \cite{Myronenko_2010_CPD}, we define $P(j \mid \mathbf{x}^i) = \frac{ P(j)  p(\mathbf{x}^i \mid j)}{p(\mathbf{x}^i)}$ as the correspondence probability between points $\mathbf{x}^i$ and $\mathbf{y}^j$. Using the i.i.d data assumption, the negative log-likelihood is given by:
 
 \begin{equation} 
 \label{eq:reparam_xconditionaly}
\begin{split}
\mathcal{L}(\theta) = \mathcal{L}(\mathbf{R},s,\mathbf{t},\sigma,\alpha) &= - \sum\limits_{i = 1}^N log \left( \sum\limits_{j = 1}^M P(j) p(\mathbf{x}^i \mid j) \right )
\end{split}
\end{equation}
Where,
 \begin{equation} 
 \label{eq:xi_given_yj}
p(\mathbf{x}^i \mid j) = \frac{\alpha\ exp\big( - \frac{1}{2\sigma^2}  \supnorm*{\mathbf{x}_p^i -  s \mathbf{R} \ \subsup{1}{y}{j}{p} - \mathbf{t}}^2 + \alpha ({\mathbf{x}_n^i})^T \mathbf{R} \subsup{1}{y}{j}{n} \big)}{2 \pi (exp(\alpha) - exp(-\alpha)) {(2 \pi \sigma^2}) ^ {3/2}}
\end{equation}


The Expectation Maximization (EM) algorithm is used to estimate the parameters that minimize the negative log-likelihood. The algorithm alternates between the E-step and the M-step until convergence. In the E-step, the posterior probability of the mixture components are computed as shown below:
\begin{equation}
\label{eq:E-step}
\resizebox{0.5\textwidth}{!}{$  P^{k}( j \mid \mathbf{x}^i) = \frac{exp\big( - \frac{1}{2\sigma^2}  \supnorm*{\mathbf{x}_p^i -  \mathcal{T}_p(\subsup{1}{y}{j}{p} ,\theta^{(k-1)} )}^2  + \alpha ({\mathbf{x}_n^i})^T   \mathcal{T}_n(\subsup{1}{y}{j}{n} ,\theta^{(k-1)} ) \big) }{  \sum\limits_{m = 1}^M  exp\big( - \frac{1}{2\sigma^2}  \supnorm*{\mathbf{x}_p^i -  \mathcal{T}_p(\subsup{1}{y}{m}{p} ,\theta^{(k-1)} )}^2  + \alpha ({\mathbf{x}_n^i})^T   \mathcal{T}_n(\subsup{1}{y}{m}{n} ,\theta^{(k-1)} ) \big)} $}
\end{equation}
 Where $\mathcal{T}_p(\subsup{1}{y}{j}{p} ,\theta^{(k-1)} )  =  s^{(k-1)} \mathbf{R}^{(k-1)} \subsup{1}{y}{j}{p} - \mathbf{t}^{(k-1)}$ and $\mathcal{T}_n(\subsup{1}{y}{j}{n} ,\theta^{(k-1)} )  =  \mathbf{R}^{(k-1)} \subsup{1}{y}{j}{n}$. The superscript $(k-1)$ indicates that the parameter values being used are the estimates from the previous iteration of the EM algorithm. In the M-step, the parameters need to be estimated by minimizing complete-data negative log-likelihood $Q$, which is given by:
\begin{equation}
\label{eq:Q_theta}
Q = - \sum\limits_{i=1}^N \sum\limits_{j=1}^M P^{k}( j \mid \mathbf{x}^i) log \big(  P(j) p(\mathbf{x}^i \mid j)  \big )
\end{equation}
Considering only terms that depend on $\theta$, we can rewrite $Q$ as:
\begin{equation}
\label{eq:E-step}
\begin{split}
Q(\theta) =  \bigg( \sum\limits_{i=1}^N \sum\limits_{j=1}^M  P^{k}( j \mid \mathbf{x}^i) \Big (  \frac{1}{2\sigma^2}  \supnorm*{\mathbf{x}_p^i -  s \mathbf{R} \ \subsup{1}{y}{j}{p} - \mathbf{t}}^2 - \\ \alpha ({\mathbf{x}_n^i})^T \mathbf{R} \subsup{1}{y}{j}{n} \Big )  \bigg)+ \frac{3N}{2} \log(\sigma^2) - N\log(\alpha) \ + \\ N\log \big( exp(\alpha) -exp(-\alpha) \big )
\end{split}
\end{equation}

Like in \cite{Myronenko_2010_CPD}, taking the partial derivative of $Q(\theta)$ with respect to $\mathbf{t}$ and setting it to zero gives:
$$
\mathbf{t} = \frac{1}{N} \big ( \mathbf{X}_p^T\mathbf{P}\mathbf{1} - s \mathbf{R} \mathbf{Y}_p^T\mathbf{P}\mathbf{1} \big ) \\
$$
where $\mathbf{P}$ is a M $\times$ N matrix with elements $ {p_{ji} = P^{k}( j \mid \mathbf{x}^i) }$.
Substituting $\mathbf{t}$ back into the equation (\ref{eq:Q_theta}) and setting ${\hat{\mathbf{X}} = \mathbf{X}_p -  \frac{1}{N} \mathbf{1} \big( {\mathbf{X}_p^T\mathbf{P}\mathbf{1}} \big)^T}$ and ${\hat{\mathbf{Y}} = \mathbf{Y}_p -  \frac{1}{N} \mathbf{1} \big( s\mathbf{R} {\mathbf{Y}_p^T\mathbf{P}\mathbf{1}} \big)^T}$, we get:

\begin{equation*}
\begin{split}
Q(\theta) &=  \frac{1}{2\sigma^2}  \big [  tr \big ( \hat{\mathbf{X}}^T d \big( \mathbf{P}^T \mathbf{1}\big) \hat{\mathbf{X}} \big)  - 2s\ tr \big(   \hat{\mathbf{X}}^T \mathbf{P}^T \hat{\mathbf{Y}} \mathbf{R}^T\big) \\ 
&+ \ s^2 tr \big ( \hat{\mathbf{Y}}^T d \big( \mathbf{P} \mathbf{1}\big) \hat{\mathbf{Y}} \big)\big]  + \alpha \ tr \big(   \mathbf{X}_n^T \mathbf{P}^T \mathbf{Y}_n \mathbf{R}^T\big) \\ &+ \frac{3N}{2} \log(\sigma^2) - N\log(\alpha) + N\log \big( exp(\alpha) -exp(-\alpha) \big )
\end{split}
\end{equation*}
Let $\mathbf{A} =  \hat{\mathbf{X}}^T \mathbf{P}^T \hat{\mathbf{Y}}$ and $\mathbf{B} =  \mathbf{X}_n^T \mathbf{P}^T \mathbf{Y}_n $. Using the invariance of trace under cyclic matrix permutation, the fact that $\mathbf{R}$ is orthogonal, we can rewrite ${Q(\theta)}$ as:
\begin{equation}
\label{eq:after_t}
\begin{split}
Q(\theta) &= \frac{1}{2\sigma^2}  \big [ tr \big ( \hat{\mathbf{X}}^T d \big( \mathbf{P}^T \mathbf{1}\big) \hat{\mathbf{X}} \big) - 2s\ tr \big( \mathbf{A}^T \mathbf{R} \big ) \\
& + s^2 tr \big ( \hat{\mathbf{Y}}^T d \big( \mathbf{P} \mathbf{1}\big) \hat{\mathbf{Y}} \big)\big] + \alpha \ tr \big(  \mathbf{B}^T \mathbf{R} \big) + \frac{3N}{2} \log(\sigma^2) \\
& - N\log(\alpha) + N\log \big( exp(\alpha) -exp(-\alpha) \big )
\end{split}
\end{equation}

The form of equation (\ref{eq:after_t}) does not allow us to directly apply the method in  \cite{Myronenko_2010_CPD} to estimate $\mathbf{R}$. Therefore, we use the BFGS algorithm to compute the parameters. For $\mathbf{R}$ to be a proper rotation matrix, it has to satisfy two constraints. I.e., $\mathbf{R}$ has to be an orthogonal matrix whose determinant is one. To avoid having to deal with these constraints during optimization, a reparameterization of $\mathbf{R}$ is necessary. We use the quaternion based parameterization of $\mathbf{R}$ based on \cite{terzakis2012recipe}. In \cite{terzakis2012recipe}, the authors describe two different ways to parameterize a rotation matrix. We use the second method which stereographically projects a 3D hyperplane onto the 4D unit quaternion sphere. Such a parameterization not only removes the constraints but also provide rational expressions for the derivative of R, which becomes significant when iterative methods like BFGS are employed during optimization. The relevant equations from \cite{terzakis2012recipe} are provided below.

The rotation matrix corresponding to an unit quaternion $\mathbf{q} = [q_0,q_1,q_2,q_3]$ is given by:
\begin{equation}
\resizebox{0.5\textwidth}{!}{$\mathbf{R}(\mathbf{q}) = \begin{bmatrix} 
q_0^2 + q_1^2 - q_2^2 - q_3^2 & 2(q_1q_2 - q_0q_3) & 2(q_1q_3 + q_0q_2) \\
2(q_1q_2 + q_0q_3) & q_0^2 - q_1^2 + q_2^2 - q_3^2 & 2(q_2q_3 - q_0q_1) \\
2(q_1q_3 - q_0q_2) & 2(q_2q_3 + q_0q_1) & q_0^2 - q_1^2 - q_2^2 + q_3^2 
\end{bmatrix}
$}
\end{equation}
The partial derivatives of the rotation matrix with respect to the components of the quaternion is given by:

\begin{equation}
\begin{split}
\frac{\partial \mathbf{R}(\mathbf{q})}{\partial q_0} &= 2\begin{bmatrix} 
q_0 & -q_3 & q_2 \\
q_3 & q_0 & -q_1 \\
-q_2 & q_1 & q_0
\end{bmatrix} \\
\frac{\partial \mathbf{R}(\mathbf{q})}{\partial q_1} &= 2\begin{bmatrix} 
q_1 & q_2 & q_3 \\
q_2 & -q_1 & -q_0 \\
q_3 & q_0 & -q_1
\end{bmatrix} \\
\frac{\partial \mathbf{R}(\mathbf{q})}{\partial q_2} &= 2\begin{bmatrix} 
-q_2 & q_1 & q_0 \\
q_1 & q_2 & q_3 \\
-q_0 & q_3 & -q_2
\end{bmatrix}\\
\frac{\partial \mathbf{R}(\mathbf{q})}{\partial q_3} &= 2\begin{bmatrix} 
-q_3 & -q_0 & q_1 \\
q_0 & -q_3 & q_2 \\
q_1 & q_2 & q_3
\end{bmatrix}
\end{split}
\end{equation}

As mentioned earlier, a stereographic projection is employed to project each point, $[x,y,z]$, on a 3D hyperplane onto a 4D unit quaternion sphere (see Figure 1 in \cite{terzakis2012recipe}). This mapping provides a way to represent a unit quaternion using a 3D vector. Given a point, $[x,y,z]$, on the hyperplane, the corresponding unit quaternion is given by:

\begin{equation}
\begin{split}
\beta^2 &= x^2 + y^2 + z^2 \\
\mathbf{q} &= \frac{1}{(\beta^2 + 1)} \bigg ( 2x,2y,2z,1-\beta^2 \bigg)
\end{split}
\end{equation}

The projection of a unit quaternion $\mathbf{q} = [q_0,q_1,q_2,q_3]$ onto the hyperplane is given by:
\begin{equation}
\begin{split}
\beta^2 &= \frac{1 - q_3}{1 + q_3} \\
\big ( x,y,z\big) &= \frac{\beta^2 + 1}{2} \big ( q_0, q_1, q_2\big )
\end{split}
\end{equation}

The partial derivatives of the quaternions with respect to the point $\mathbf{\psi} = (x,y,z)$ is given by:

\begin{equation}
\begin{split}
\frac{\partial \mathbf{q}(\mathbf{\psi})}{\partial x} &= \frac{1}{(\beta^2 + 1)^2}\begin{bmatrix} 
 2(\beta^2 + 1) - 4x^2 \\
-4xy  \\
-4xz \\
 -4x
\end{bmatrix} \\
\frac{\partial \mathbf{q}(\mathbf{\psi})}{\partial y} &= \frac{1}{(\beta^2 + 1)^2}\begin{bmatrix}  
 -4xy  \\
 2(\beta^2 + 1) - 4y^2 \\
-4yz \\
 -4y
\end{bmatrix} \\
\frac{\partial \mathbf{q}(\mathbf{\psi})}{\partial y} &= \frac{1}{(\beta^2 + 1)^2}\begin{bmatrix}  
 -4xz  \\
-4yz \\
 2(\beta^2 + 1) - 4z^2 \\
 -4z
\end{bmatrix} \\ 
\end{split}
\end{equation}

The partial derivative of the rotation matrix, $\mathbf{R}$, with respect to $\mathbf{\psi}$ can now be written as:
\begin{equation}
\begin{split}
\frac{\partial \mathbf{R}(\mathbf{q}(\mathbf{\psi}))}{\partial x} &= \sum_{j = 0}^{3} \frac{\partial \mathbf{R}(\mathbf{q})}{\partial q_j} \frac{\partial q_j}{\partial x} \\
\frac{\partial \mathbf{R(}\mathbf{q}(\mathbf{\psi)})}{\partial y} &= \sum_{j = 0}^{3} \frac{\partial \mathbf{R}(\mathbf{q})}{\partial q_j} \frac{\partial q_j}{\partial y} \\
\frac{\partial \mathbf{R}(\mathbf{q}(\mathbf{\psi}))}{\partial z} &= \sum_{j = 0}^{3} \frac{\partial \mathbf{R}(\mathbf{q})}{\partial q_j} \frac{\partial q_j}{\partial z} \\
\end{split}
\end{equation}

The partial derivatives required for optimizing $Q(\theta)$ using the BFGS algorithm are provided below:

 \begin{equation}
\begin{split}
K_1 &= tr \big ( \hat{\mathbf{Y}}^T d \big( \mathbf{P} \mathbf{1}\big) \hat{\mathbf{Y}} \big) \\ 
K_2 &= tr \big ( \hat{\mathbf{X}}^T d \big( \mathbf{P}^T \mathbf{1}\big) \hat{\mathbf{X}} \big) \\
\frac{\partial Q(\theta)}{\partial \sigma} &= \frac{-1}{\sigma^3} \big ( K_2 - 2s \ tr(\mathbf{A}^T \mathbf{R} ) + s^2 K_1 \big) + \frac{3 N}{\sigma} \\ 
\frac{\partial Q(\theta)}{\partial s} &= \frac{-1}{\sigma^2} tr(\mathbf{A}^T \mathbf{R} ) + \frac{s}{\sigma^2} K_1 \\
\frac{\partial Q(\theta)}{\partial \alpha} &= -tr(\mathbf{B}^T \mathbf{R} ) - \frac{N}{\alpha} + N \frac{exp(\alpha) + exp(-\alpha)}{exp(\alpha) - exp(-\alpha)} \\
\frac{\partial Q(\theta)}{\partial x} &= \frac{-s}{\sigma^2} tr(\mathbf{A}^T \frac{ \partial \mathbf{R}}{\partial x} ) -\alpha \ tr(\mathbf{B}^T \frac{ \partial \mathbf{R}}{\partial x} ) \\
\frac{\partial Q(\theta)}{\partial y} &= \frac{-s}{\sigma^2} tr(\mathbf{A}^T \frac{ \partial \mathbf{R}}{\partial y} ) -\alpha \ tr(\mathbf{B}^T \frac{ \partial \mathbf{R}}{\partial y} ) \\
\frac{\partial Q(\theta)}{\partial z} &= \frac{-s}{\sigma^2} tr(\mathbf{A}^T \frac{ \partial \mathbf{R}}{\partial z} ) -\alpha \ tr(\mathbf{B}^T \frac{ \partial \mathbf{R}}{\partial z} ) \\
\label{eq:final_gradients}
\end{split}
\end{equation}

The steps involved in the registration algorithm are summarized in algorithm \ref{alg:estimate_rst}.

\begin{figure*}[htp]
\centering
\begin{subfigure}[b]{0.24\textwidth}
	\includegraphics[width=\textwidth]{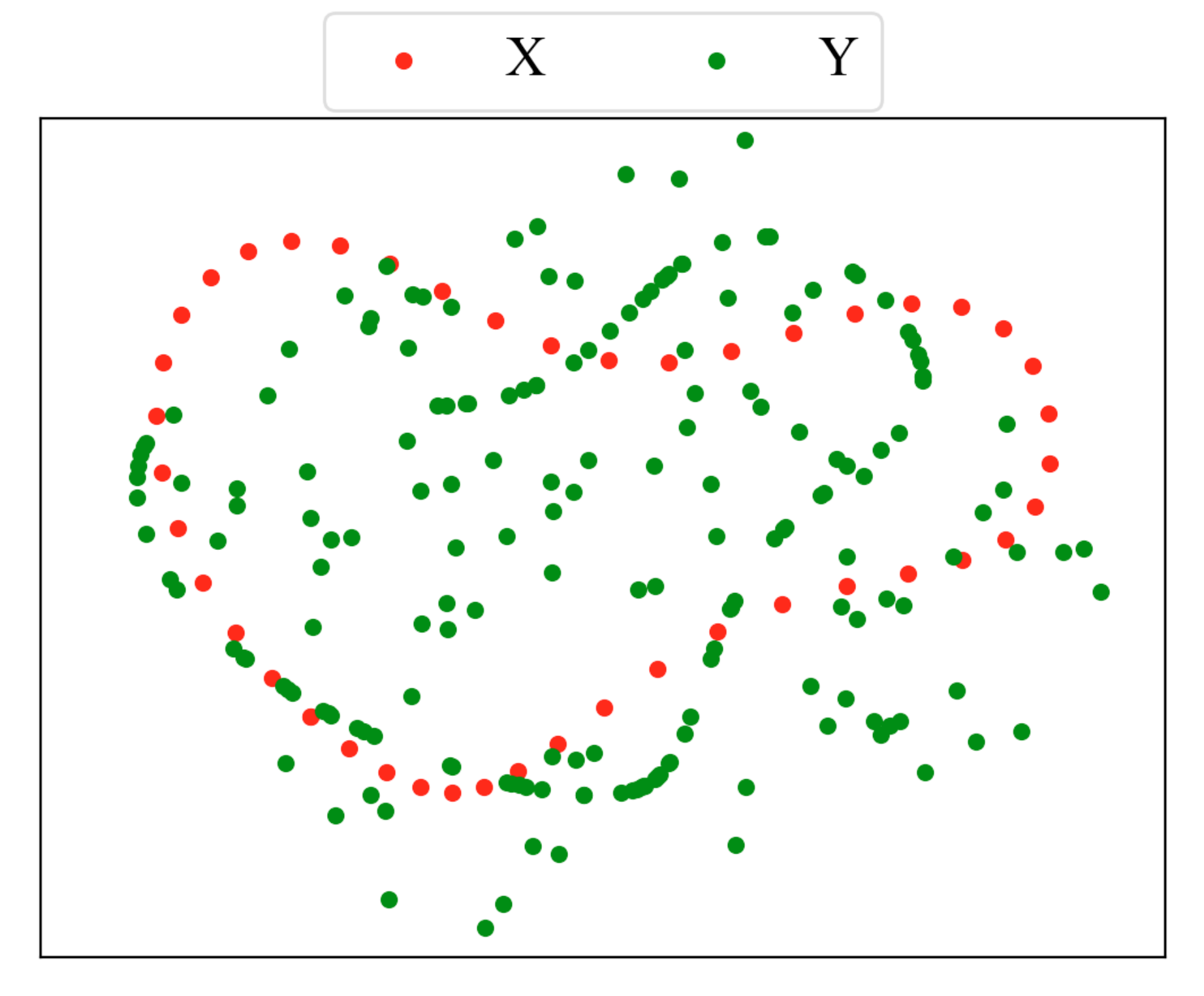}
	\caption{}
\end{subfigure}	
\begin{subfigure}[b]{0.24\textwidth}
	\includegraphics[width=\textwidth]{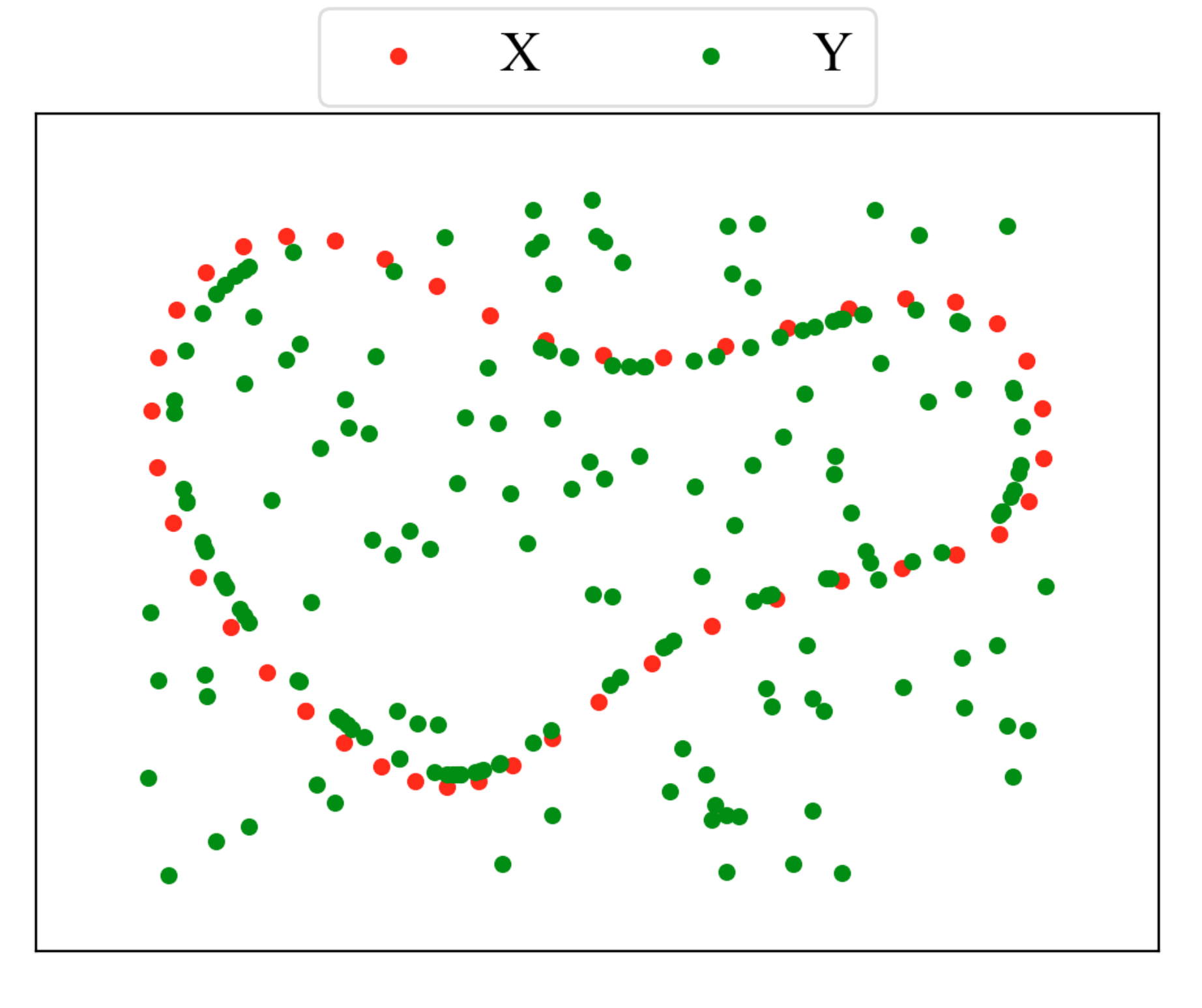}
	\caption{}
\end{subfigure}	
\begin{subfigure}[b]{0.24\textwidth}
\centering
	\includegraphics[width=\textwidth]{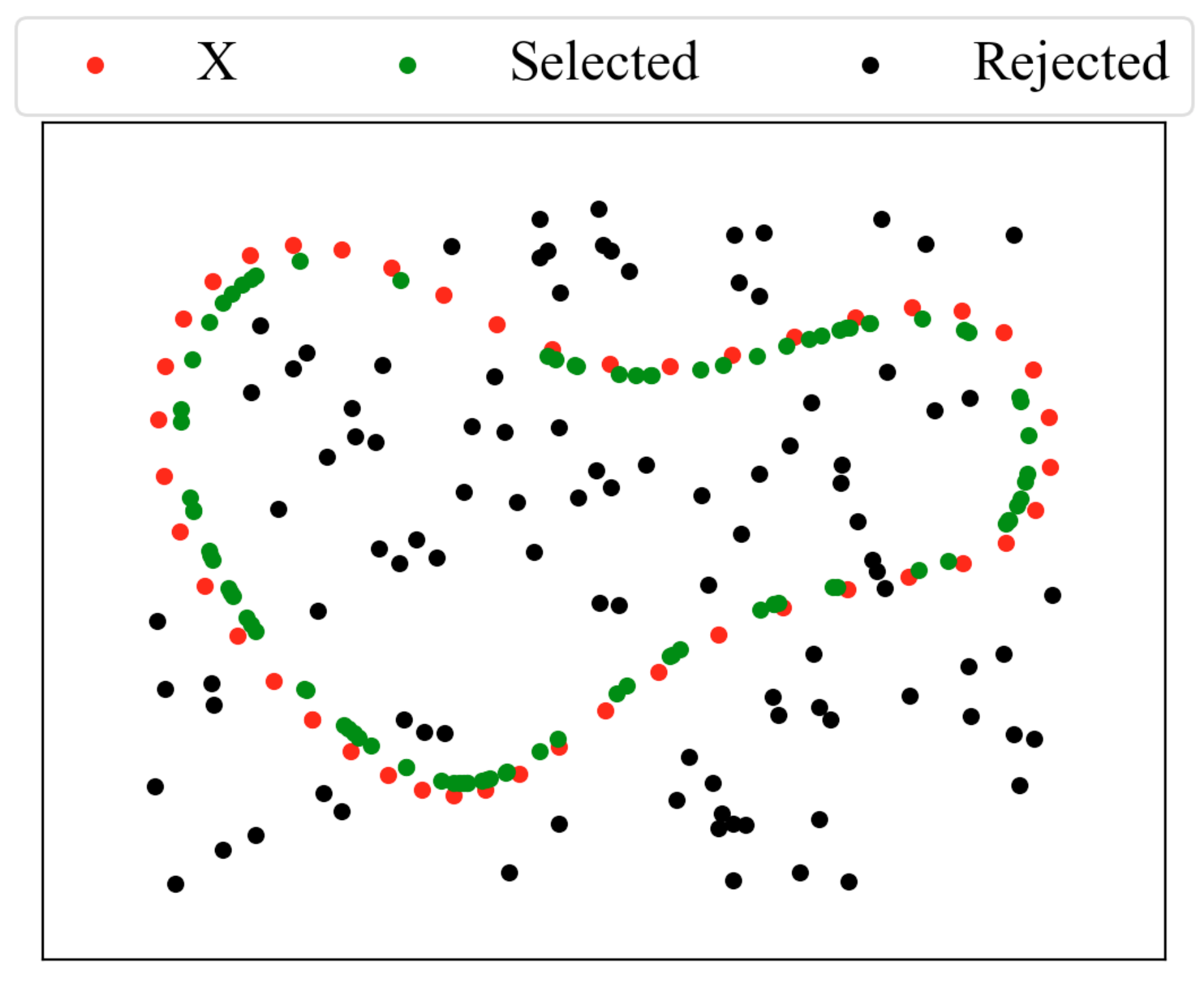}
	\caption{}
\end{subfigure}
\begin{subfigure}[b]{0.24\textwidth}
\centering
	\includegraphics[width=\textwidth]{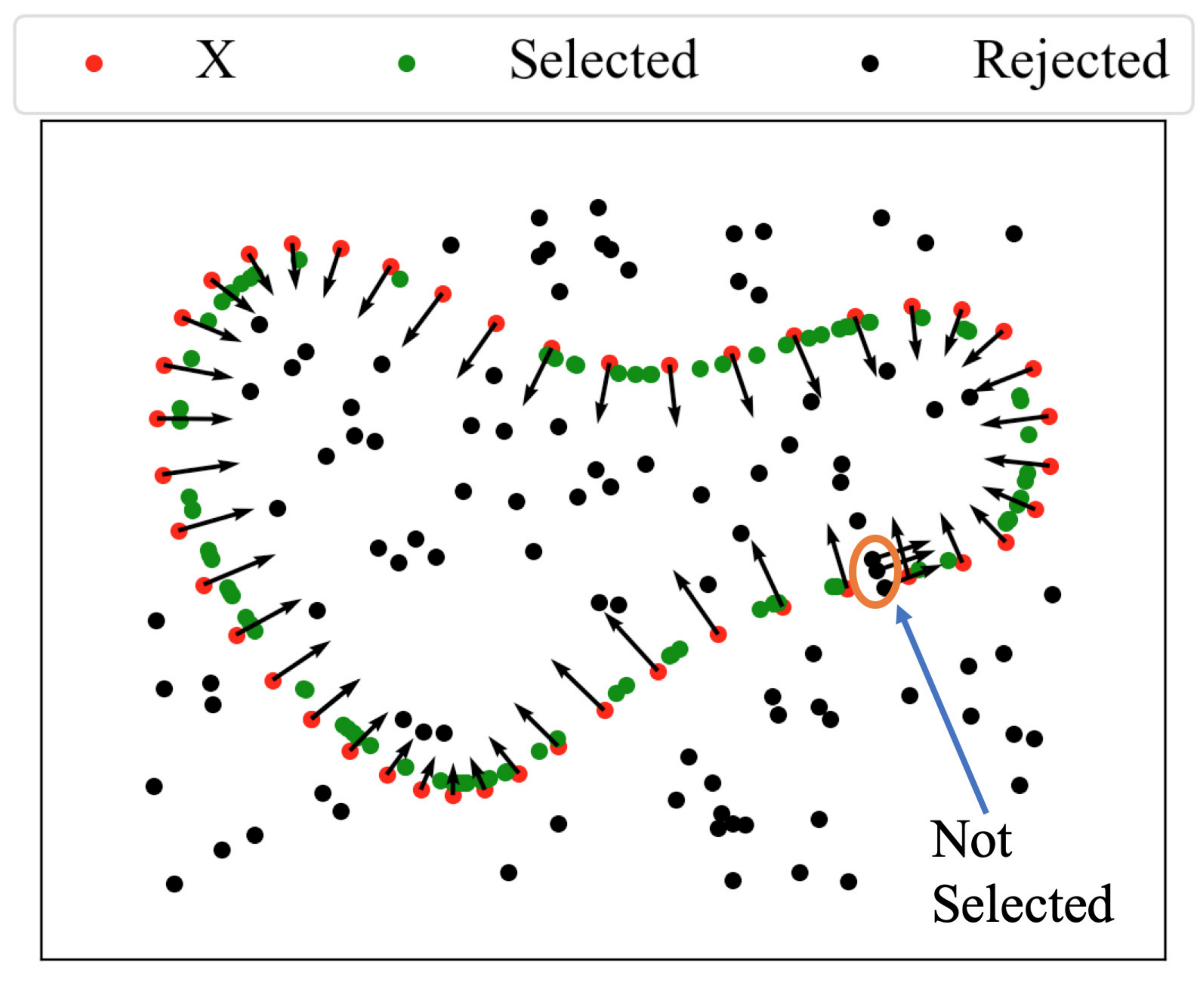}
	\caption{}
\end{subfigure}
\caption{Illustration of point selection after registration. (a) The two point sets, $\mathbf{X}$ representing the cluster for which we are seeking a match (cluster 0 in our example) and $\mathbf{Y}$ representing the neighboring points, before registration. (b) The transformed $\mathbf{Y}$, $\mathcal{T}(\mathbf{y}^j) = ( s \mathbf{R}\mathbf{y}^j_p + \mathbf{t},  \mathbf{R}\mathbf{y}^j_n )$, after registration. (c) Chosen and rejected points, after threshold based selection, shown in different colors. (d) Having a member of $\mathbf{X}$ close by, after transformation, is not enough to be selected, the point normals should also closely match. The point normals of all the red points are shown. Also shown are the point normals of three green points (within the circle) that were rejected. The point normal orientations of the green points are too different from that of its neighboring red points, and hence these points are not chosen.}
\label{fig:point_selection}
\end{figure*}

\subsubsection*{ Growing Parts Using Registration} \label{sec:specific_registration}
In method 2, like in method 1, given an initial cross-sectional cluster, we grow parts by finding its neighboring cross-sectional cluster. Method 2 uses registration to find the neighboring cross-sectional cluster. From the definition of a part as a GC, we know that two cross-sectional clusters of a part are related to each other by a similarity transformation. In method 2, we use registration to estimate this transformation. For instance, there exists a rotation $\mathbf{R}$, uniform scale factor $s$ and a translation $\mathbf{t}$ that can transform  cluster 0, in Fig. \ref{fig:growing_stages}, to cluster 1. Looking at the two cross-sectional clusters, we can see that the rotation matrix in this case would be close to the identity matrix, the scale factor would be close to, but a little less than, one and the translation vector would point from the center of cluster 0, towards the center of cluster 1.
The registration algorithm takes two point sets as its input. One of the point sets, $\mathbf{X}$, would be the cluster for which we are seeking a match. In our example, this would be cluster 0. The idea is to now provide points in the neighborhood of cluster 1 to the registration algorithm as the second point set, $\mathbf{Y}$. The registration algorithm would then find matches for points in cluster 0 in its neighborhood leading to the detection of cluster 1. In method 1, we did consider a group of planes whose orientation is close to the orientation of the cross-sectional plane of cluster 0. The second point set is formed by all points that are part of the inlier set of any of these planes (belonging to the set $\mathbf{A}_{\theta} \bigtimes \mathbf{A}_{\phi}$). As illustrated in Fig. \ref{fig:nxt_seed_point} (b), these inlier points of various planes in set $\mathbf{A}_{\theta} \bigtimes \mathbf{A}_{\phi}$ will contain points in cluster 1. Some additional point, not part of cluster 1, are also contained in the inlier sets. But the registration algorithm is capable of dealing with the presence of some additional points.

Given cluster 0 and the points in its neighborhood, the registration algorithm estimates the parameters, $\theta$. After registration, we still have to choose points from the second point set $\mathbf{Y}$ to form the neighboring cluster, cluster 1. Fig. \ref{fig:point_selection}, illustrates the procedure of identifying points belonging to neighboring cluster, post registration. A 2D illustration is used for simplicity. As shown, the point $ \mathbf{y}^j  \in \mathbf{Y}$ is chosen only if it has at least one point  $\mathbf{x}^i \in \mathbf{X}$ sufficiently close to it and if the point normal orientations $ \mathbf{y}^j$  and $ \mathbf{x}^i $ are sufficiently close. Points $\mathbf{x}^i$ and $\mathbf{y^j}$ are considered sufficiently close if $d_{eucl}(\mathbf{x}^i_p, \mathbf{y^j_p}) \leq  1.5\sigma$, where $\sigma$ refers to the standard deviation of the GMM just estimated using the registration process. The point normal orientations are considered sufficiently close if the angle between them is less than $20^\circ$.

As described above, the registration process can be used to detect cluster 1 given cluster 0. This process can be repeated to find a matching cross-sectional cluster for cluster 1, resulting in the detection of cluster 2. This process of growing a part stops when a good match cannot be found for a cross-sectional cluster. As mentioned above, $P(j \mid \mathbf{x}^i) $ is defined as the correspondence probability between points in the two point sets. For each point in $\mathbf{X}$, we can find the best match in $\mathbf{Y}$ using the correspondence probability, $P(j \mid \mathbf{x}^i) $. The average angular difference between the best matches is used to decide whether the growing process needs to be terminated. If the average angular difference between best matches is greater than $15^\circ$, the growing process terminates.

As depicted in Fig. \ref{fig:why_register}, method 2 is better equipped for dealing with junctions and therefore is the default method to grow parts. But when the cross-sectional cluster is sparse, the registration process cannot be relied on. The point normal estimation is itself unreliable in such a scenario. Therefore, method 1 is employed only when the cross-sectional cluster is sparse. In practice, if the cross-sectional cluster has less than hundred points in it, method 1 is employed.

\section{Optimal Part Selection} \label{sec:optimal_part_selection}
In the previous section, methods to grow parts were described. The point cloud was clustered into $M$ clusters and one part per cluster was grown. Given these $M$ parts, we first assign costs to each part and then select the best subset of parts that cover the whole point cloud based on these scores. The overall cost has four components. Below, we first describe each of these components. The process of combining these components to obtain a single cost per part is then described. And finally we explain how the optimization process, which selects the best parts, is set up.

\subsection{Cost Components}\label{sec:cost_components}

The first component is the registration cost. For a good part, the adjacent cross-sectional clusters would be a good match for each other. Remember, the relationship between adjacent cross-sectional clusters is defined by a similarity transform and via the registration process, we have estimated the best similarity transform parameters. After transformation (with the parameters estimated by the registration algorithm), the two point sets should be a good match to each other as illustrated in Fig. \ref{fig:point_selection} (b). Once we perform the registration between adjacent clusters of a part, for each point in the point set $\mathbf{X}$, we can find the best match in point set $\mathbf{Y}$ using the correspondence probability, $P(j \mid \mathbf{x}^i) $. The average angular difference between the best matches is used as the registration cost. Therefore, the registration cost would be low if the corresponding points in the two point sets have similar orientation. If a part is made up of $k$ cross-sectional clusters, we would obtain $k-1$ registration costs from pairs of adjacent clusters. To obtain the registration cost, we take the mean of these $k-1$ costs.

The second component, referred to as the fit cost, measures the goodness of fit of the skeleton to the points that constitute the part. This cost is computed in the same manner as the cost, $c(\theta,\phi)$, in algorithm \ref{alg:cross-sectional_plane}. As explained before, and as depicted in Fig. \ref{fig:gc_illustration} (c), the point normals are expected to be perpendicular to the normal of the local cross-sectional plane. The fit cost is designed to penalize any deviations to this. To compute the fit cost, for a part with $k$ cross-sectional clusters, we compute the cost, $c(\theta,\phi)$, for each cross-sectional cluster and take their mean value.

\begin{figure}[htp]
\centering
	\includegraphics[width=0.35\textwidth]{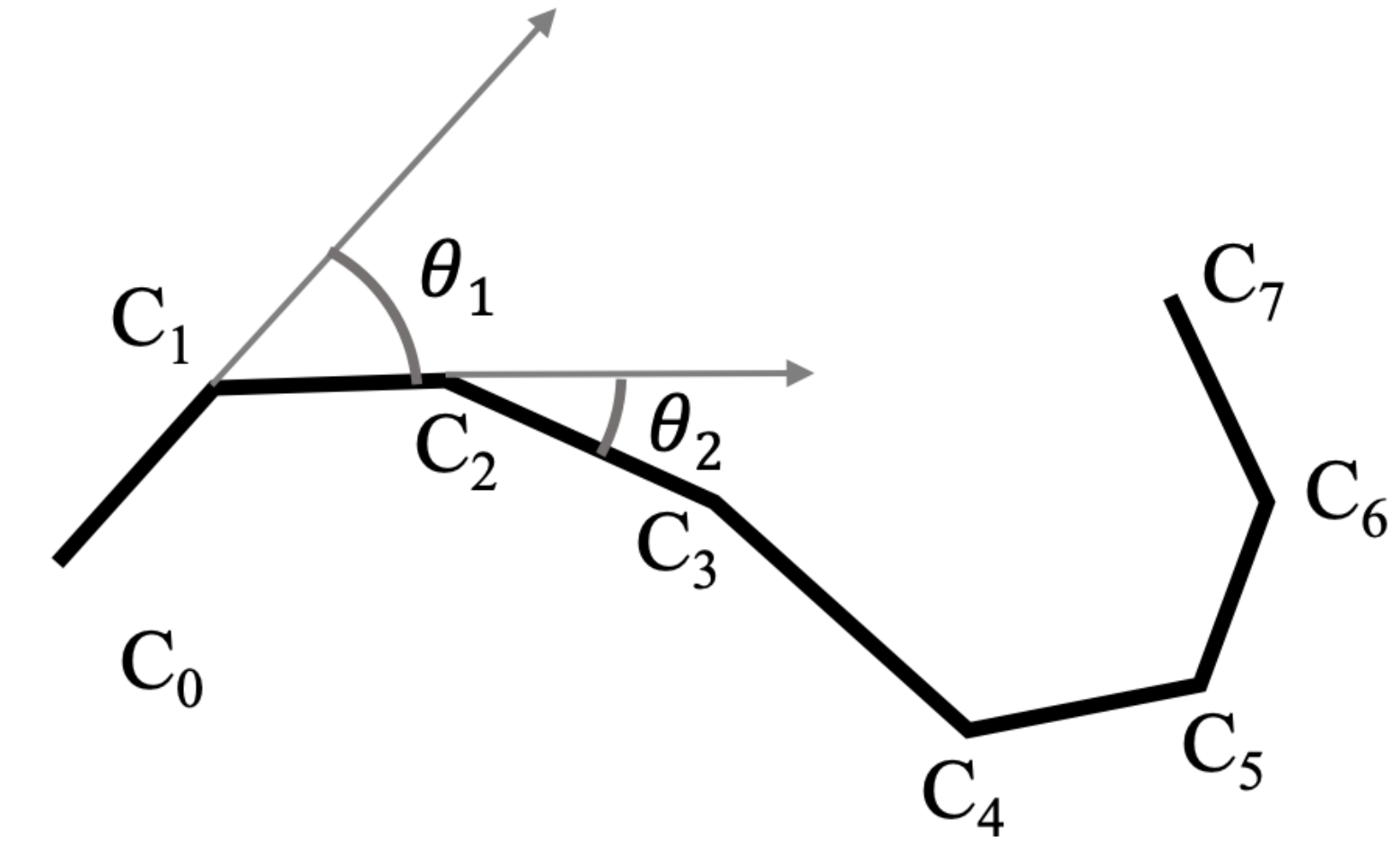}
\caption{Skeletons are formed by joining cluster centers. The black lines represent the skeleton. The length of the skeleton is computed as the sum of the euclidean distances between cluster centers when traveling from one end ($C_0$) to the other end ($C_7$) of the part. The turning angles at cluster centers $C_1$ and $C_2$ are also shown.}
\label{fig:skel_cost_demo}
\end{figure}

The third component of the overall cost is the length of the part. The length here refers to the length of the skeleton. Since, skeletons are represented by cluster centers of the cross-sectional clusters, the length is computed as the sum of the euclidean distances between the centers as you travel from one end to the other end of the part. For instance, in Fig. \ref{fig:skel_cost_demo}, $ \sum\limits_{i=0}^6 d_{eucl}(C_i,C_{i+1})$ represents the length of the skeleton. Longer parts are preferred over shorter parts. 

Since, we prefer smooth curves as skeleton, the fourth component is the total turning angle of the skeleton. Fig. \ref{fig:skel_cost_demo}, shows how turning angles are computed. In the example shown, the total turning angle cost would be $ \frac{\sum\limits_{i=1}^6 \theta_i}{6}$.

To obtain the overall cost, each individual cost component is first normalized separately by subtracting the mean and dividing by the standard deviation. Let $c_{reg}^i$, $c_{fit}^i$, $c_{len}^i $  and $c_{ang}^i $ represent the normalized registration, fit, length and turning angle costs of part $i$ respectively. The overall cost for part $i$ is then given by: $c_{ovr}^i = c_{reg}^i + c_{fit}^i - c_{len}^i + c_{ang}^i$. Note, during optimization we minimize the cost and since we prefer longer parts over short parts, we subtract the length component instead of adding it.

\subsection{Optimal Parts Selection}
Once we assign a cost to each part, the optimal subset of parts are chosen by minimizing the following constrained binary integer program.
\begin{equation}
\begin{aligned}
& minimize &\pmb{c}^T\pmb{x}\\
& \text{subject to}
\end{aligned}
\label{eq:optimization}
\end{equation}

\vspace{-1.4em}
\begin{align*}
 \pmb{x^{T} a} &\geq \left(\dfrac{k_1}{100}\right)\,N \tag{$C_1$}\label{cstr:c1}\\
 \pmb{x^{T} Q\ x} &\leq \left(\dfrac{k_2}{100}\right)\,N \tag{$C_2$}\label{cstr:c2}\\
\end{align*}

\vspace{-2.4em}
\begin{align*}
& \text{where,} &\\
& \pmb{x} =  (x^1,x^2, \cdots, x^M) \in \{0,1\}^{M \times 1}, \\  
& \pmb{c} =  (c_{ovr}^1,c_{ovr}^2, \cdots, c_{ovr}^M) \in  \R^{M \times 1} \\ 
& \pmb{a} \in \mathbb{N}^{M \times 1} , \ \pmb{Q} \in \mathbb{N}^{M \times M}  \\
  \end{align*}

\begin{figure}[htp]
\centering
\begin{subfigure}[b]{0.4\textwidth}
	\includegraphics[width=\textwidth]{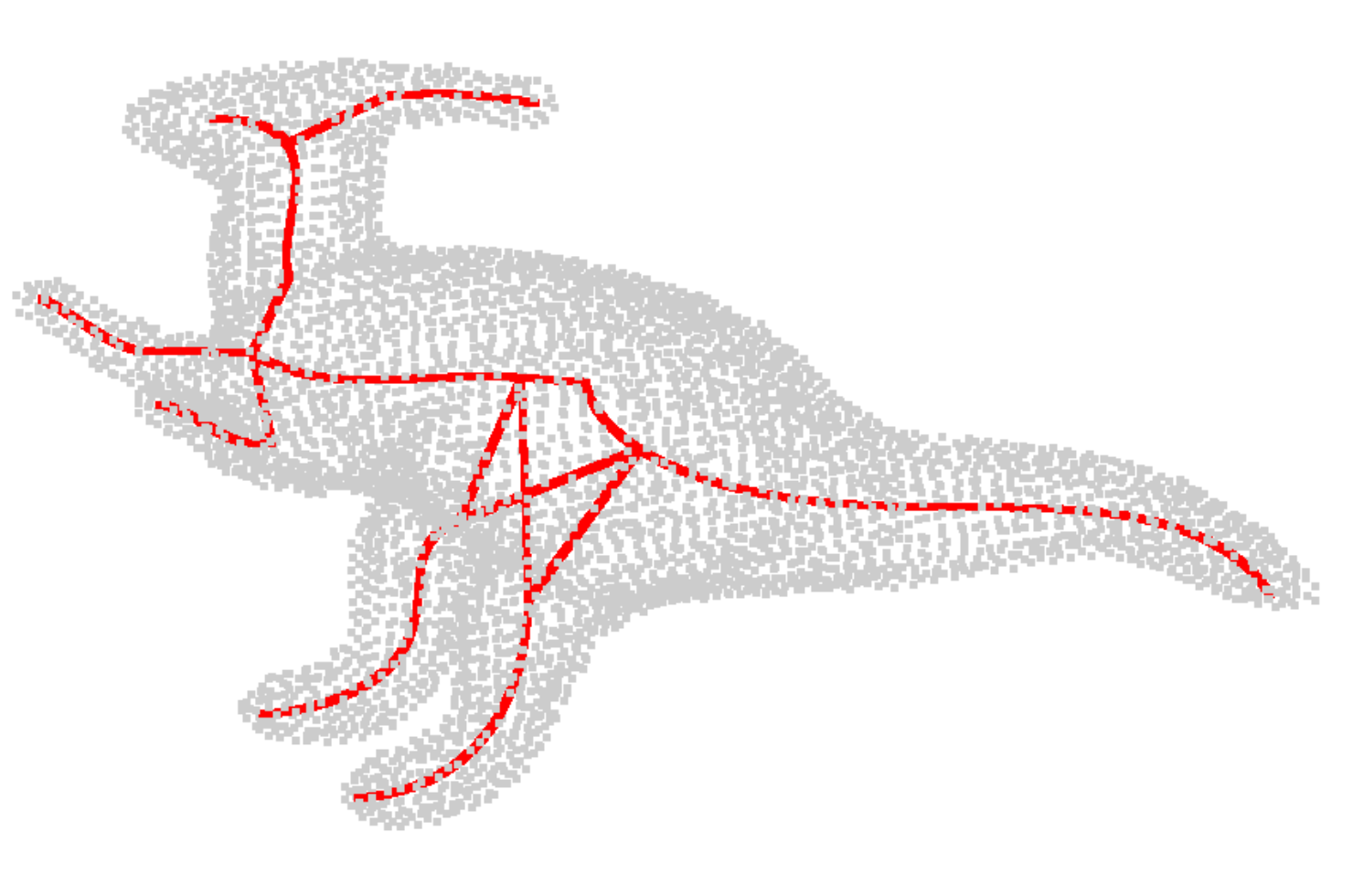}
	\caption{}
\end{subfigure}	
\begin{subfigure}[b]{0.44\textwidth}
\centering
	\includegraphics[width=\textwidth]{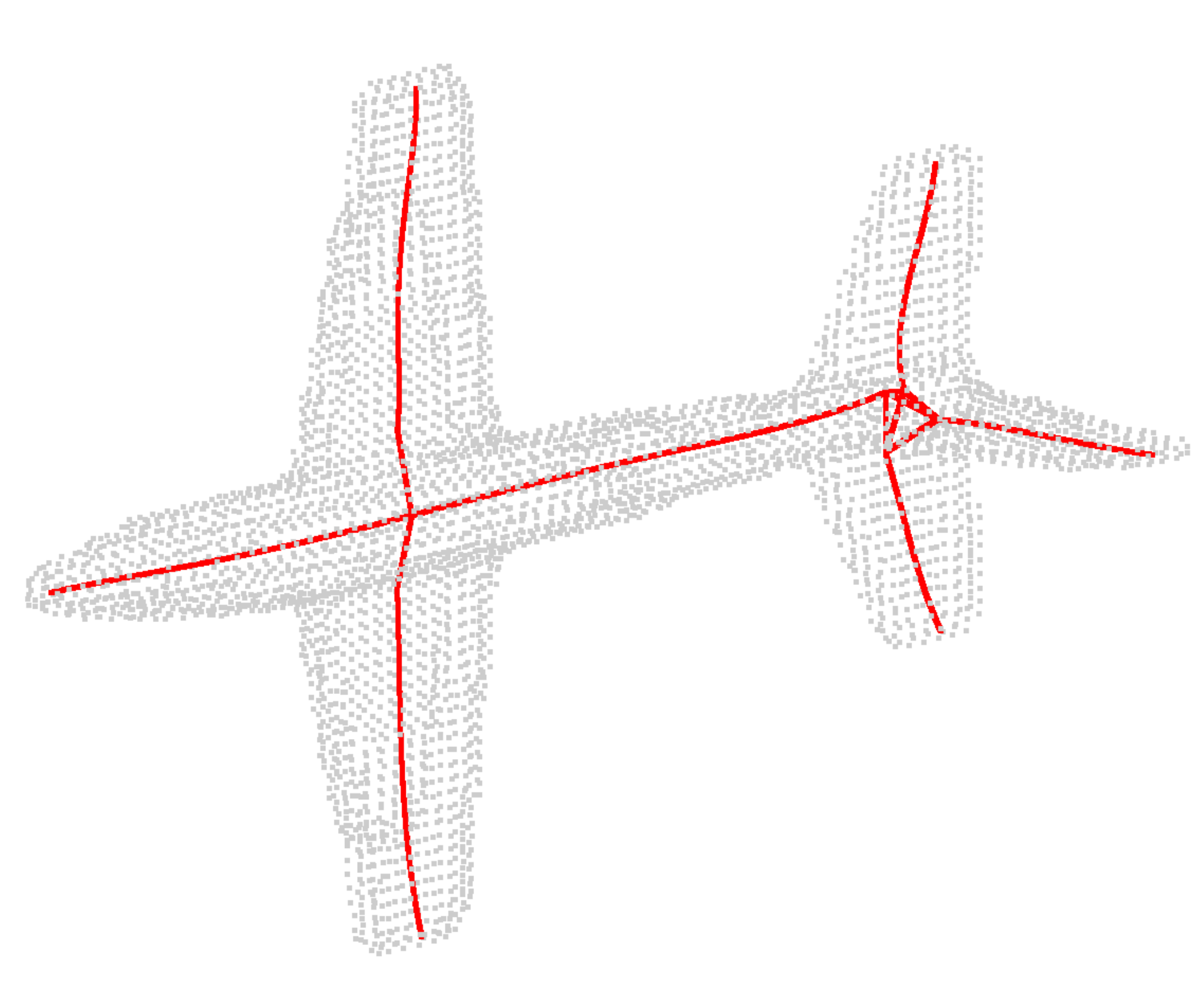}
	\caption{}
\end{subfigure}
\caption{Linking all parts that can be potentially linked according to $G_{cnct}$ can lead to unnecessary additional connections. (a)  The legs can potentially connect to both the torso and the tail parts. (b) The many links near the tail of the airplane are unnecessary.}
\label{fig:link_all}
\end{figure}

\begin{figure*}[htp]
\centering
\begin{subfigure}[b]{0.38\textwidth}
	\includegraphics[width=\textwidth]{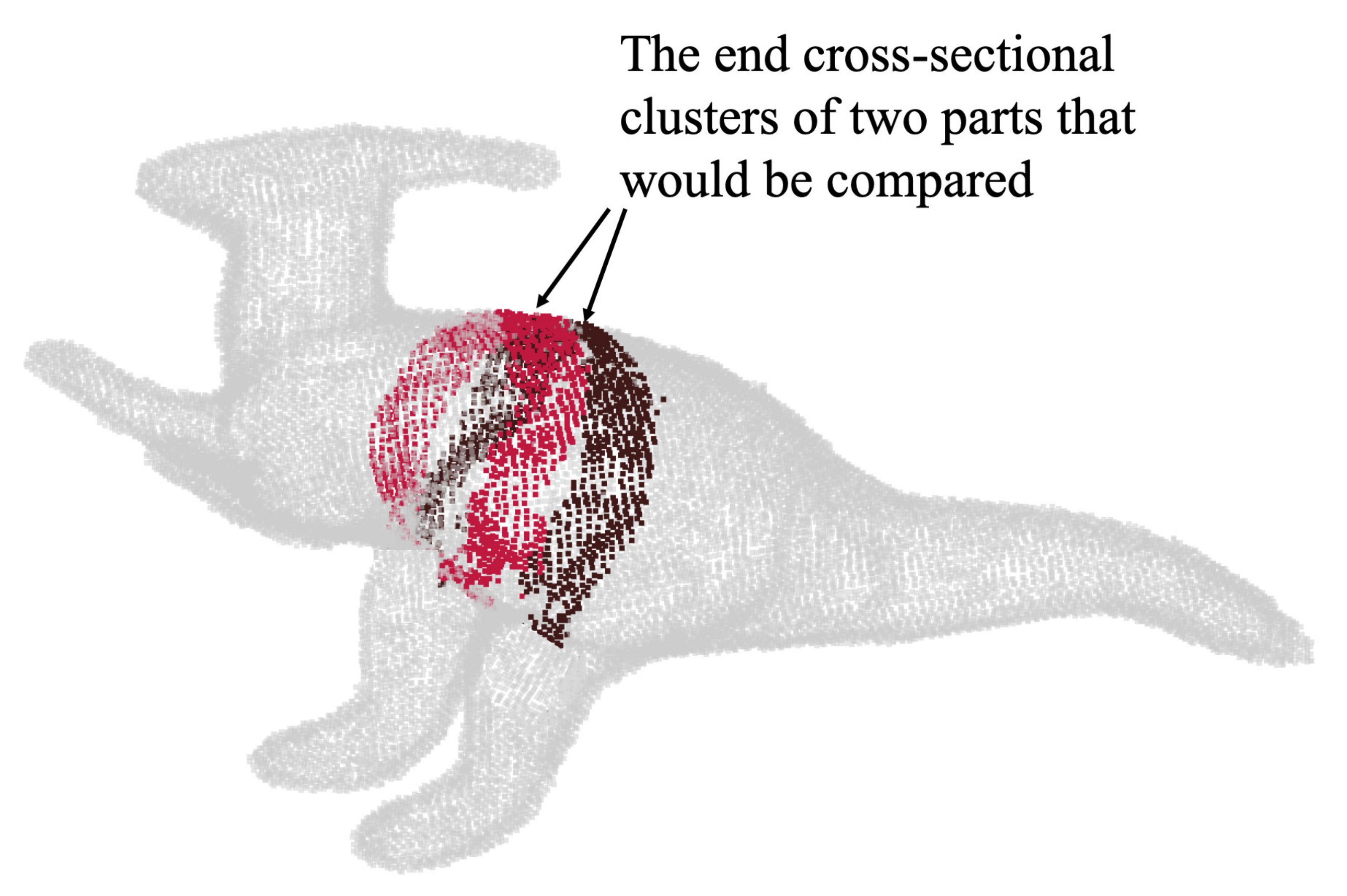}
	\caption{}
\end{subfigure}	
\begin{subfigure}[b]{0.38\textwidth}
\centering
	\includegraphics[width=\textwidth]{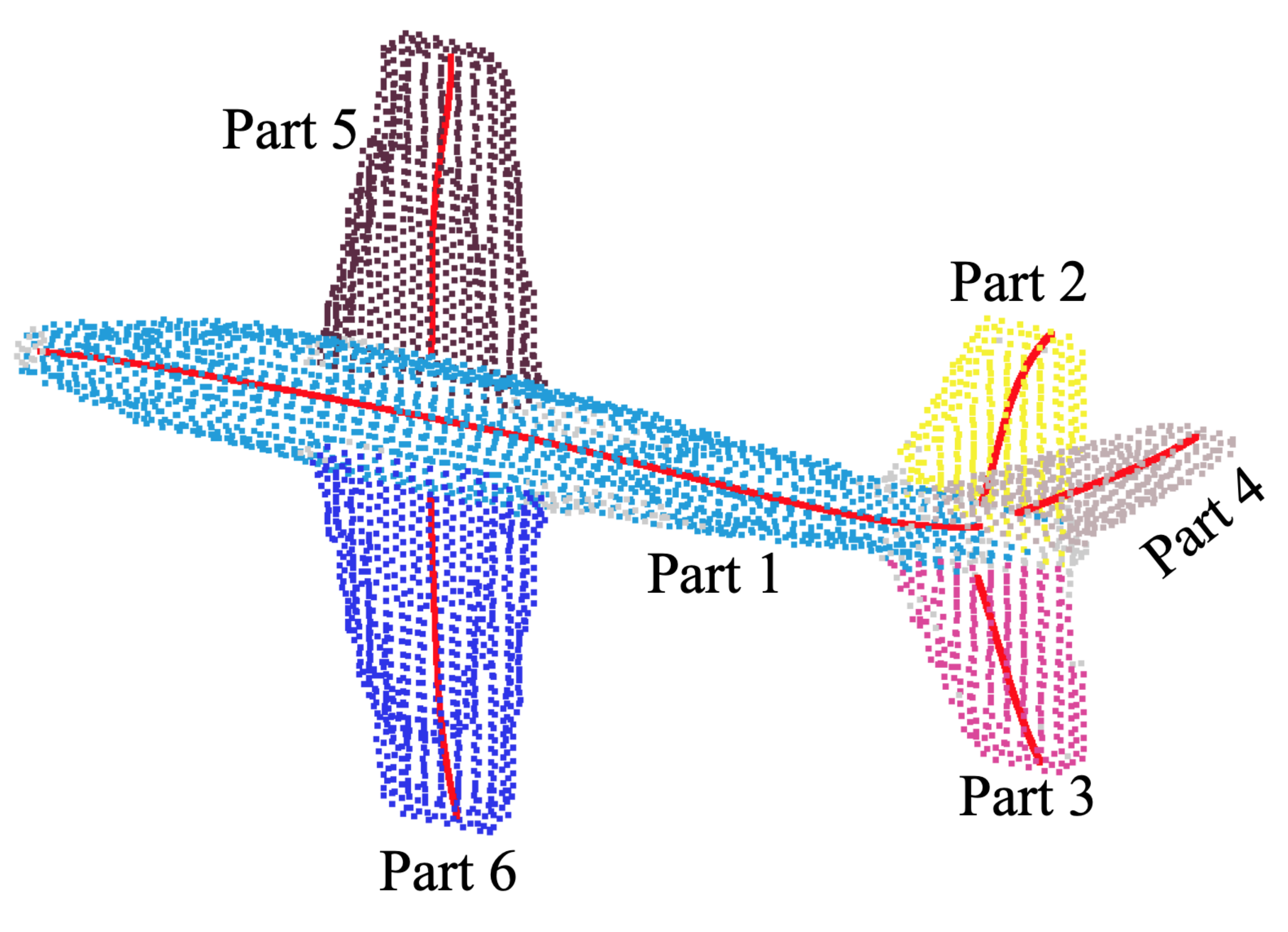}
	\caption{}
\end{subfigure}
\begin{subfigure}[b]{0.23\textwidth}
\centering
	\includegraphics[width=\textwidth]{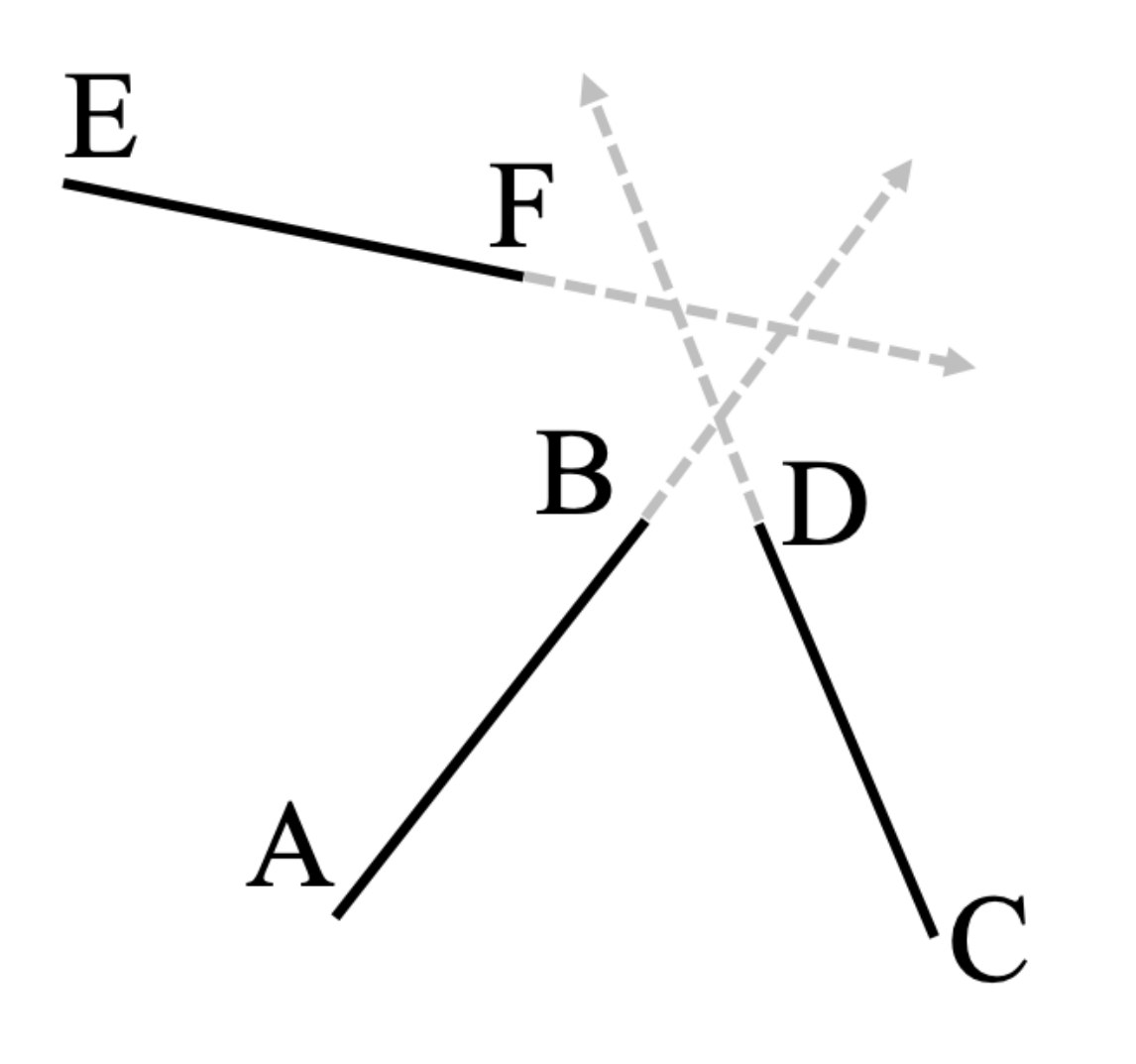}
	\caption{}
\end{subfigure}
\caption{Steps for linking parts. (a) The cross-sectional clusters at the end of parts are compared to see if the two part skeletons needs to be combined. (b) All parts identified by the algorithm are shown for a point cloud of an airplane. Parts 1, 2, 3 and 4 form a clique where the end points of the corresponding parts meet. In this scenario, the algorithm looks for a common point to connect the four skeletons together. (c) AB, CD and EF represent three part skeletons that form a clique of size three.  Rays $\protect \overrightarrow{ AB} $, $\protect \overrightarrow{ CD} $ and $\protect \overrightarrow{ EF} $ are obtained by extending the corresponding skeletons.}
\label{fig:link}
\end{figure*}

The optimization is carried out over the vector $\pmb{x}$, and individual components of $\pmb{x}$ indicate whether a particular part is selected or rejected. I.e., $x^i = 0$ implies that part $i$ is rejected and $x^i = 1$ implies that it is selected. Vector $\pmb{c}$ represents the costs associated with individual parts. Therefore, by minimizing $\pmb{c}^T\pmb{x}$, we are minimizing the overall cost. The trivial and uninteresting solution for $\pmb{x}$ is the vector of all zeros. But we are interested in solutions that cover almost all the points in the point cloud. Note, a point in the cloud is said to be covered if the optimization process selects a part, of which the point is a member. In order to force the optimization process to cover most of the points in the point cloud, we impose the constraint \ref{cstr:c1}. The $i^{th}$ component of vector $\pmb{a}$ gives the number of points belonging to part $i$. Therefore, imposing constraint \ref{cstr:c1} would ensure that at least $k1\%$ of the total number of points in the cloud are covered. The total number of points in the cloud is indicated by $N$. But if two of the selected parts share some points, these points would be counted twice. Note, a point in the point cloud can be a member of multiple parts. In Fig. \ref{fig:overview} (a), for example, two parts (in the top-left corner) covering the tail portion have substantial overlap. If part $i$ covers $n_i$ points, part $j$ covers $n_j$ points and if the two parts share $n_s$ points, then the total number of points covered by selecting both parts $i$ and $j$ is $n_i + n_j - n_s$. To avoid too much overlap between parts, constraint \ref{cstr:c2} is imposed. The element $q_{ij}$ of $\pmb{Q}$, represents the number of points that parts $i$ and $j$ share. Thus constraint  \ref{cstr:c2} ensures that  the sum of pair-wise overlap (in terms of number of points) between parts is less than $k_2 \%$ of the total number of points in the cloud. To see this, note that, the element $q_{ij}$ is relevant only when both parts $i$ and $j$ are selected, i.e., when $x^i =1$ and $x^j = 1$. Note, the lower triangular part of $\pmb{Q}$ is zero to ensure that the pair-wise overlaps are not penalized twice. In short, the constraints ensure that most points in the point cloud are covered with little overlap between parts and the objective function ensures that this is done at a low cost (meaning the best parts are selected). 

The constrained binary integer program, described above, is solved using the Gurobi solver~\cite{VJ-gurobi}. Since, the value of $M$, the number of candidate parts, is usually small (150 or less), the optimization algorithm converges fast, within a few seconds in most cases. The result of the optimization procedure is depicted in Fig. \ref{fig:overview} (b). Once the individual parts that cover the point cloud are identified, the next step is to link them to obtain a skeletal representation of the shape. This procedure is explained in the next section.


\section{Linking Part Skeletons} \label{sec:linking_parts}
The first step in linking skeletons is to determine potential connections among the various parts. For instance, the legs of the dinosaur in Fig. \ref{fig:overview} (b) could be linked to the parts representing the torso or the tail but not to the parts representing the horn or neck of the dinosaur. I.e., a part can only be linked to a subset of other parts and the first step determines this subset of potential links for each part. To determine if two parts can be potentially linked to each other, a very simple observation is used. Two parts, say part A and part B, can be potentially linked only if there is at least one point in A who is the neighbor of some point in B. To determine if any two points in the point cloud are neighbors, the connectivity matrix $ G_{cnct} $, described in section \ref{sec:locally_adaptive_threshold}, is used. 

Linking all the parts that can be potentially linked would lead to strange additional connections like the one shown in Fig. \ref{fig:link_all}. Note, the parts extracted for the dinosaur and the airplane are shown in  Fig. \ref{fig:overview} (b) and Fig. \ref{fig:link} (b) respectively. In the case of the dinosaur, the problem is that parts that could be combined, stay as separate parts. The skeletons for parts representing the tail and the torso of the dinosaur can be combined to form a single smooth curve. This process, of combining skeletal curve where it is possible, will be helpful later when we attempt to connect the legs of the dinosaur to the body. On the other hand if the tail and torso remain as separate parts, interconnecting the two legs, the torso and the tail could be confusing. Therefore, combining skeletons wherever it is possible would be a good first step. To decide if two parts can be combined, we check if the cross-sectional clusters at the end are a good match. This is done by registering the end cross-sectional clusters from the two parts. The end cross-sectional clusters for the torso and the tail of the dinosaur are shown in Fig. \ref{fig:link} (a). After registering the two cross-sections, the registration cost (the same as the one defined in section \ref{sec:cost_components}) is computed. Keep in mind that a lower registration cost means that the points in one cross-sectional cluster, after registration, were able to find, near it, a similarly oriented match from the other cross-sectional cluster. If the registration cost, measured in terms of angular difference, is less than a threshold of  $35^\circ$ and the scale factor difference from one is less than 0.5, it is decided to combine the two parts. These thresholds were chosen after experimenting with a few values.

In the next step,  we deal with parts that form connected components. If a part is only connected to one other part, there is no confusion as to how to link the parts. In Fig. \ref{fig:link} (b), for example, part 5 has only part 1 as a potential link. Therefore, all that needs to be done is to connect par 5 to part 1. The case with part 6 is also the same. But when there are connected components, it is not immediately clear as to how to interconnect them. For instance, in Fig. \ref{fig:link} (b), parts 1, 2, 3 and 4 form a connected component. The ideal way to interconnect them would be to find a junction point, and connect all parts to that single point forming one junction. To find the junction point, rays formed by extending the part skeletons are considered. Fig. \ref{fig:link} (c), shows three part skeletons and the rays obtained by extending them. For each ray, the point that minimizes the sum of distances to the other rays is found. In the example in Fig. \ref{fig:link} (c), for ray $\protect \overrightarrow{AB} $, we find the point on $\protect \overrightarrow{AB} $ that minimizes the sum of distances to rays $\protect \overrightarrow{ CD} $ and $\protect \overrightarrow{ EF} $. The same is done for rays $\protect \overrightarrow{CD} $ and $\protect \overrightarrow{EF} $. Among these points (we get one point per ray), the point with minimum sum of distances to the other rays is chosen as the junction point. 
 
\begin{figure}[htp]
\centering
	\includegraphics[width=0.25\textwidth]{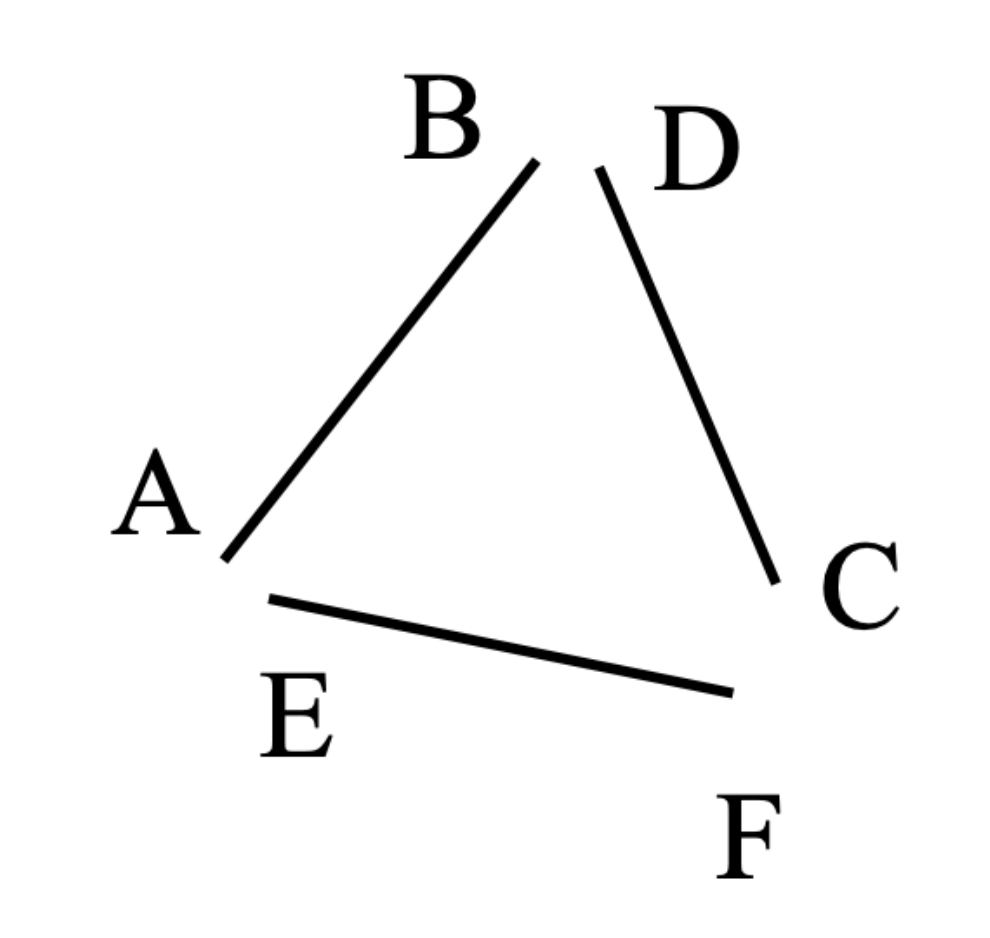}
\caption{An example of a clique of size three for which finding a junction point to interconnect the parts would not be appropriate. The key point is to see if it's the same end point of a part that connects to all other parts in the clique.}
\label{fig:link_triangle_clique}
\end{figure}

\begin{figure*}[htp]
\centering
	\includegraphics[width=0.95\textwidth]{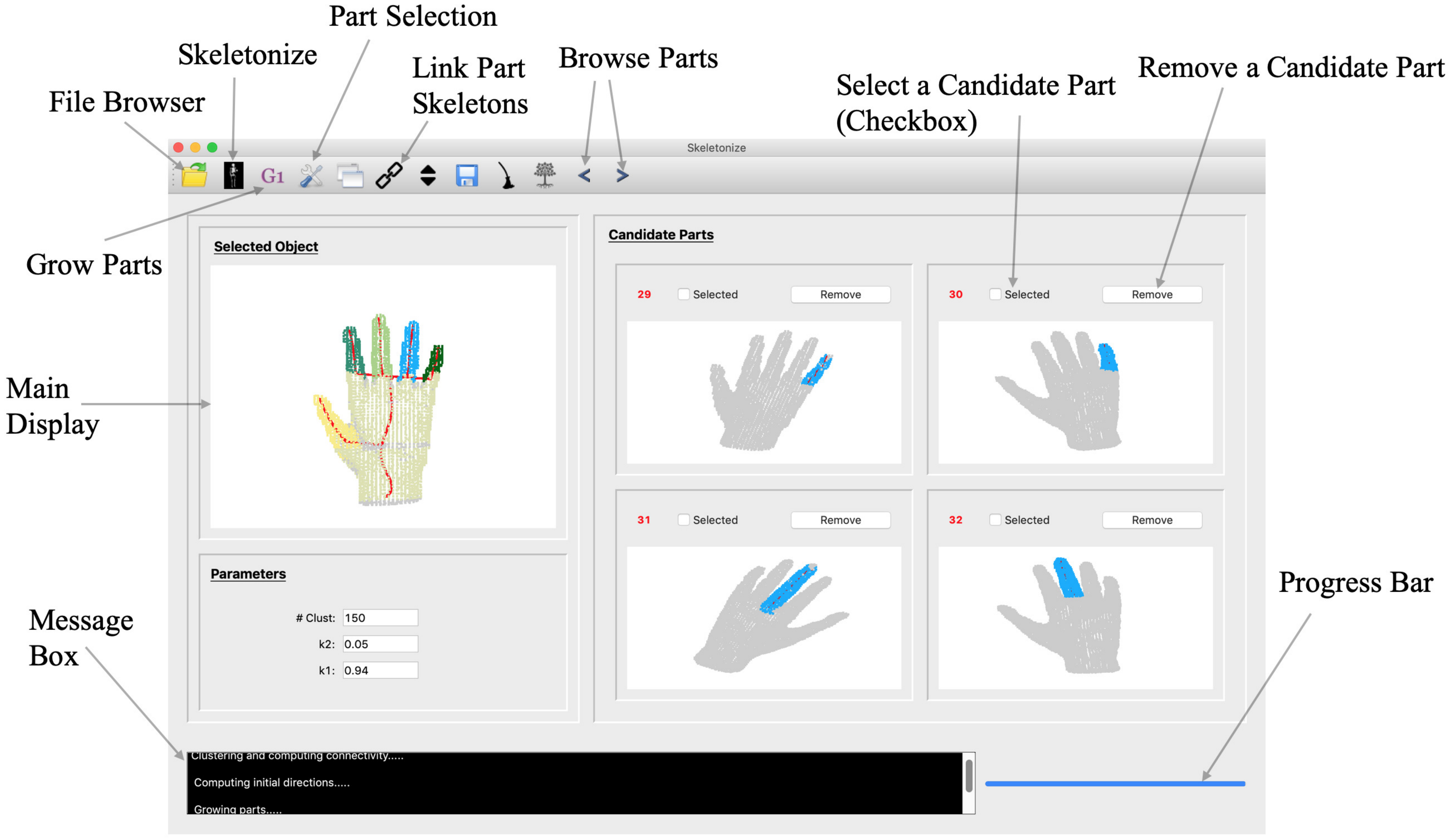}
\caption{The Graphical User Interface}
\label{fig:GUI}
\end{figure*}

Finding a junction point to connect together all the members of the connected component would not be appropriate in the case when both end points of a part are involved in the clique formation. An example of such a scenario is shown in Fig \ref{fig:link_triangle_clique}. Here, each end point of a part has only one potential link and therefore liking these parts together is straightforward. Hence, finding a junction point to link all the members of the connected component, is only appropriate when for each part involved, it is the same end point that connects to all the other parts in the component. Once, the issues depicted in Fig. \ref{fig:link_all} are addressed, all remaining potential connections are made to obtain the final skeleton.



\section{User Interface} \label{sec:GUI_Skel}
One significant advantage of a part based skeleton extraction algorithm is the ease with which users can interact with the algorithm. For point clouds that are not noisy, user interaction is unnecessary. But the part based algorithm can handle very noisy point clouds with little user interaction. To demonstrate the ease of interaction, a graphical user interface (GUI) was developed. Fig. \ref{fig:GUI}, shows the GUI with some annotations. Since the GUI is not the focus of this work, only a brief overview is provided and only the relevant features are explained. To begin the processing, the user can use the file browser to load the point cloud. Once the point cloud is loaded, the main display would show the rotating point cloud. Note, after each processing step the main display will update to show the corresponding results. The toolbar button labelled \Quote{Grow Parts} can be used to grow parts from initial cross-sections. I.e., this button executes steps described in sections \ref{sec:locally_adaptive_threshold}, \ref{sec:init_cross-sections} and \ref{sec:growing_parts}. The grown parts are then displayed on the two-by-two grid display on the right side of the main display. Note, all the displays (the main display and the grid displays) allow user interaction. I.e., the users can play/pause the rotation by double clicking on the display and also manually rotate the display using the mouse/touchpad. The button labelled \Quote{Part Selection}, can then be used to run the optimization routine that selects the optimal subset of parts. And finally, the \Quote{Link Part Skeletons} button can be used to link the individual part skeletons to obtain the final skeleton. The user may also choose to simply click the \Quote{Skeletonize} button to run all the steps (of generating parts, selecting optimal parts and linking selected parts) at once. The message box and the progress bar will provide the user with feedback on the status of the program execution. 

The two buttons labelled \Quote{Browse Parts}, can be used to browse through all the parts generated. The left and right arrows can be used to view the previous and next four parts respectively. After the parts are generated (by using the \Quote{Grow Parts} button), the users, instead of running the optimization routine, may choose to manually select the parts that they think are the best parts. For this purpose, each part shown in the grid display is accompanied with a checkbox that will allow the user to select that part. The users can instead run the optimization routine, which selects the best subset of parts, and then choose to correct mistakes, if any, made by the routine. For all parts that the optimization routine selects, the corresponding checkbox in the grid display would be checked. The user can browse through all the parts, after the optimization routine is run, and decide to reject the parts chosen by the routine by unchecking the corresponding checkbox. The user can also add to the currently selected list of parts by checking the checkbox corresponding to any unselected part. This procedure, allows the user to easily modify the selection of the optimizing routine with a few clicks.  The user also has the option to permanently remove noisy parts from the available list of parts by clicking the \Quote{Remove} button.


\section{Results} \label{sec:results_skel}

\begin{figure}[htp]
\centering
	\includegraphics[width=0.45\textwidth]{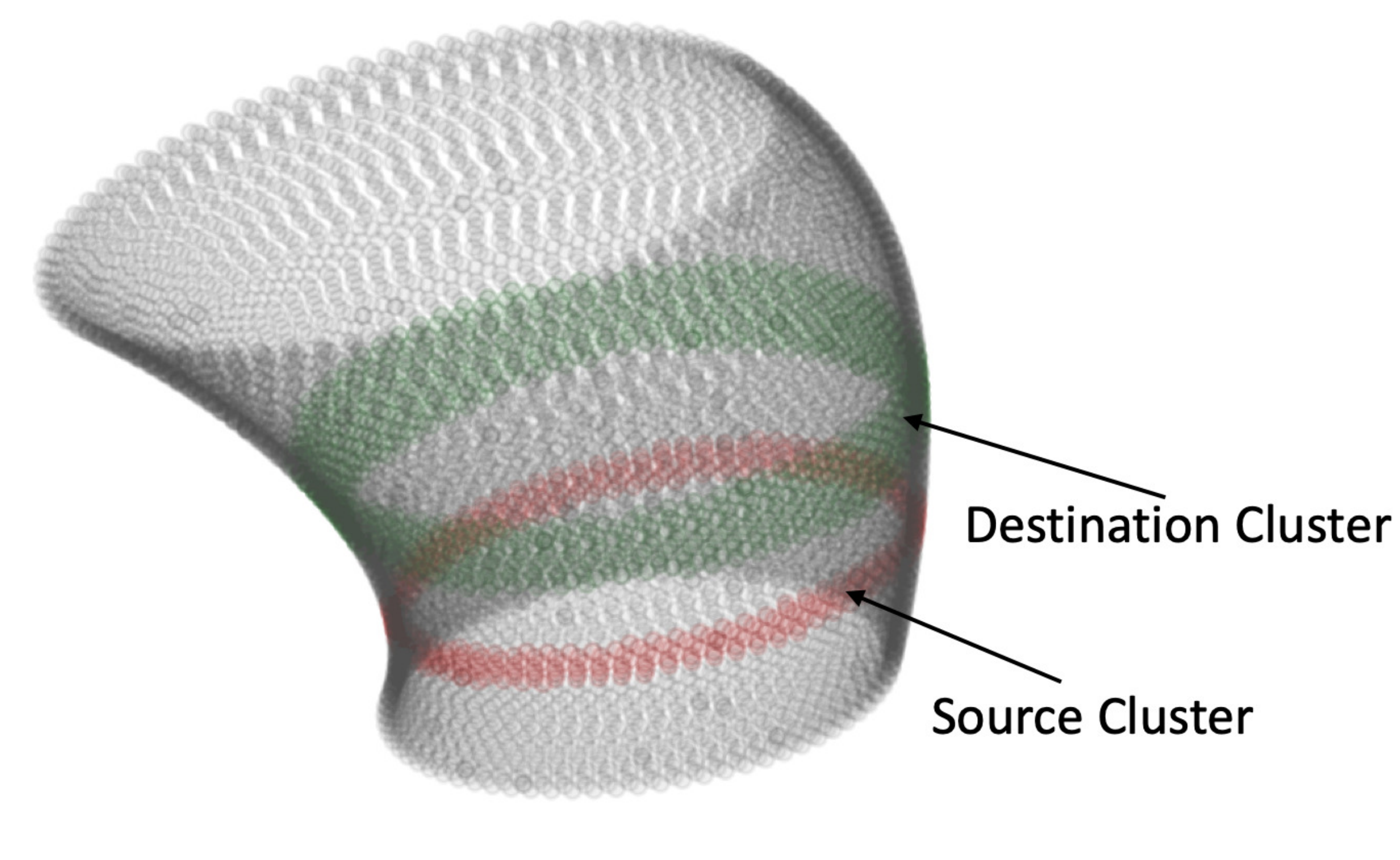}
\caption{ A synthetically generated GC.}
\label{fig:gc_pc}
\end{figure}

The results are divided into two sections. We first present the results for comparison of the our newly proposed registration algorithm with that of Myronenko and Song \cite{Myronenko_2010_CPD}. Next, the results of using the proposed algorithm for skeleton extraction are presented. 

\subsection{Registration Results} \label{sec:reg_results}
As we described above, we use the registration algorithm to grow parts from initial cross-sectional clusters. Therefore, it is appropriate to use a similar setting to test the newly proposed registration algorithm. I.e., we first generate synthetic point clouds that represent GCs and then measure the accuracy of the registration algorithms in registering a random cross-sectional cluster with another randomly chosen neighborhood. For instance, in Fig. \ref{fig:gc_pc}, the point cloud corresponding to a GC is shown. To measure the accuracy of the registration algorithms, we register the source cluster with the destination cluster. Since, the point cloud was generated by us, we can measure the accuracy by comparing the registration results with the ground truth.

\begin{figure}[htp]
\begin{subfigure}[b]{0.255\textwidth}
	\includegraphics[width=\textwidth]{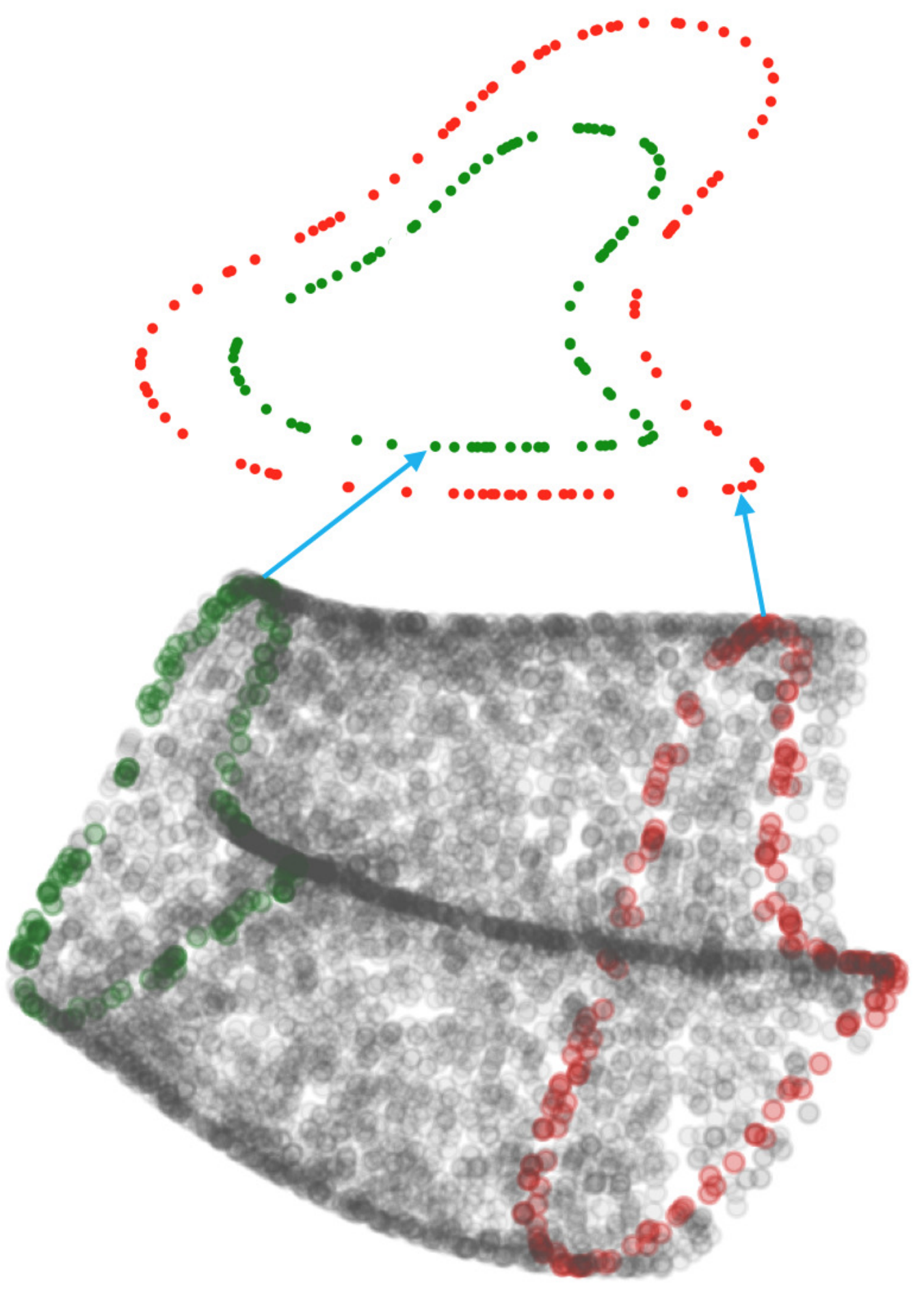}
	\caption{Random sampling}
\end{subfigure}	
\begin{subfigure}[b]{0.225\textwidth}
\centering
	\includegraphics[width=\textwidth]{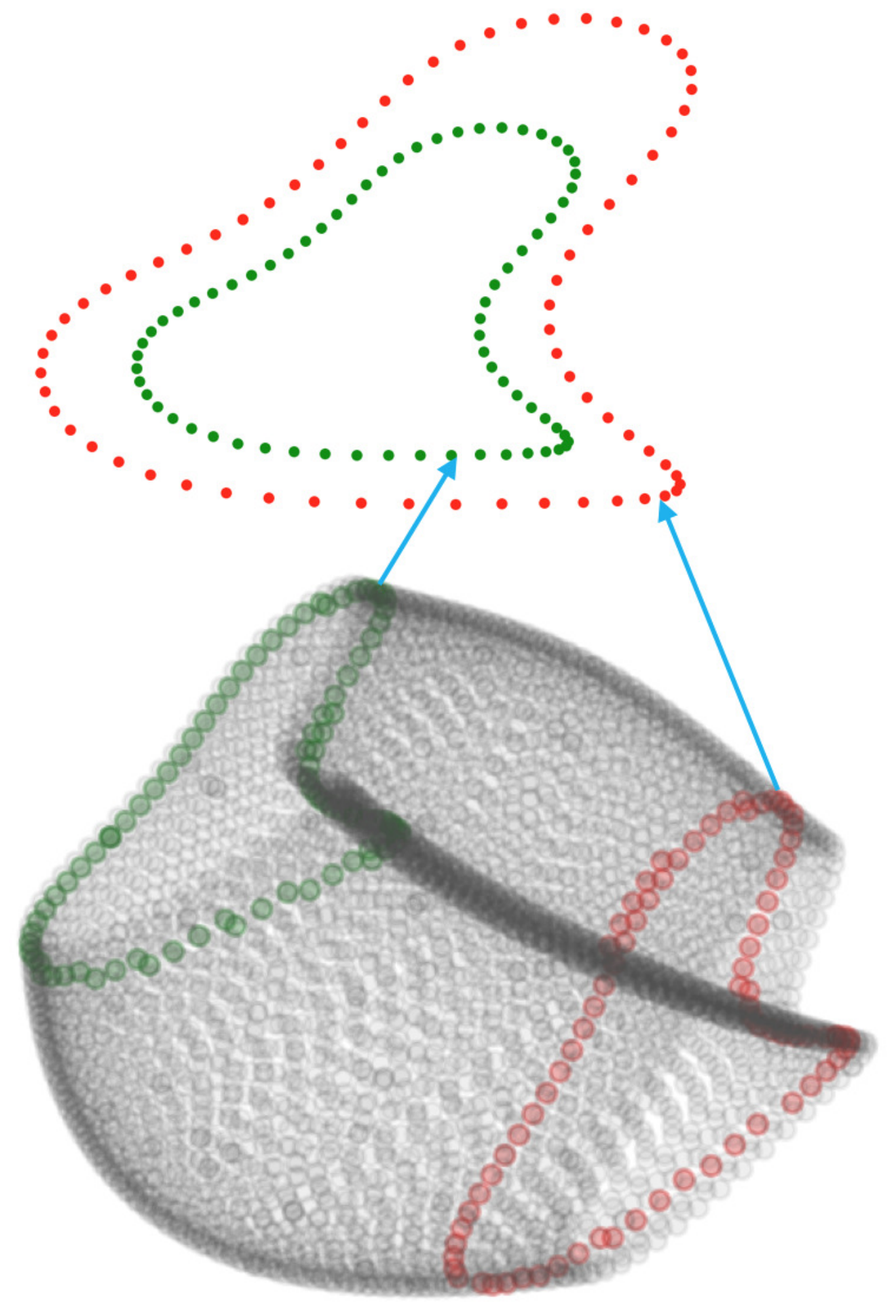}
	\caption{Regular sampling}
\end{subfigure}
\caption{ The random sampling of the point cloud could lead to inaccuracies in the registration if position alone is used to evaluate the fit.}
\label{fig:gc_contour_sampling}
\end{figure}

There are a couple of challenges that the registration algorithm has to overcome in order to accurately estimate the registration parameters. Understanding these challenges is important in understanding why it is appropriate to test the registration algorithms in the setting described above. One of the two challenges is related to the issue of random sampling. I.e., the points on a point cloud can be thought of as random samples from the surface of the shape. Or equivalently, they are random samples from a scaled rotated and translated 2D cross-sectional contour. Samples from two such contours from a point cloud are shown in Fig. \ref{fig:gc_contour_sampling}. The point clouds are shown at the bottom with two contours highlighted, and the highlighted contours are shown at the top. The point cloud in  Fig. \ref{fig:gc_contour_sampling} (a) is obtained by transforming (i.e., scaling, rotating and translating) 2D contours which are sampled at random, while the point cloud in Fig. \ref{fig:gc_contour_sampling} (b) is obtained by transforming 2D contours which are sampled at regular intervals. It is evident that, in the case of contours in Fig. \ref{fig:gc_contour_sampling} (a), due to random sampling, a simple scaling of points in one of the contours would not produce the other point set exactly. Therefore, using just positional information to estimate the registration parameters would lead to inaccuracies in the estimate as we would demonstrate later.  We would expect the registration algorithm to be able to handle point clouds like the one shown in  Fig. \ref{fig:gc_contour_sampling} (a). Adding the orientation information in such a scenario will make the registration process more robust and stable.

To understand the second challenge, keep in mind the process of searching for a match, described in section \ref{sec:method_2}. Starting from an initial cross-sectional cluster, we search for a similar cluster in its neighborhood. The second cluster, therefore, is obtained from the neighborhood of the initial cluster and almost always has a lot more points than the initial cluster for which we seek a match. For example, in Fig. \ref{fig:gc_pc}, the destination cluster is wider and has a lot more points than the source cluster. Looking just at the positional information of the points, to register the source cluster with the destination cluster, could lead to wrong estimates for the rotation parameter. Here again, penalizing large variations in point normal orientation, in addition to penalizing positional disparity, would significantly improve the estimates of the registration parameters. This is especially true if the neighborhood cluster has a lot more points than the initial cluster. Through experiments described below, we show that adding orientation information can effectively deal with these issues. Note, the random sampling issue and the issue of trying to find a match in a larger point cloud are issues encountered when registration is employed to grow parts. Since, the primary use of registration for us is in growing parts, it is only appropriate to test the registration algorithms in such a setting.The process of generating the point clouds is described next, followed by the description of the experiments and their results.

\subsubsection{Point Cloud Generation}
\begin{figure}[htp]
\centering
\begin{subfigure}[b]{0.24\textwidth}
	\includegraphics[width=\textwidth]{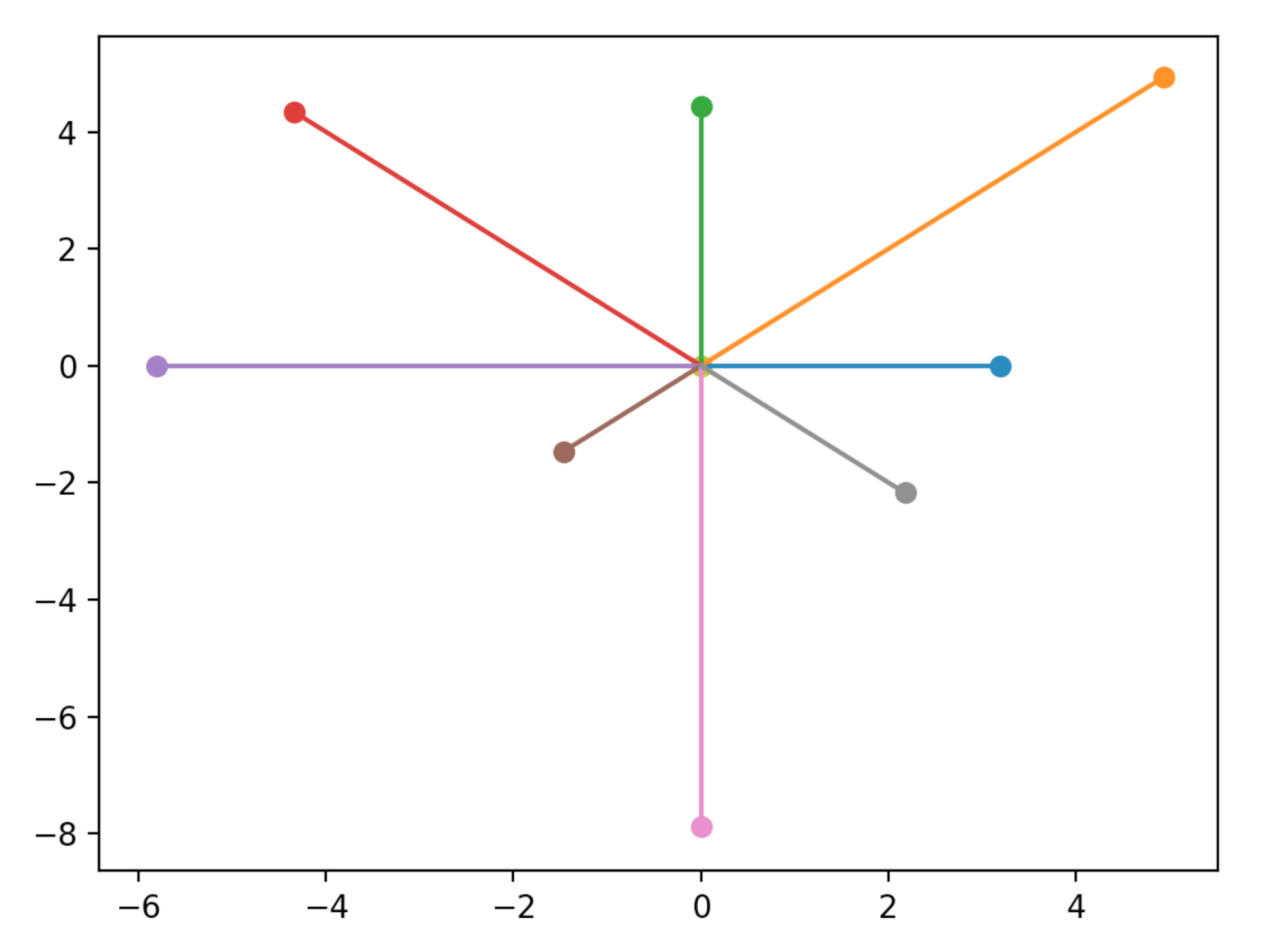}
	\caption{}
\end{subfigure}	
\begin{subfigure}[b]{0.24\textwidth}
\centering
	\includegraphics[width=\textwidth]{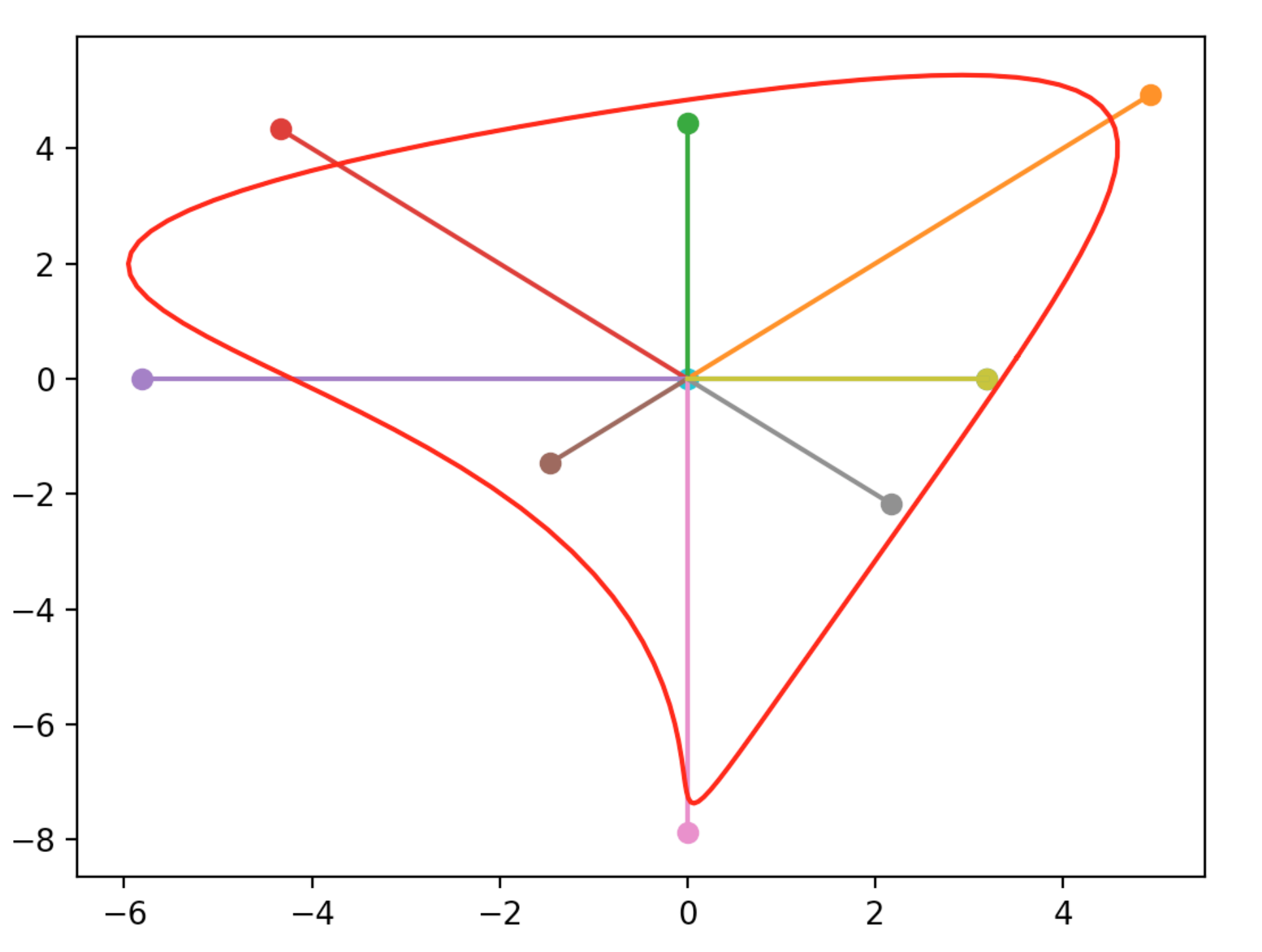}
	\caption{}
\end{subfigure}
\caption{Generating a random 2D cross-sectional contour. (a) Generate a set of points at equal angular interval of $\pi/4$ radians whose distance from the origin is random. (b) Fit a smooth closed contour (shown in red) to these points to obtain the 2D cross-sectional contour.}
\label{fig:2D_contour}
\end{figure}

As shown in Fig. \ref{fig:gc_illustration}, a scale function, a 2D cross-sectional contour and a 3D axis are the three components necessary to generate a GC. The following parametric equation is used to generate the 3D axis of the GC:
$$
( \ x(t),y(t),z(t) \ ) = ( \ C_1 cos(t), C_2 sin(t), C_3 t \ )
$$
Where, $C_1, C_2$ and $C_3$ are random real numbers in the interval $(0,50)$. To obtain a smooth scale function, a sinusoidal curve with a random phase shift is used. Since scale values cannot be negative, a constant offset is added to the phase shifted sinusoid, to ensure that all the values are positive. In fact, the offset added is such that all values in the scale function are greater than or equal to one. To generate the 2D cross-sectional contour, we first generate a set of eight points as shown in Fig. \ref{fig:2D_contour}. The angular intervals between the points are equal and their distances to the origin is randomly chosen. We then fit a smooth, periodic, spline to these points to obtain the 2D cross-sectional contour. 

To obtain a GC, the 3D axis is sampled at regular intervals. In Fig. \ref{fig:gc_illustration}, the red blobs on the 3D axis represent this discretization. The key idea in the construction of a GC is to place a rotated and scaled 2D cross-sectional contour at each sample point along the 3D axis. In fact, the 2D cross-sectional contour is translated in addition to rotation and scaling and the translation is determined by the 3D axis. The smooth scale function, defined above, is used to scale the cross-sectional contour along the 3D axis. To obtain a smooth rotation function along the axis, we employ the Frenet-Serret frame. I.e., at each point on the 3D axis, the tangent, normal, and binormal unit vectors of the axis, determines the rotation that needs to be applied at that point. Therefore, defining the 3D axis simultaneously defines the rotation that needs to be applied to the 2D cross-sectional contour along the 3D axis. Fig. \ref{fig:gc_illustration} (a) shows the GC generated when rotated and scaled 2D cross-sectional contours are placed at the sample points along the 3D axis. Since, we are interested in a point cloud representation of the GC, the 2D contour is sampled before applying the scaling, rotation and translation transformations. Note, the number of samples chosen for the 2D contour and the 3D axis would determine the sparsity/density of the point cloud. 

\subsubsection{Experiments and Results}

\begin{figure*}[htp]
\centering
\begin{subfigure}[b]{0.49\textwidth}
	\includegraphics[width=\textwidth]{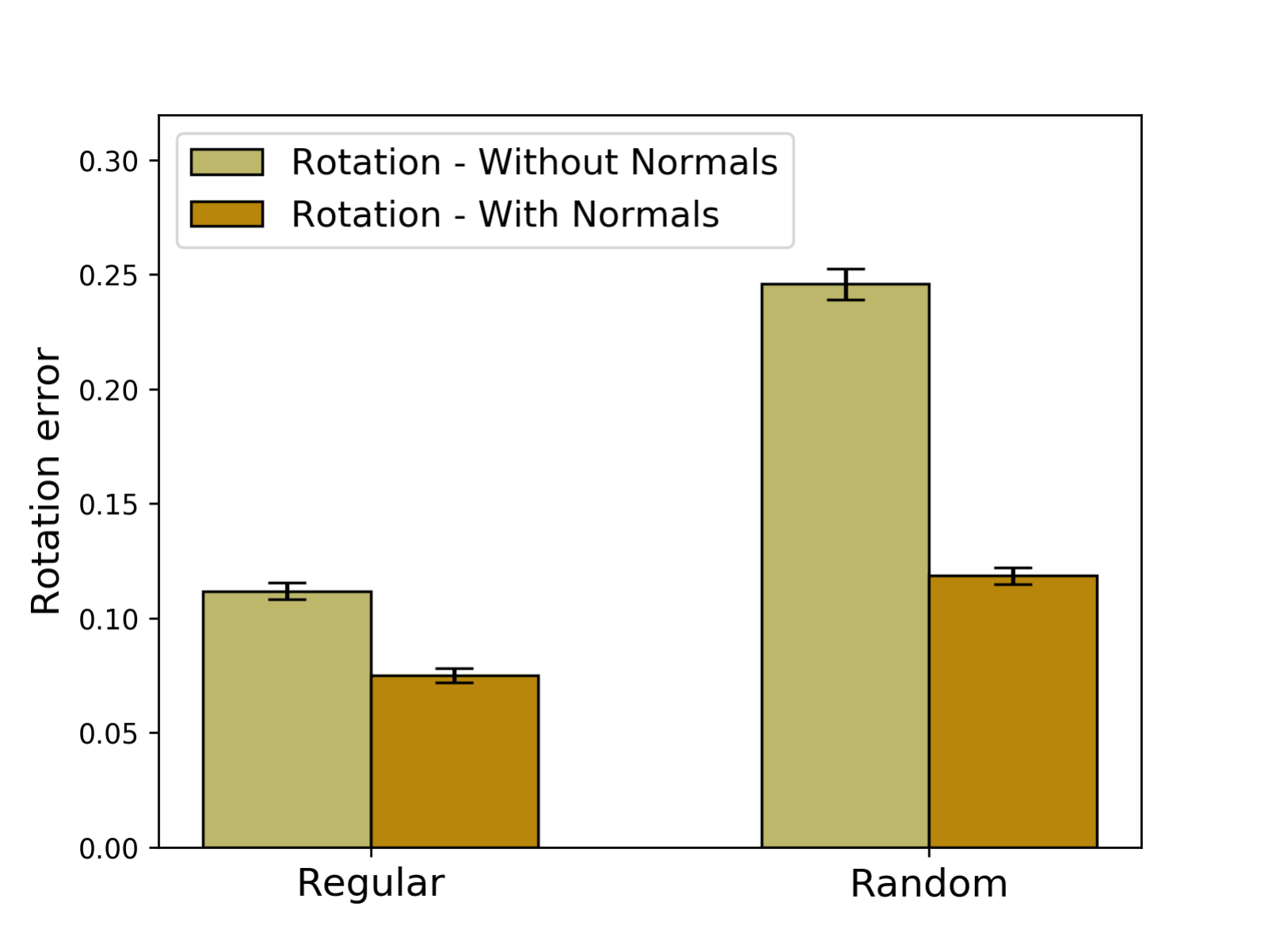}
	\caption{}
\end{subfigure}	
\begin{subfigure}[b]{0.49\textwidth}
\centering
	\includegraphics[width=\textwidth]{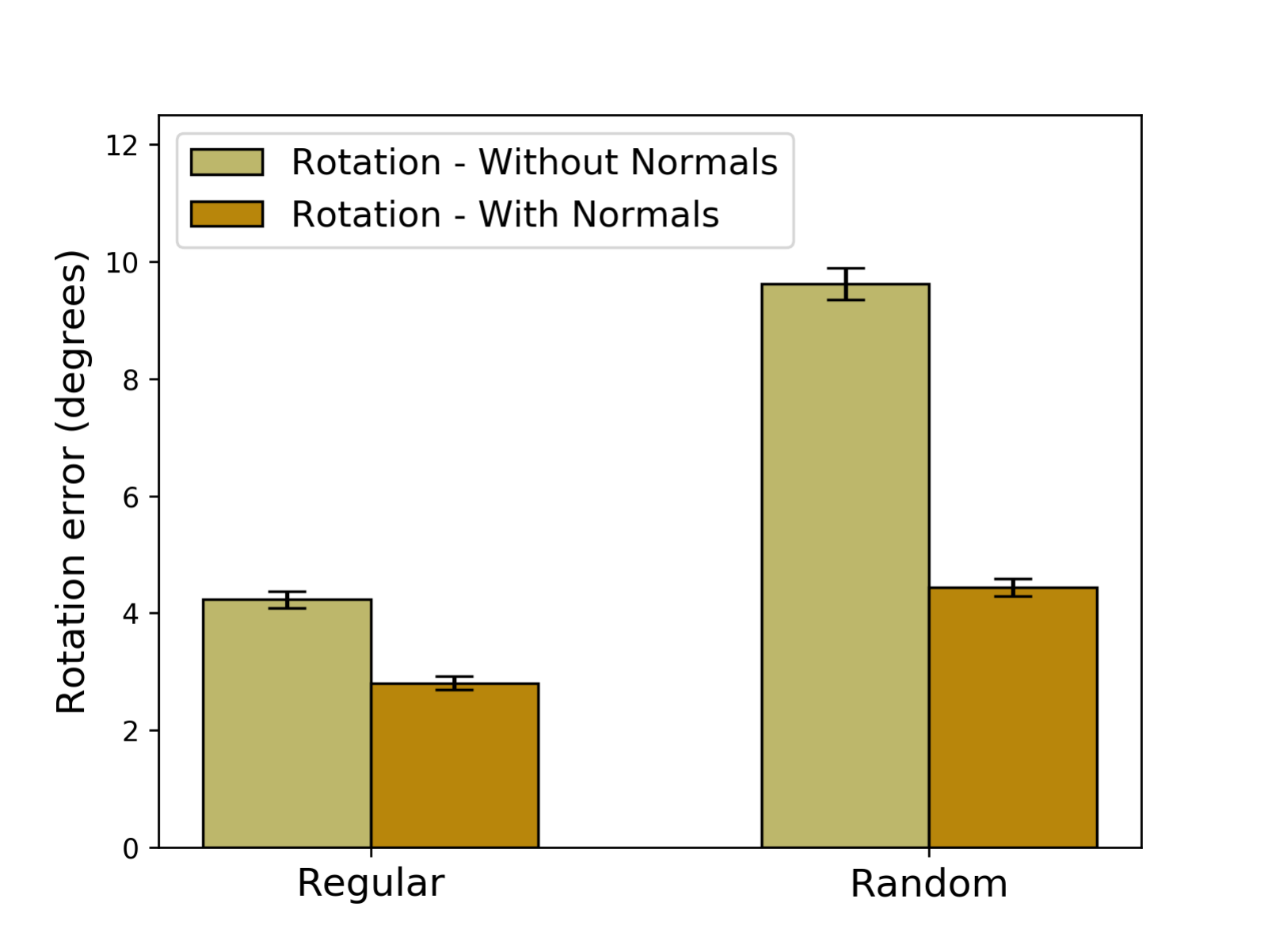}
	\caption{}
\end{subfigure}
\begin{subfigure}[b]{0.49\textwidth}
\centering
	\includegraphics[width=\textwidth]{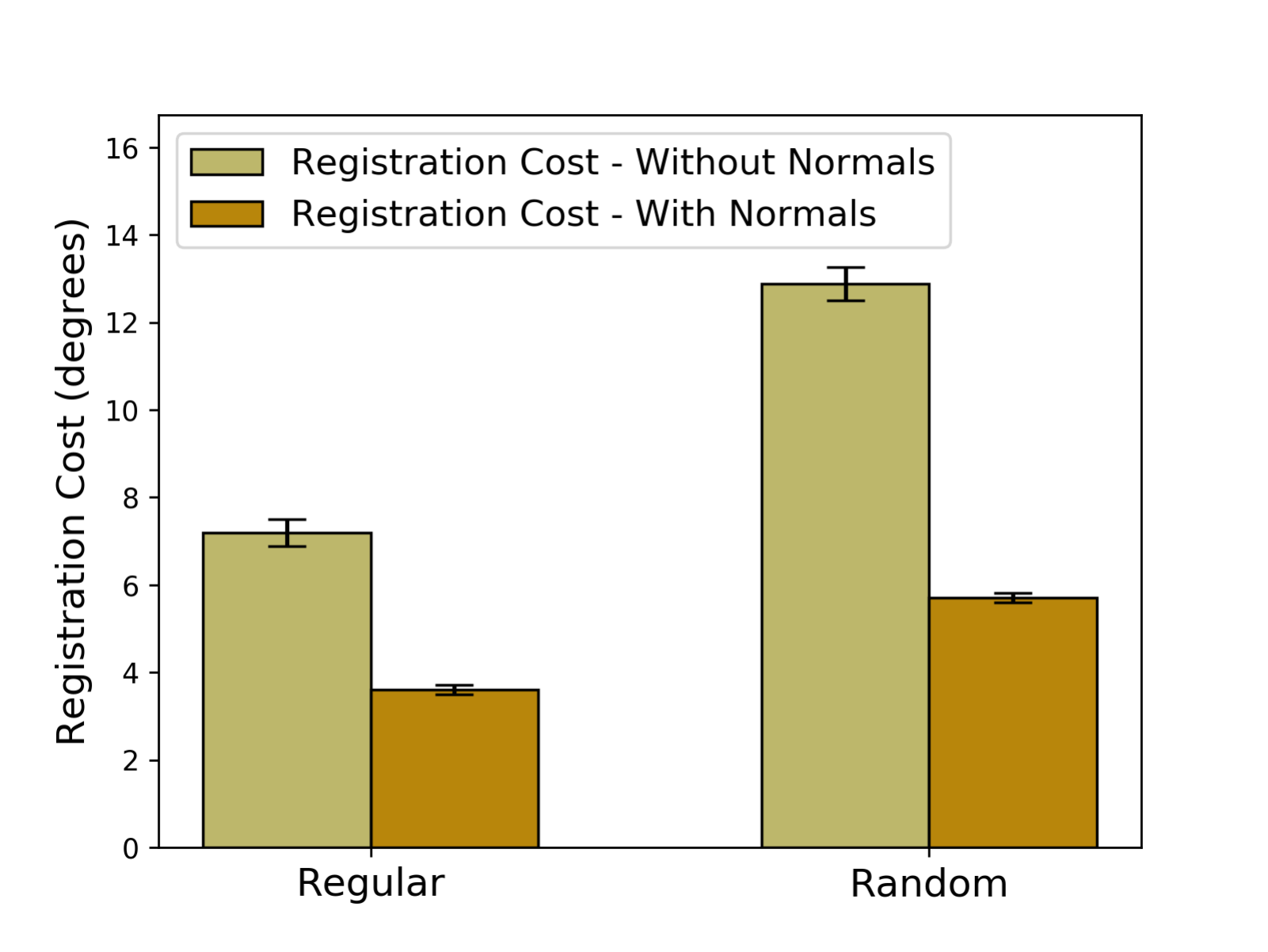}
	\caption{}
\end{subfigure}
\begin{subfigure}[b]{0.49\textwidth}
\centering
	\includegraphics[width=\textwidth]{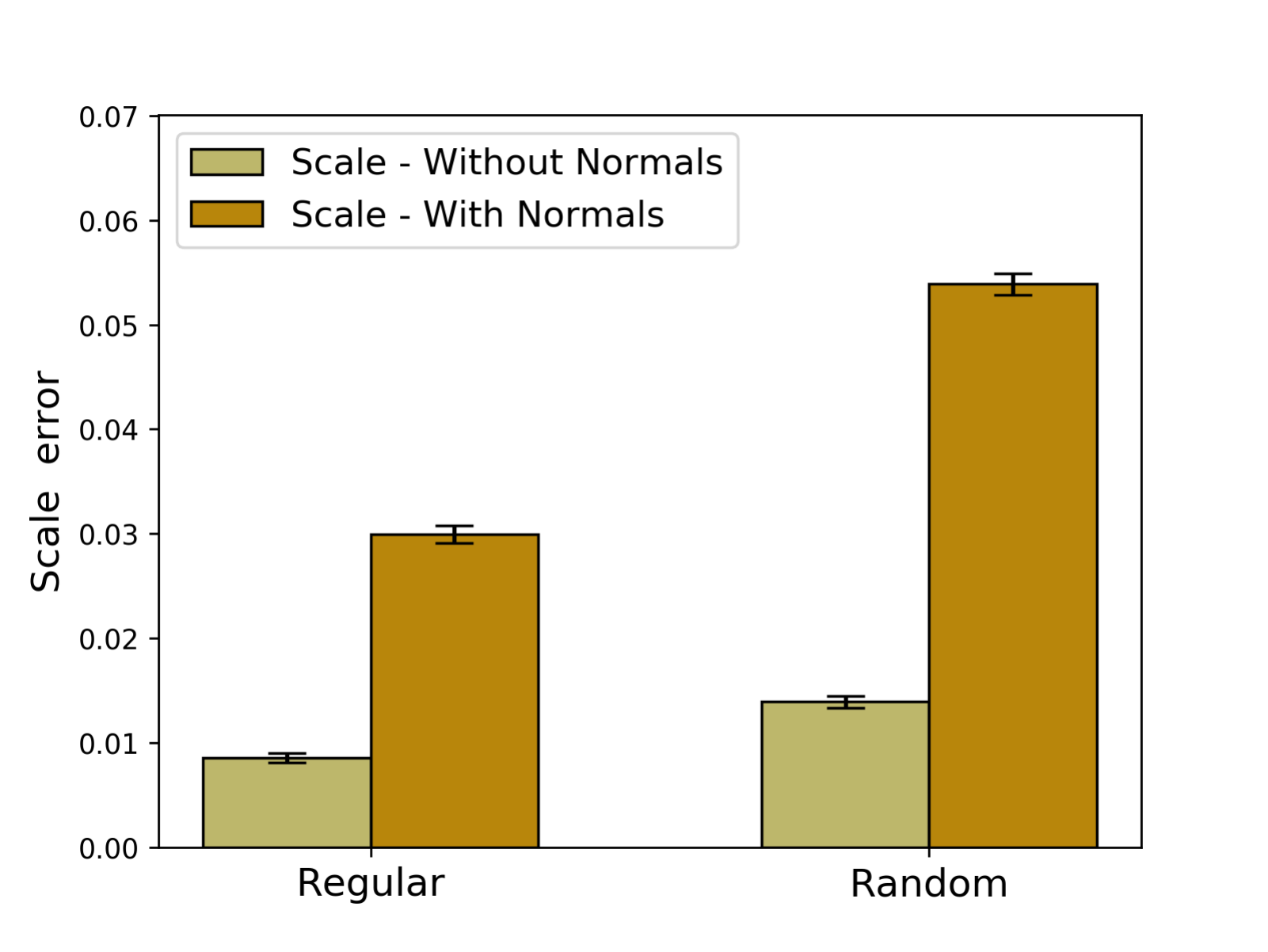}
	\caption{}
\end{subfigure}

\caption{In the legend, \Quote{With Normals} refers to the proposed method and  \Quote{Without Normals} refers to the method in  \cite{Myronenko_2010_CPD}. (a) Error in estimated rotation, expressed in terms of the Frobenius norm of the difference in rotation matrices. (b) Error in the estimated orientation of the cross-sectional plane, expressed as the angular difference between the plane normals in degrees. (c) Comparison of the registration costs (same as the one defined in section \ref{sec:cost_components}). (d) Error in estimation of the scaling parameter.}
\label{fig:reg_bar_plots}
\end{figure*}

As mentioned in section \ref{sec:reg_results} and as depicted in Fig. \ref{fig:gc_pc}, to compare the registration algorithms, we register a randomly chosen source cluster with a randomly chosen destination cluster within the GC. We consider two different cases for comparing the registration accuracies. In one case we sample the 2D cross-sectional contour, before transforming it, randomly (as shown in Fig. \ref{fig:gc_contour_sampling} (a)) and in the other case we sample the contour at regular intervals  (as shown in Fig. \ref{fig:gc_contour_sampling} (b)). 


Fig. \ref{fig:reg_bar_plots}, shows the results of comparing the two algorithms for both the cases. The results were obtained by performing five hundred registrations. I.e., a random GC point cloud was generated and a randomly chosen source cluster, within the GC, was registered with a randomly chosen destination cluster. This process was repeated five hundred times to obtain the results. Note, in the plots, \Quote{With Normals} refers to our method as it utilizes the point normal information. Similarly, the case in which the 2D contour is sampled at regular intervals is referred to as \Quote{Regular} and the other case is referred to as \Quote{Random}. 

In Fig. \ref{fig:reg_bar_plots} (a), the mean error, from the five hundred registrations, in estimating the rotation matrix is shown. In  \cite{huynh2009metrics}, the author discusses various metrics employed to measure the difference between rotation matrices. We use the one based on the Frobenius norm which is given by: $$err_{rot}( \pmb{R}_1,  \pmb{R}_2) \ = \ \| \pmb{I}_3 - \pmb{R}_1 \pmb{R}_2^T \|_F$$ where $\pmb{R}_1$ and $\pmb{R}_2$ are the two rotation matrices we want to compare. The value of this metric will be in the range $[0, 2\sqrt{2}]$. As shown in Fig. \ref{fig:reg_bar_plots} (a), the proposed method that uses the normal information does better than the one in \cite{Myronenko_2010_CPD} in both cases. The difference is much higher in the random case. This is expected because an error function based just on positional information is less capable of dealing with the random sampling. Since, it is a little difficult to interpret the metric values for the rotation error, we also plot the error in the estimated orientation of the cross-sectional plane in Fig. \ref{fig:reg_bar_plots}  (b). These plots resemble the plots for the rotation error, but are easier to interpret as the error is measured in terms of the angle between the actual normal and the estimated normal of the cross-sectional plane and is expressed in degrees. In the context of growing parts, the accurate estimation of the cross-sectional plane orientation is crucial because an error in this estimate will accumulate as more registrations are performed to grow the part.

The comparisons for the registration cost, defined in section \ref{sec:cost_components}, are presented in Fig. \ref{fig:reg_bar_plots} (c). Remember, the registration cost is mean difference in the point normal orientation between a point in the source cluster and its best match (found using the correspondence probability) in the destination cluster. Also keep in mind that, as described in section \ref{sec:specific_registration}, if this cost is above the threshold of $15^\circ$, the part growth algorithm will decide that it is unable to find a good match in the neighborhood, and will terminate the growth process. Therefore, keeping this cost low is essential to obtain fully grown parts. As shown in Fig. \ref{fig:reg_bar_plots} (c), proposed algorithm outperforms the other method in both the cases, significantly so in the case of random sampling. Also note that the rise in this cost, when comparing the regular case with the random case, is moderate for our registration method. The use of point normal information leads to a much more stable algorithm.

Fig. \ref{fig:reg_bar_plots} (d), shows the error in scale estimate. As it can be seen, the other method (i.e., the method in  \cite{Myronenko_2010_CPD}) has a smaller scale error compared to our method. To understand this result better, keep in mind the observation that the estimated value of the parameter,  $\sigma$,  for our method is almost always much larger than the value estimated by the other method. For the other method, that only uses the positional information, to lower the cost, the points in the source cluster has to get as close as possible to the points in the destination cluster, and at the same time keep the value of $\sigma$ as small as possible. This can be done by tolerating bigger differences in point normal orientation. But for our method, since it uses both the orientation as well as positional information, getting closer to points in the destination cluster at the expense of larger differences in point normal orientation is not desirable. Moreover, the error in scale estimates is only about five percent on average in the worst case, which is not a significant error and does not have much consequences in the context of growing parts. This is true because the estimated parameters will decide which points in the neighborhood of a cross-sectional cluster would form the adjacent cluster. Therefore, the error is scale estimates is tolerable, as long as the error does not prevent the algorithm from picking the right neighboring points. Keep in mind that the value of the estimated $\sigma$ is used to determine the neighbors (see the explanation of method 2 in page \pageref{sec:specific_registration} ). This makes small errors in the scale estimate insignificant in the context of growing parts, especially because of the higher $\sigma$ value. As mentioned earlier, the error in rotation parameter estimates will accumulate as we perform multiple registrations to grow parts and therefore it is very significant to get it right. This again shows us why using the point normal orientation information is important in the context of using registrations to grow parts.

And finally, to demonstrate the ability of our algorithm to handle the second challenge of registering the source cluster against a much larger destination cluster, mentioned in section \ref{sec:reg_results}, we ran the following experiment.  Like before, we performed five hundred random registration on synthetically generated GC point clouds. For each GC point cloud, we chose a random source cluster and two different neighboring destination cluster from within the GC. One of the neighboring cluster being much larger (more number of points included) than the other. We registered the source cluster against the two neighboring destination clusters for both methods. We then looked at the proportion of cases, from among the five hundred pairs of registrations for each method, where the registration with a larger destination cluster lead to larger errors in the estimated rotation parameter. For our registration method, in 16.2\% of the cases, using a larger destination cluster (which is equivalent to looking for a match in a larger neighborhood in the context of growing parts) lead to larger errors in the estimation of the rotation parameter. This proportion was 72.2\% in the case of the other method. This simply shows that our method is more robust to variations in the neighborhood size as we search for a matching cross-sectional cluster while growing parts.



\begin{figure*}[!t]
        \centering
        \begin{tabular}{ccccccc}
            \begin{minipage}[t]{0.18\paperwidth}
            \centering
            \includegraphics[keepaspectratio, width=1.0\linewidth]{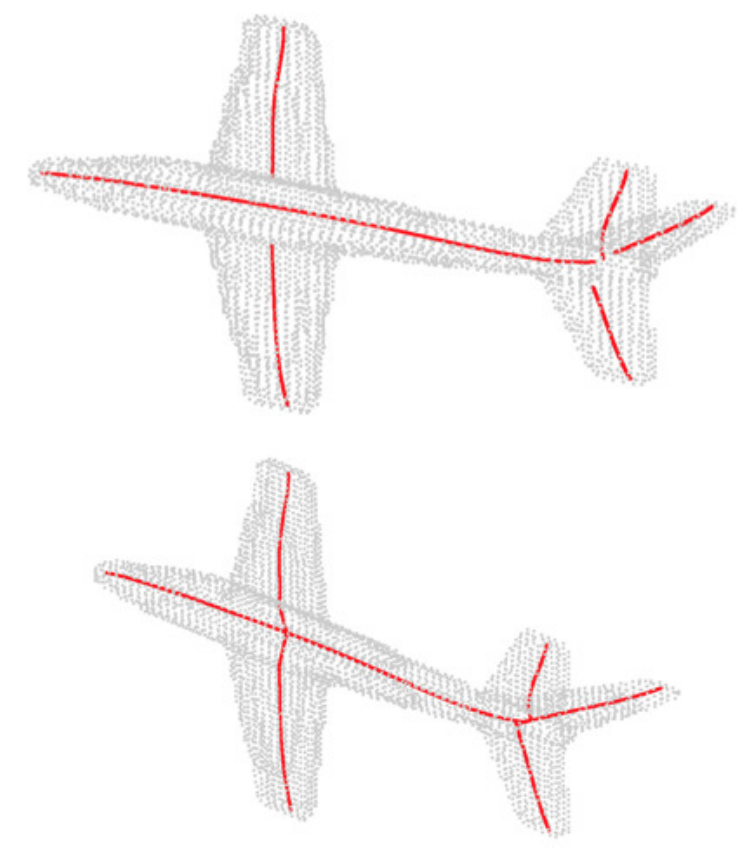}
            \end{minipage}
 & 
	\begin{minipage}[t]{0.18\paperwidth}
            \centering
            \includegraphics[keepaspectratio, width=0.60\linewidth]{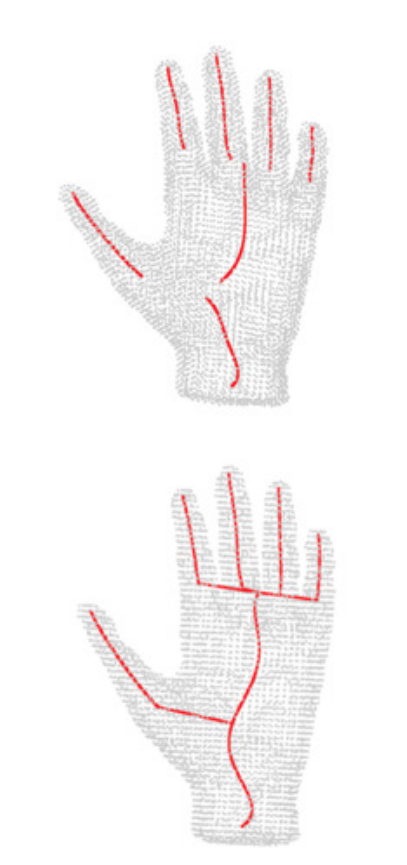}
            \end{minipage}
 &
 	\begin{minipage}[t]{0.18\paperwidth}
            \centering
            \includegraphics[keepaspectratio, width=0.60\linewidth]{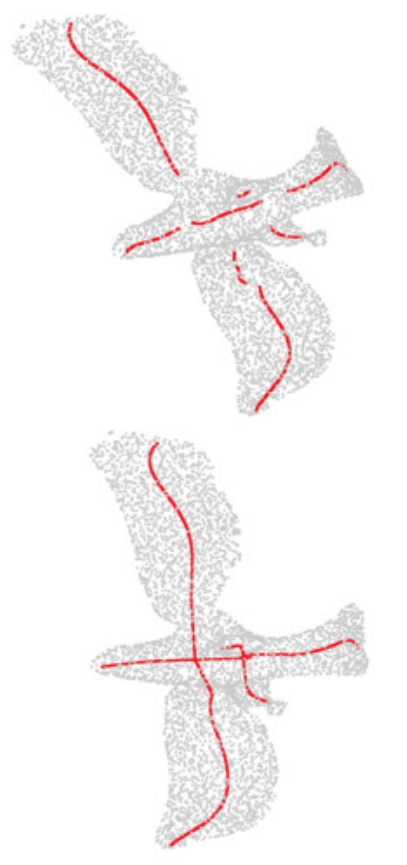}
            \end{minipage}
 & 
        \begin{minipage}[t]{0.18\paperwidth}
            \centering
            \includegraphics[keepaspectratio, width=1.1\linewidth]{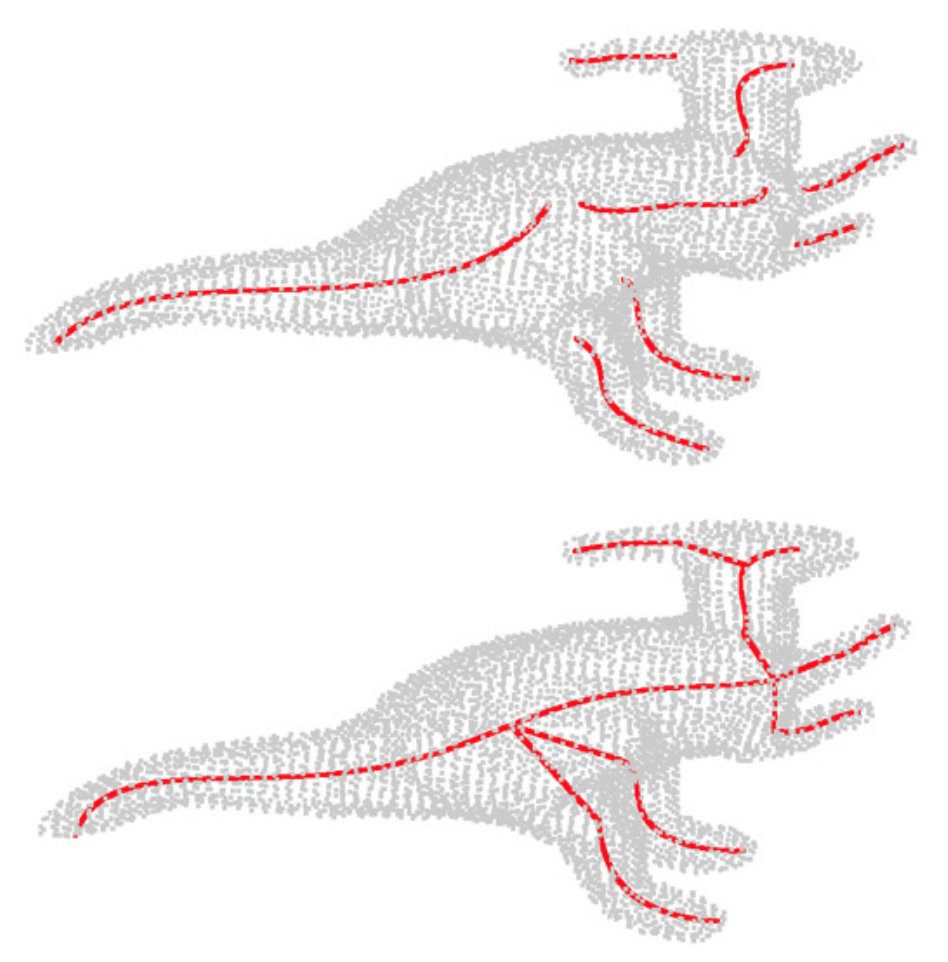}
            \end{minipage}
 \\
            (a) & (b) &  (c) & (d)  \\
            
 \\
      
            \begin{minipage}[t]{0.18\paperwidth}
            \centering
            \includegraphics[keepaspectratio, width=1.0\linewidth]{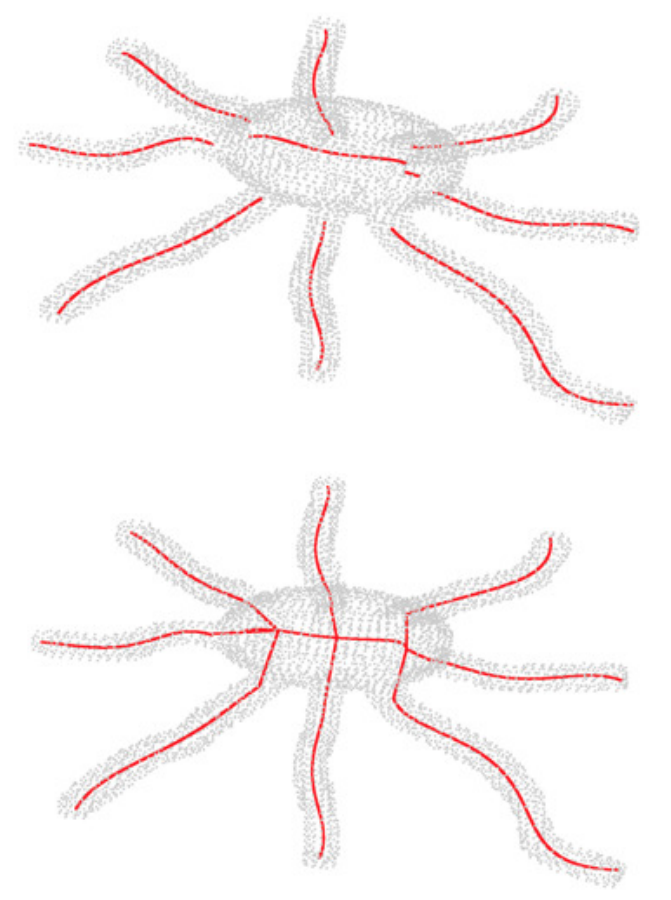}
            \end{minipage}
 & 
	\begin{minipage}[t]{0.18\paperwidth}
            \centering
            \includegraphics[keepaspectratio, width=0.65\linewidth]{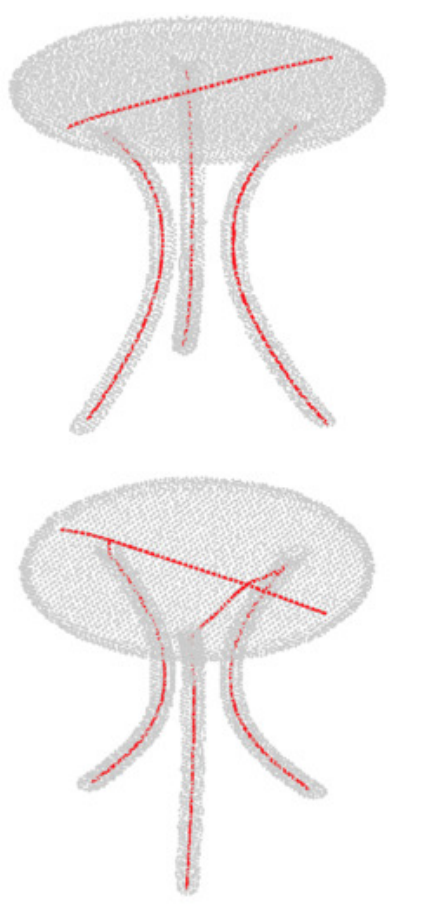}
            \end{minipage}
 &
 	\begin{minipage}[t]{0.18\paperwidth}
            \centering
            \includegraphics[keepaspectratio, width=0.65\linewidth]{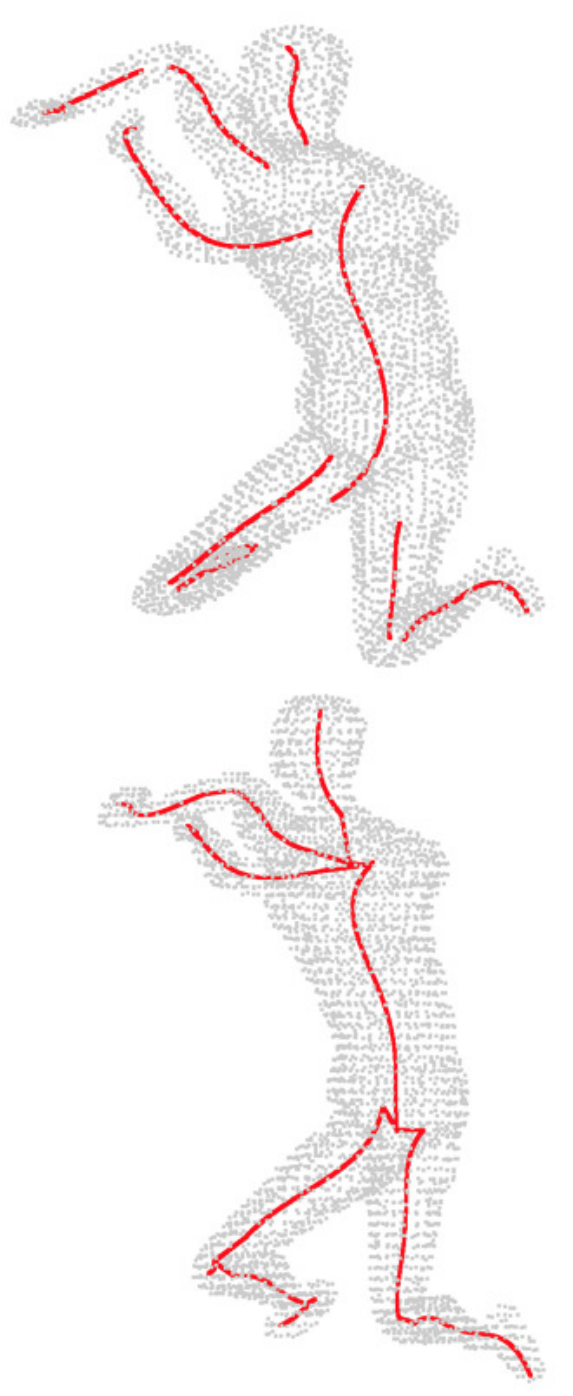}
            \end{minipage}
 & 
        \begin{minipage}[t]{0.18\paperwidth}
            \centering
            \includegraphics[keepaspectratio, width=1.0\linewidth]{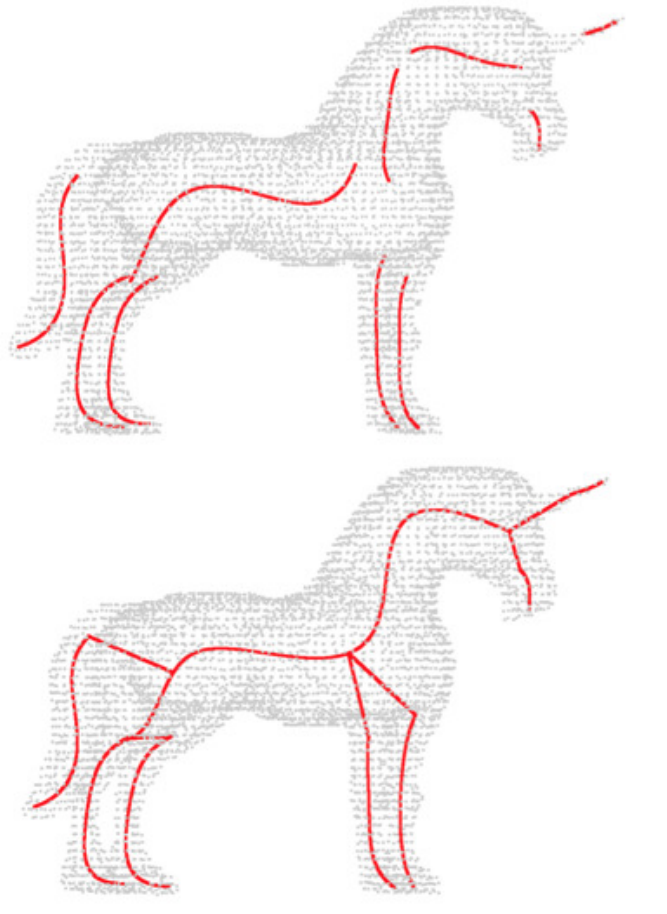}
            \end{minipage}
            
  \\
            (e) & (f) &  (g) & (h)  \\
   \\
      
            \begin{minipage}[t]{0.18\paperwidth}
            \centering
            \includegraphics[keepaspectratio, width=1.0\linewidth]{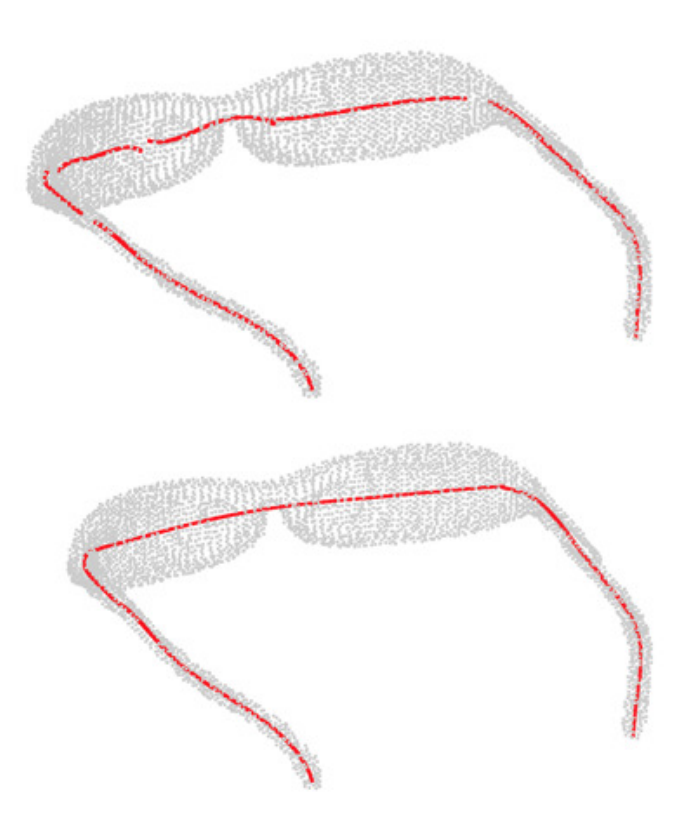}
            \end{minipage}
 & 
	\begin{minipage}[t]{0.18\paperwidth}
            \centering
            \includegraphics[keepaspectratio, width=0.7\linewidth]{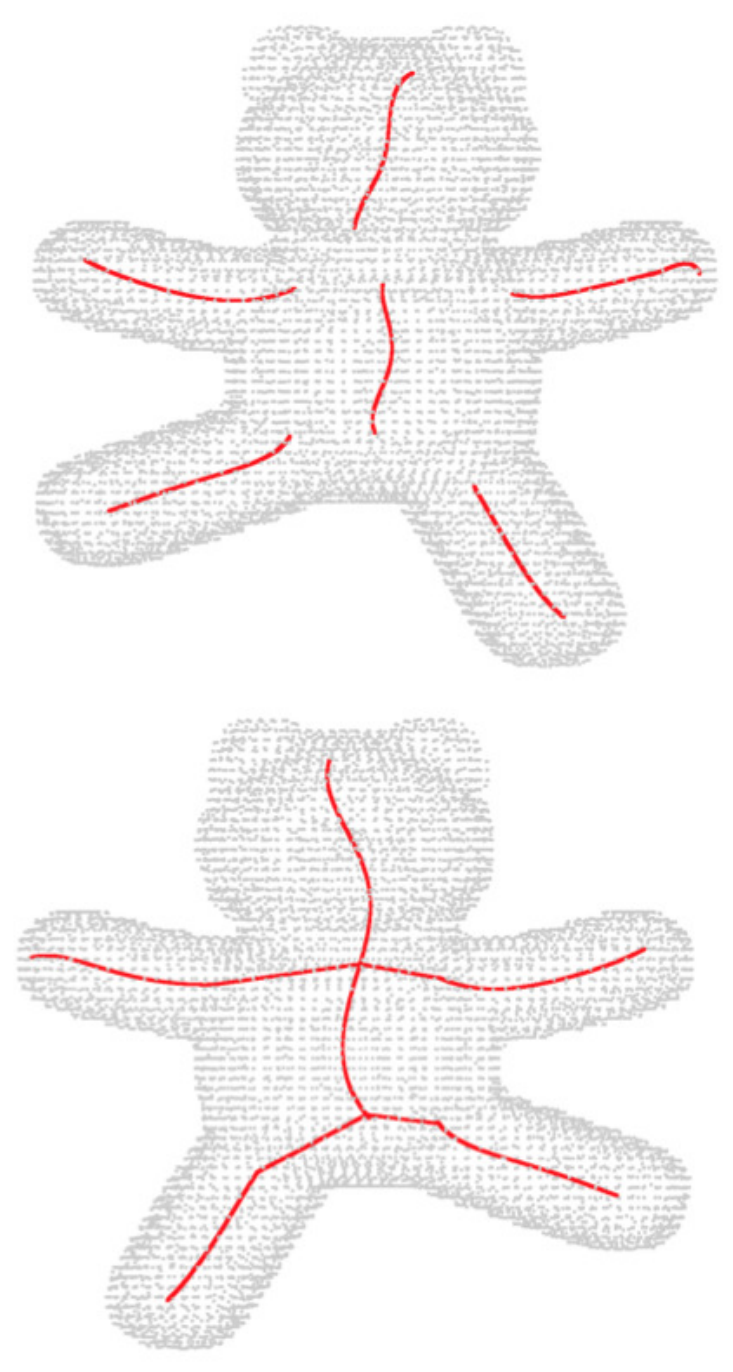}
            \end{minipage}
 &
 	\begin{minipage}[t]{0.18\paperwidth}
            \centering
            \includegraphics[keepaspectratio, width=0.70\linewidth]{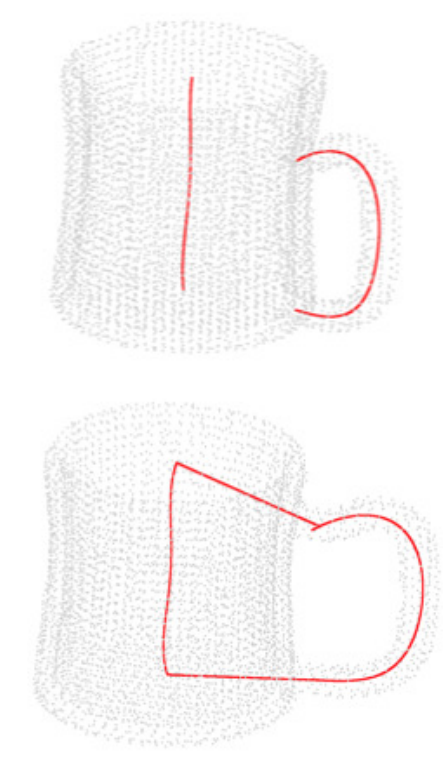}
            \end{minipage}
 & 
        \begin{minipage}[t]{0.18\paperwidth}
            \centering
            \includegraphics[keepaspectratio, width=1.2\linewidth]{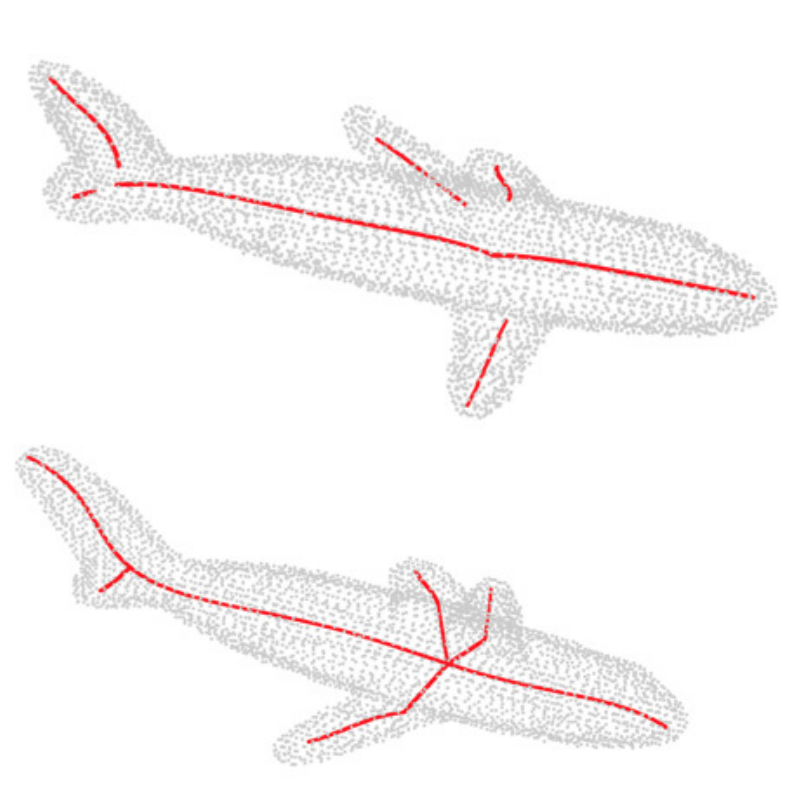}
            \end{minipage}
            
  \\
            (i) & (j) &  (k) & (l)               
            
\\
        \end{tabular}
        \caption{Results for part identification and skeleton extraction. For each shape the individual skeletons corresponding to the parts identified by the algorithm are shown at the top and the linked final skeleton is shown at the bottom. }
    \label {fig:skeleton_results} 
    \end{figure*}

\subsection{Skeleton Extraction Results} \label{sec:skel_results}
Skeleton extraction results for a variety of shapes are shown in Fig. \ref{fig:skeleton_results}. For each shape the individual skeletons for the parts identified by the algorithm are shown at the top and the linked skeleton is shown at the bottom. An extended gallery of results for skelton extraction, along with rotating 3D animations, is available at: \url{https://web.ics.purdue.edu/~vthottat/skel/}. The models were obtained mainly from the McGill 3D shape benchmark \cite{zhang2005retrieving} and from the 3D object recognition database, ObjectNet3D \cite{xiang2016objectnet3d}. As can be seen in Fig. \ref{fig:skeleton_results}, the algorithm is capable of handling a wide variety of shapes. Both, flat objects like the circular table top (Fig. \ref{fig:skeleton_results} (f)) and spectacles (Fig. \ref{fig:skeleton_results} (i)), and rounder objects like the unicorn (Fig. \ref{fig:skeleton_results} (h)) or teddy bear (Fig. \ref{fig:skeleton_results} (j)) are handled well.

\begin{figure*}[!t]
        \begin{tabular}{M{1.25cm} | M{2.0cm} M{2.0cm} M{2.0cm} M{2.0cm} M{2.0cm} M{2.0cm} M{2.0cm}}
            \toprule \\
             & (a) & (b) & (c) & (d) & (e) & (f) & (g) \\
            \midrule
            \\        

            Cao et al. 

  &
            \begin{minipage}[t]{0.08\paperwidth}
            \centering
            \includegraphics[keepaspectratio, width=1.1\linewidth]{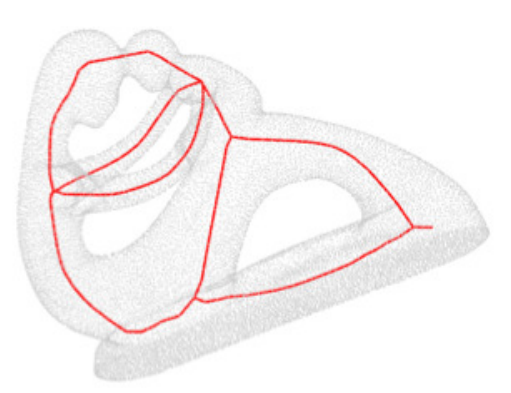}
            \end{minipage}
              &
            \begin{minipage}[t]{0.08\paperwidth}
            \centering
            \includegraphics[keepaspectratio, width=0.8\linewidth]{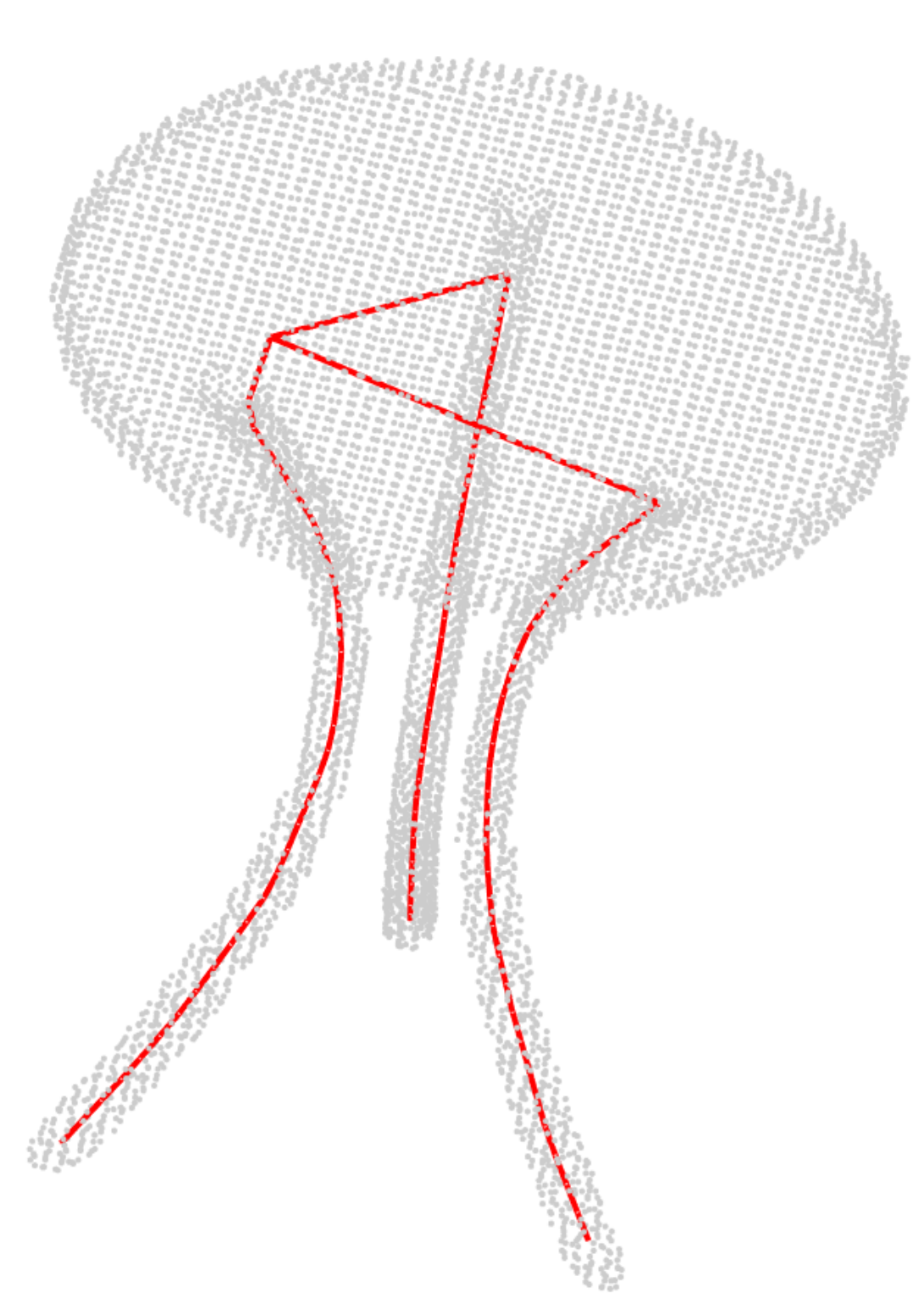}
            \end{minipage}
              &
            \begin{minipage}[t]{0.08\paperwidth}
            \centering
            \includegraphics[keepaspectratio, width=0.93\linewidth]{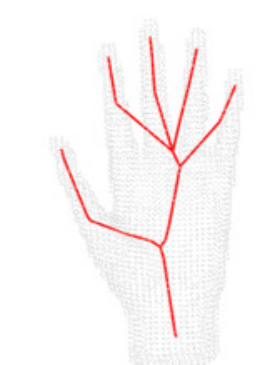}
            \end{minipage}
              &
            \begin{minipage}[t]{0.08\paperwidth}
            \centering
            \includegraphics[keepaspectratio, width=1.1\linewidth]{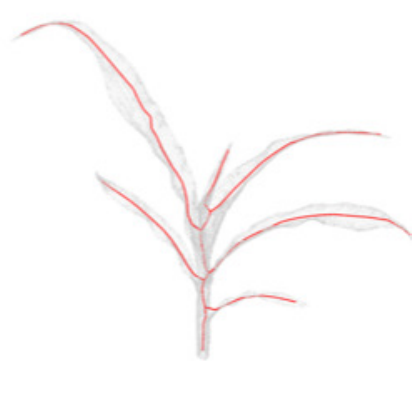}
            \end{minipage}
              &
            \begin{minipage}[t]{0.08\paperwidth}
            \centering
            \includegraphics[keepaspectratio, width=0.9\linewidth]{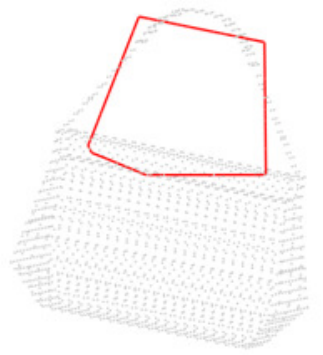}
            \end{minipage}
              &
            \begin{minipage}[t]{0.08\paperwidth}
            \centering
            \includegraphics[keepaspectratio, width=1.0\linewidth]{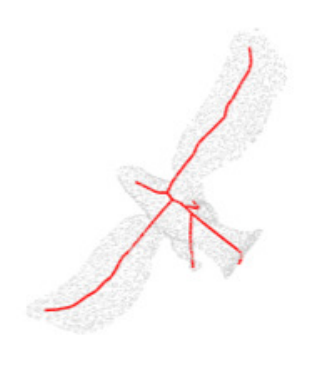}
            \end{minipage}
              &
            \begin{minipage}[t]{0.08\paperwidth}
            \centering
            \includegraphics[keepaspectratio, width=1.2\linewidth]{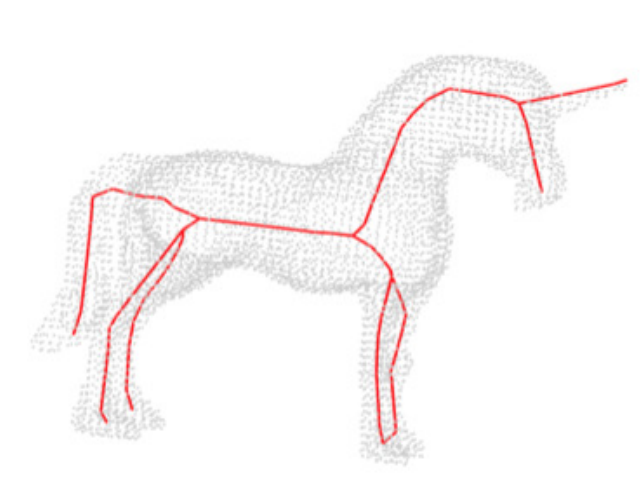}
            \end{minipage}
\\           
            Our Results       
             &
	\begin{minipage}[t]{0.08\paperwidth}
            \centering
            \includegraphics[keepaspectratio, width=1.1\linewidth]{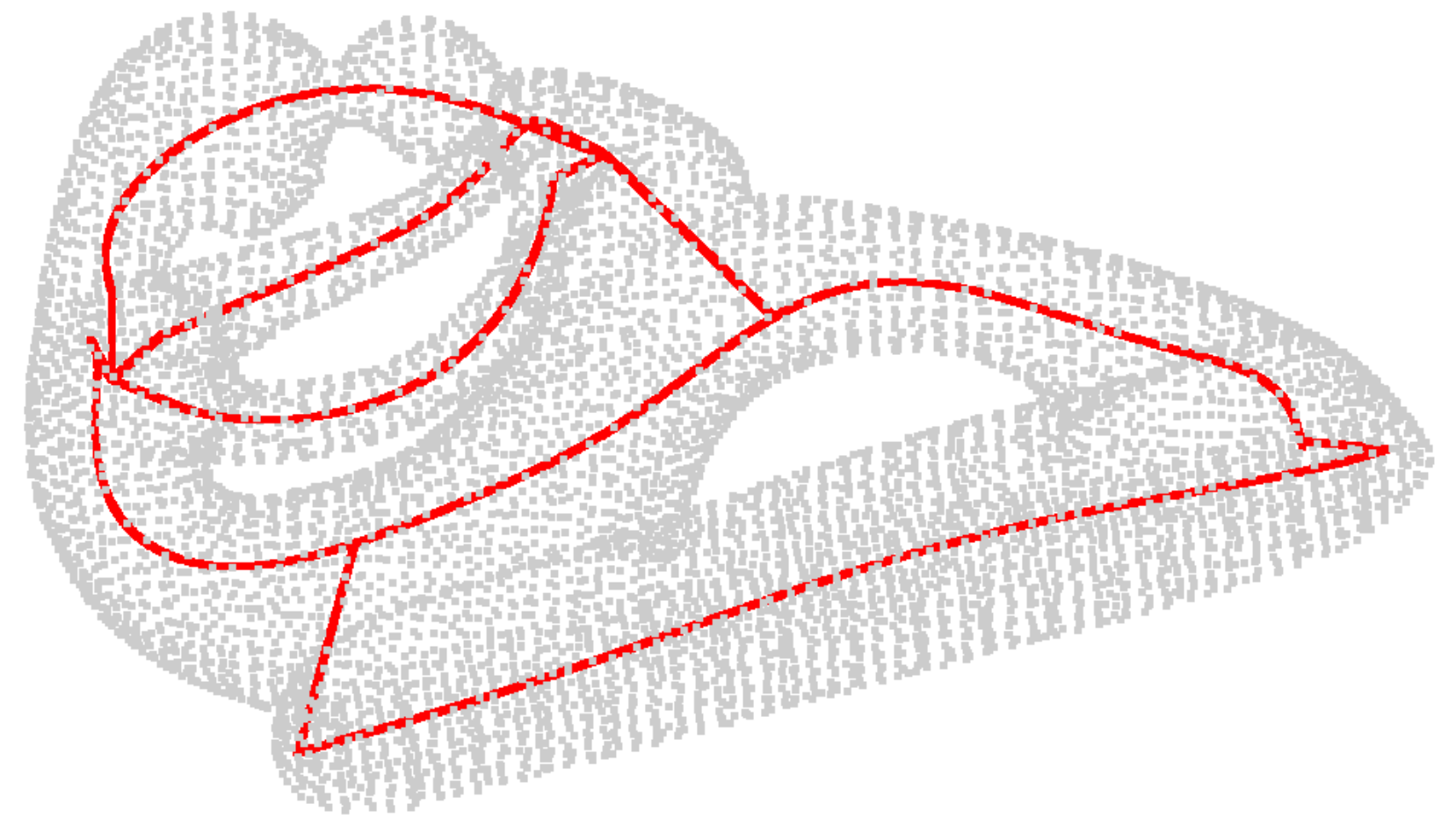}
            \end{minipage}
             &
	\begin{minipage}[t]{0.08\paperwidth}
            \centering
            \includegraphics[keepaspectratio, width=0.85\linewidth]{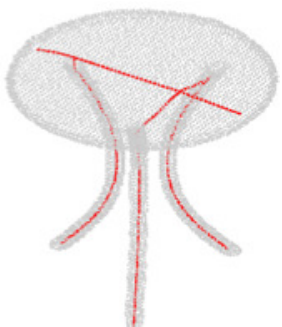}
            \end{minipage}
             &
	\begin{minipage}[t]{0.08\paperwidth}
            \centering
            \includegraphics[keepaspectratio, width=0.775\linewidth]{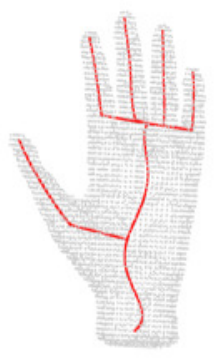}
            \end{minipage}
             &
	\begin{minipage}[t]{0.08\paperwidth}
            \centering
            \includegraphics[keepaspectratio, width=1.1\linewidth]{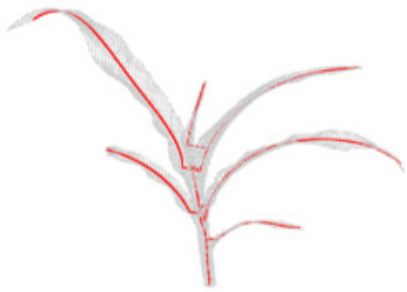}
            \end{minipage}
             &
	\begin{minipage}[t]{0.08\paperwidth}
            \centering
            \includegraphics[keepaspectratio, width=1\linewidth]{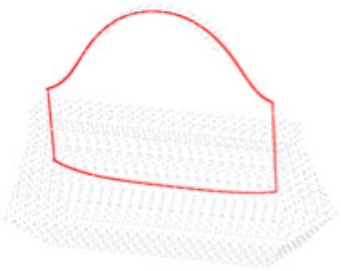}
            \end{minipage}
             &
	\begin{minipage}[t]{0.08\paperwidth}
            \centering
            \includegraphics[keepaspectratio, width=0.825\linewidth]{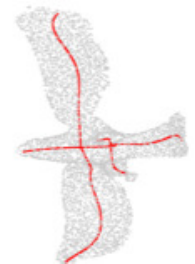}
            \end{minipage}
             &
	\begin{minipage}[t]{0.08\paperwidth}
            \centering
            \includegraphics[keepaspectratio, width=1.1\linewidth]{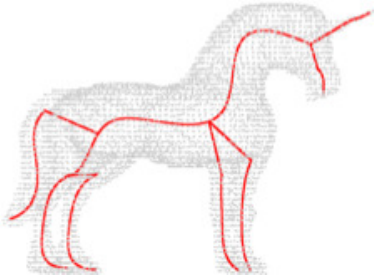}
            \end{minipage}
            
 \\
            Huang et al. 
	&
 	\begin{minipage}[t]{0.08\paperwidth}
            \centering
            \includegraphics[keepaspectratio, width=0.85\linewidth]{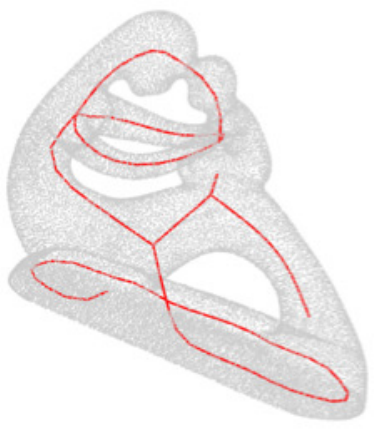}
            \end{minipage} 
            &
 	\begin{minipage}[t]{0.08\paperwidth}
            \centering
            \includegraphics[keepaspectratio, width=0.85\linewidth]{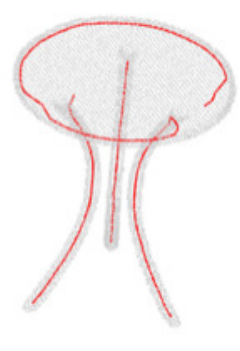}
            \end{minipage}    
            &
 	\begin{minipage}[t]{0.08\paperwidth}
            \centering
            \includegraphics[keepaspectratio, width=0.7\linewidth]{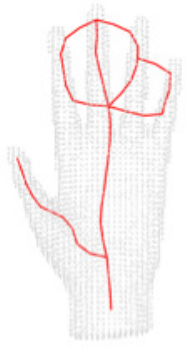}
            \end{minipage}    
            &
 	\begin{minipage}[t]{0.08\paperwidth}
            \centering
            \includegraphics[keepaspectratio, width=1.1\linewidth]{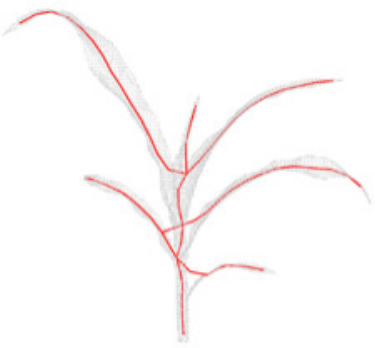}
            \end{minipage}    
            &
 	\begin{minipage}[t]{0.08\paperwidth}
            \centering
            \includegraphics[keepaspectratio, width=0.90\linewidth]{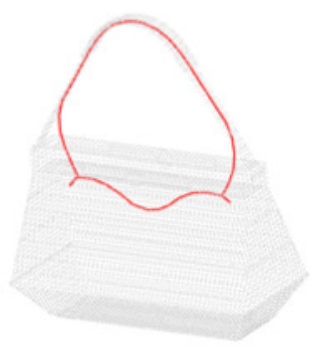}
            \end{minipage}    
            &
 	\begin{minipage}[t]{0.08\paperwidth}
            \centering
            \includegraphics[keepaspectratio, width=0.85\linewidth]{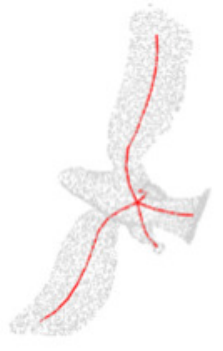}
            \end{minipage}    
            &
 	\begin{minipage}[t]{0.08\paperwidth}
            \centering
            \includegraphics[keepaspectratio, width=1.1\linewidth]{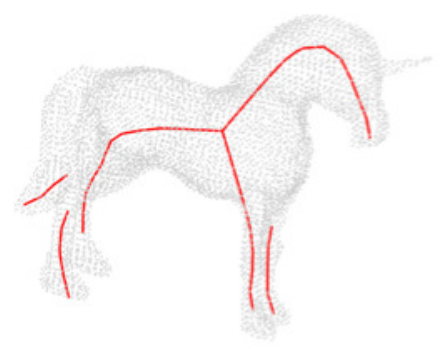}
            \end{minipage}    
\\
        \end{tabular}
        \caption{Qualitative comparison of skeletons extracted by our method with the methods by Cao et al. \cite{cao2010point}, and Huang et al. \cite{huang2013L1}.} 
        \label{fig:result_comparison}
    \end{figure*}

As shown in Fig. \ref{fig:GUI}, there are three parameters that can be tuned to obtain the desired results. The number of clusters from which candidate parts would be grown is one of the parameters. There should be at least as many clusters as there are parts. Remember, the final parts are selected from the candidate part list by the optimization algorithm (section \ref{sec:optimal_part_selection}). Therefore, the number of clusters is usually set to a much higher value. In practice, depending on the shape, we use $50 n$ clusters, where, $1 \leq n \leq 4$. The parameters $k_1$ and $k_2$ are the optimization parameters that are defined in section \ref{sec:optimal_part_selection}. The value of $k_2$, which determines the amount of overlap between parts in terms of the number of points they share, was set to $0.05$ for all the examples in Fig. \ref{fig:skeleton_results}. Keep in mind that it is possible for a point in the point cloud to be assigned to multiple parts. For example, a point at the intersection of multiple parts would be shared by all parts forming the junction. With $k_2 = 0.05$, less than five percent of all the points in the point cloud could be shared between various parts identified by the algorithm. The value of $k_1$ will decide the percentage of points covered by the parts chosen by the algorithm. Note that the optimization problem, described in section \ref{sec:optimal_part_selection} may not be feasible for all values of $k_1$. For example, it might not be possible to cover 100\% of the points in the cloud with only 5\% overlap. Since, we would like the parts extracted to cover maximum number of points, the algorithm picks the maximum possible value of $k_1$. It does so by checking the feasibility, starting at 90\% (i.e., $k_1$ = 0.90). If 90\% is feasible then it tries a higher value, else it decreases the value of $k_1$. This process continues until, the maximum feasible value for $k_1$ is found. However, sometimes picking a value a little less than maximum feasible value for $k_1$ gives better results. Hence, the user has the option to modify this value through the user interface.

We provide qualitative comparison of our results with the results from Cao et al. \cite{cao2010point}, and Huang et al. \cite{huang2013L1} in Fig. \ref{fig:result_comparison}. We use the implementations provided by the authors themselves for the comparison. We found that using the default parameter settings for methods by Cao et al. and Huang et al., do not produce good results in most cases. Hence, we manually tuned the available parameters to obtain better results and the best results that were obtained are presented in Fig. \ref{fig:result_comparison}. Note, while we and Cao et al. use point normal information in our algorithms, Huang et al. do not use this information. However, keep in mind that the input to our algorithm is just a point cloud. As described in section \ref{sec:point_normal_estimation}, we estimate the point normals as a first step of our algorithm. Since the input to all the algorithms is the same, the algorithm of Huang et al. is not at any disadvantage.

\begin{figure}[htp]
\centering
\begin{subfigure}[b]{0.24\textwidth}
	\includegraphics[width=\textwidth]{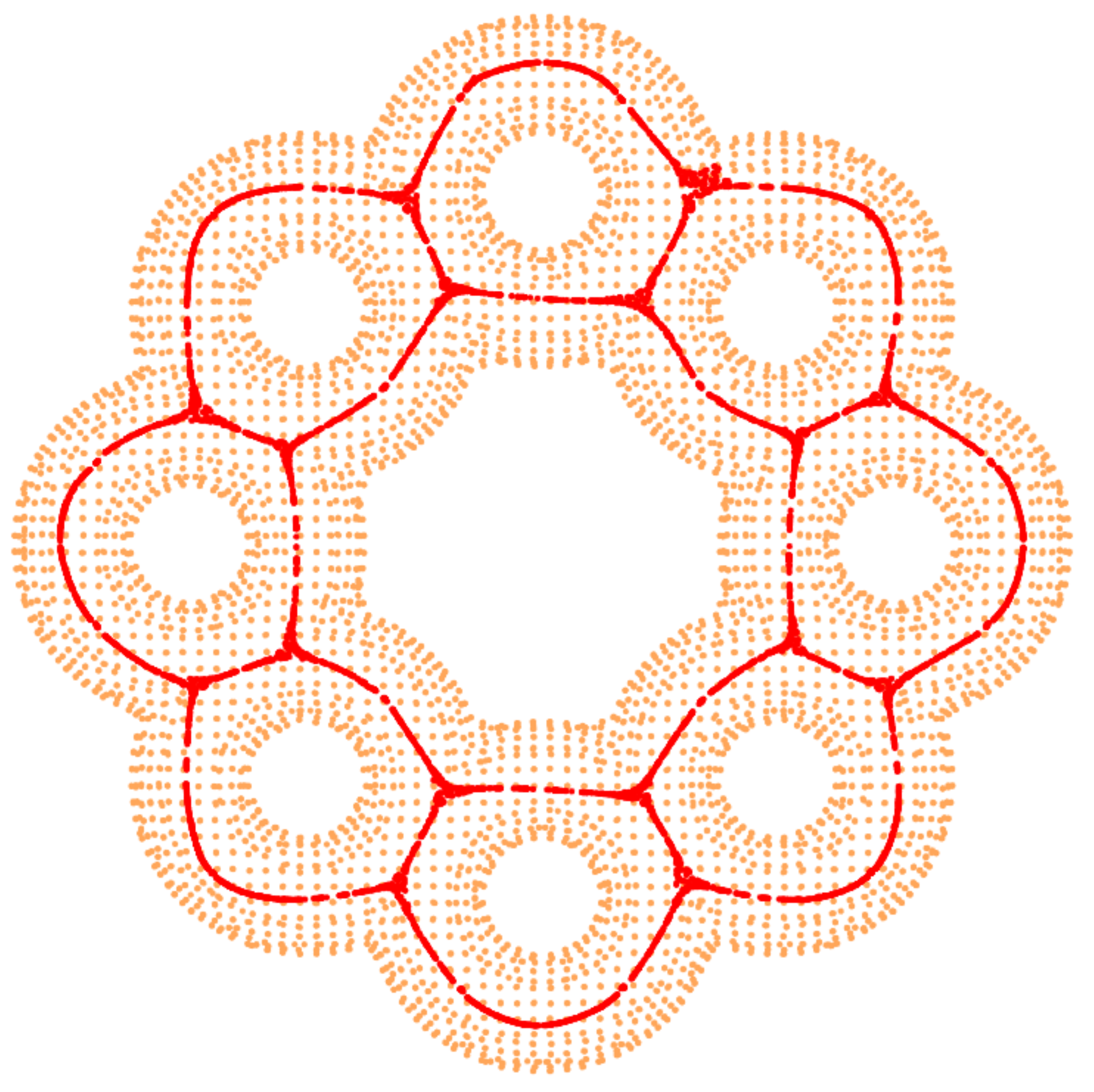}
	\caption{}
\end{subfigure}	
\begin{subfigure}[b]{0.24\textwidth}
\centering
	\includegraphics[width=\textwidth]{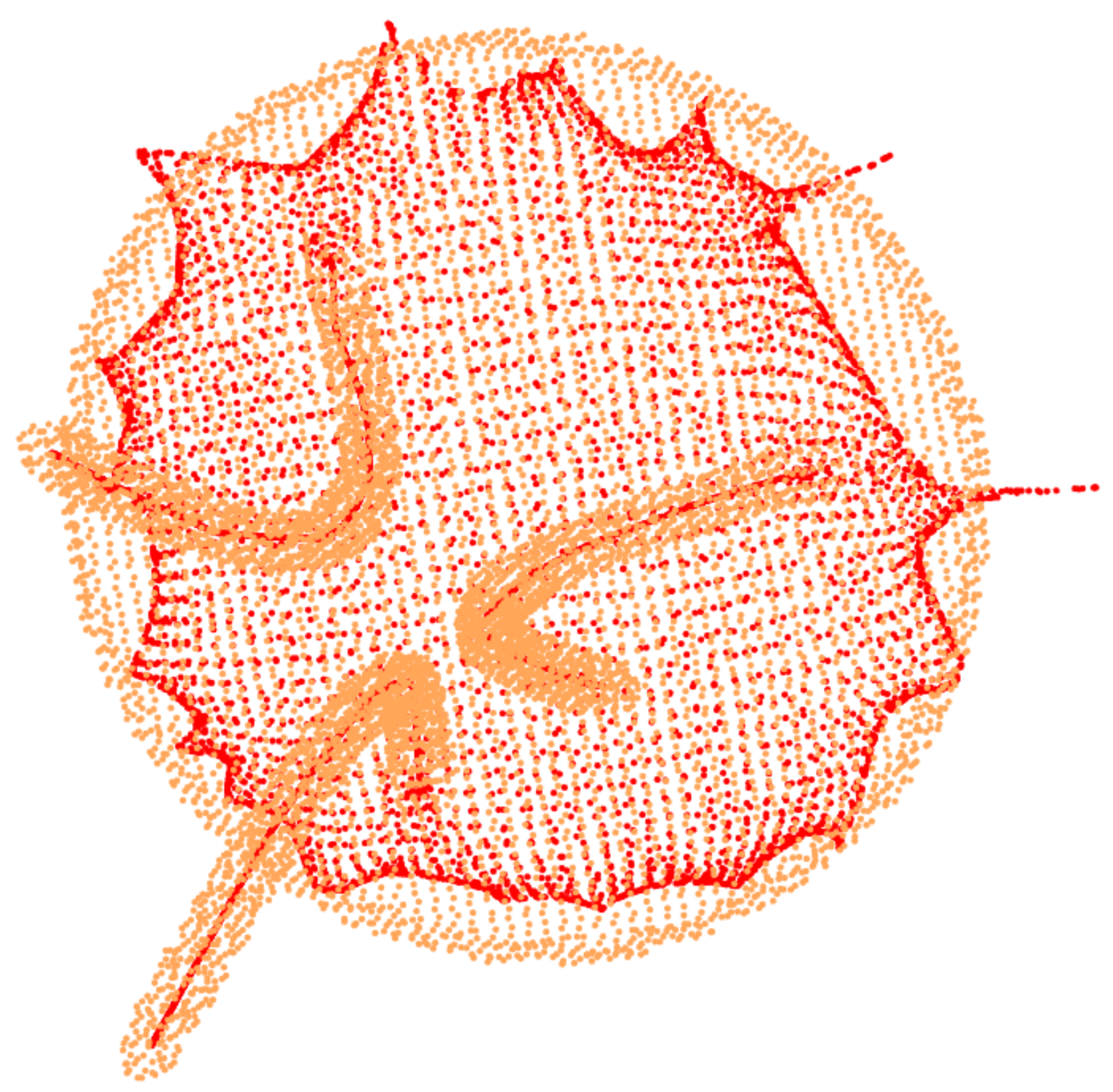}
	\caption{}
\end{subfigure}
\begin{subfigure}[b]{0.24\textwidth}
\centering
	\includegraphics[width=\textwidth]{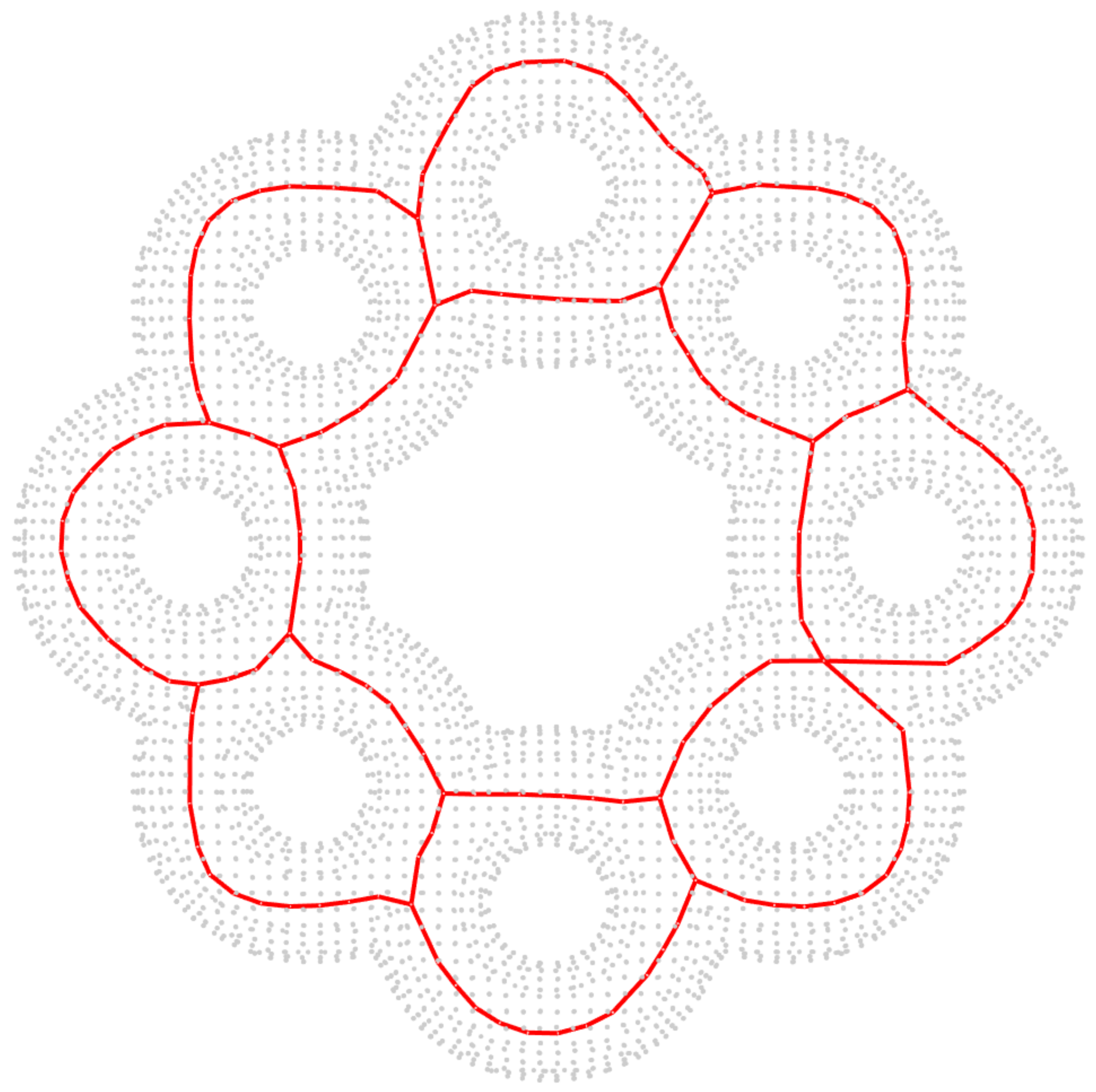}
	\caption{}
\end{subfigure}
\begin{subfigure}[b]{0.2\textwidth}
\centering
	\includegraphics[width=\textwidth]{images/compare/cao2-eps-converted-to.pdf}
	\caption{}
\end{subfigure}
\caption{The contracted points, at the end of the contraction step in the algorithm by Cao et al. \cite{cao2010point}, is shown in red in (a) and (b). The final skeletons extracted by Cao et al., for point clouds shown in (a) and (b), are shown in (c) and (d) respectively.}
\label{fig:contraction}
\end{figure}

As shown in Fig. \ref{fig:result_comparison}, the method by Huang et al. had difficulty in extracting skeletons for many of the point clouds shown. For the unicorn and the eagle ( Fig. \ref{fig:result_comparison} (g) and (f) respectively), parts of the skeleton were missing. For the hand (Fig. \ref{fig:result_comparison}  (c)), the connection between the fingers are incorrect. Note, Huang et al. draws an initial set of samples from the point cloud to start their skeleton extraction process. We observed that even for the same set of parameters but different initial samples, the skeletons extracted varied widely. The method by Cao et al. did well on most point clouds except for the table (Fig. \ref{fig:result_comparison}  (b)) and the suitcase (Fig. \ref{fig:result_comparison}  (e)). In fact, these two cases demonstrate one of the drawbacks with their approach to skeleton extraction. As shown in Fig. \ref{fig:contraction}, when the contracted points closely represent the skeleton of the shape qualitatively, like in the case of the eight inter-connected toruses (Fig. \ref{fig:contraction} (a)), the extracted skeleton (Fig. \ref{fig:contraction} (c)) would correspondingly be good. But in Fig. \ref{fig:contraction} (b), we can see that at the end of the contraction step, the contracted points do not resemble the skeleton of the round table top. Correspondingly, the extracted skeleton fails to provide a good representation for the skeleton of the table top. Note that distance field based approaches would also share this drawback. 

\begin{figure}[htp]
\centering
\begin{subfigure}[b]{0.24\textwidth}
	\includegraphics[width=\textwidth]{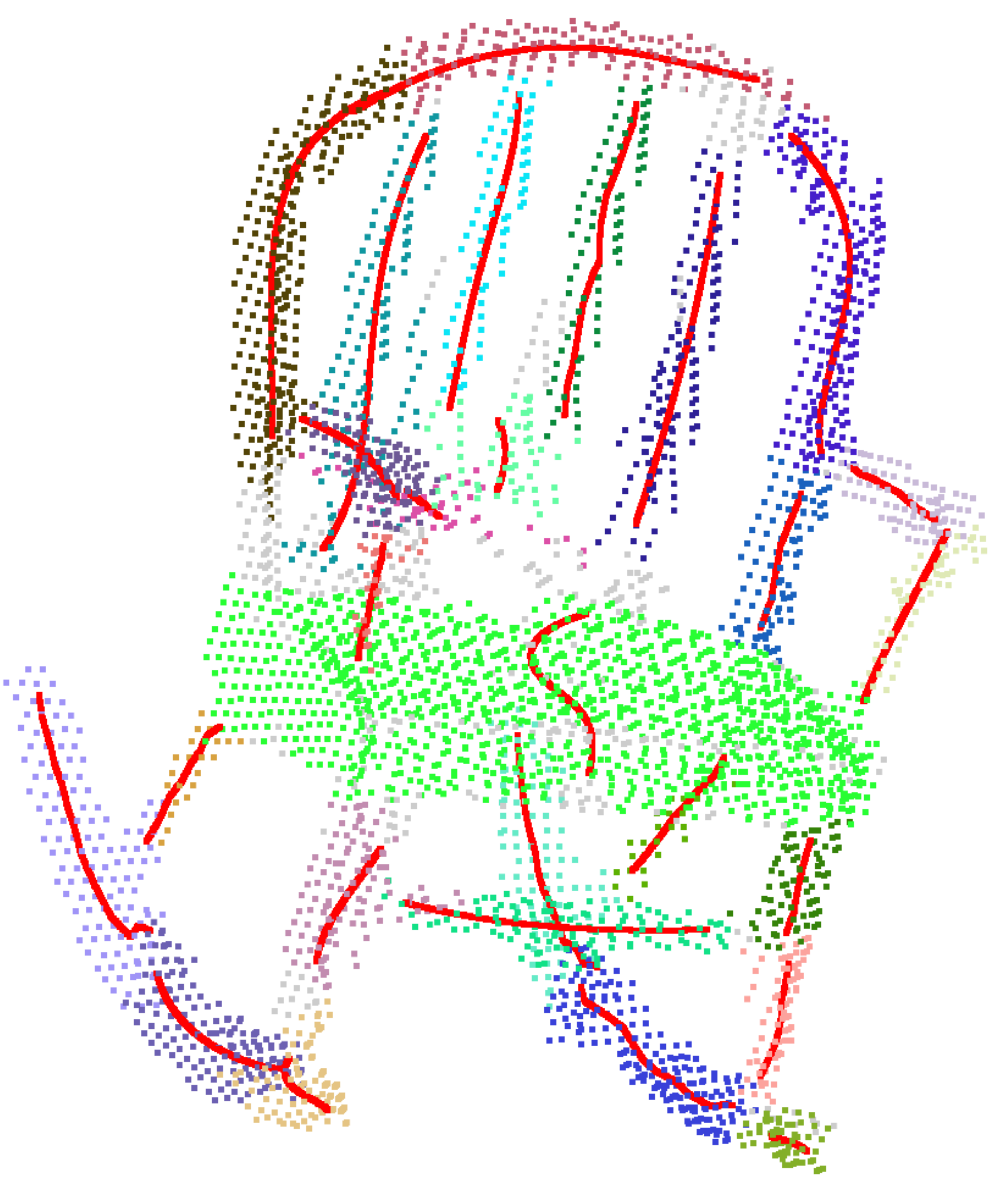}
	\caption{}
\end{subfigure}	
\begin{subfigure}[b]{0.21\textwidth}
\centering
	\includegraphics[width=\textwidth]{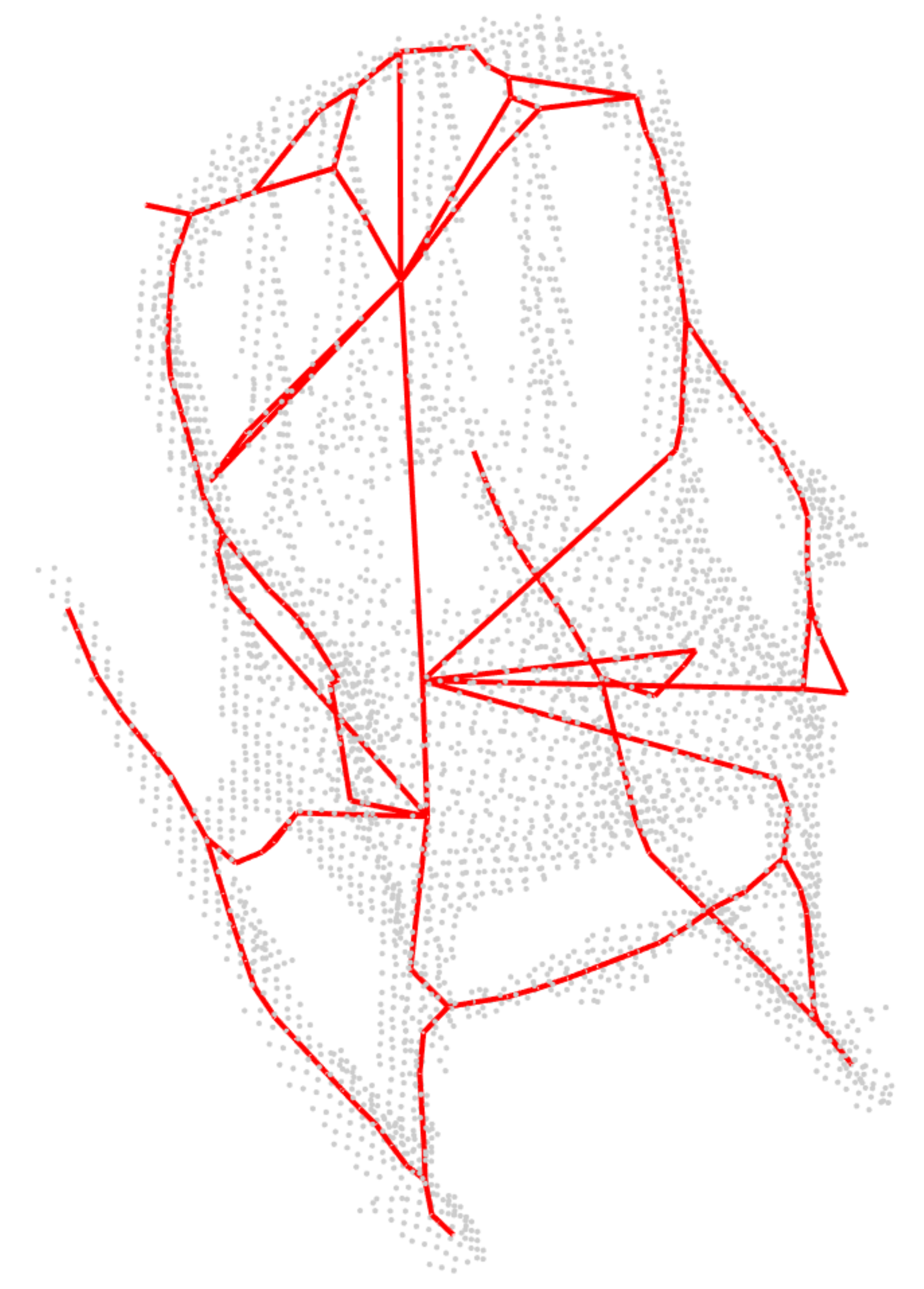}
	\caption{}
\end{subfigure}
\caption{Skeleton extraction with noisy point clouds. (a) Part skeletons extracted by our method with points belonging to different parts shown in different colors. (b) Skeleton extraction results from Cao et al.. }
\label{fig:rocking_chair}
\end{figure}

As shown in Fig. \ref{fig:result_comparison} (b), our method was able to extract a nice straight line skeleton for the table top, primarily because we are looking to extract part skeletons first. Similarly, our part based approach was able to extract centered skeletons for the point cloud shown in Fig. \ref{fig:result_comparison} (e). These examples show why decomposing objects into GCs to detect parts and part skeletons can effectively deal with a wide variety of shapes. Fig. \ref{fig:rocking_chair} compares the performance of our algorithm to that of Cao et al. in detecting skeletons from noisy point clouds. This point cloud represents a rough 3D shape reconstruction of a rocking chair. Notice that our algorithm successfully identifies the various parts within the noisy point cloud. For instance, it can be seen that our algorithm detected the seat of the rocking chair (shown in light green) as a GC. This demonstrates that using a very basic property like translational symmetry to detect GCs can be a powerful tool in identifying parts and thereby simplifying the skeleton extraction problem. This example also shows that our method can be used as the first step in denoising point clouds and in surface reconstruction.

\begin{figure}[htp]
\centering
\begin{subfigure}[b]{0.24\textwidth}
	\includegraphics[width=\textwidth]{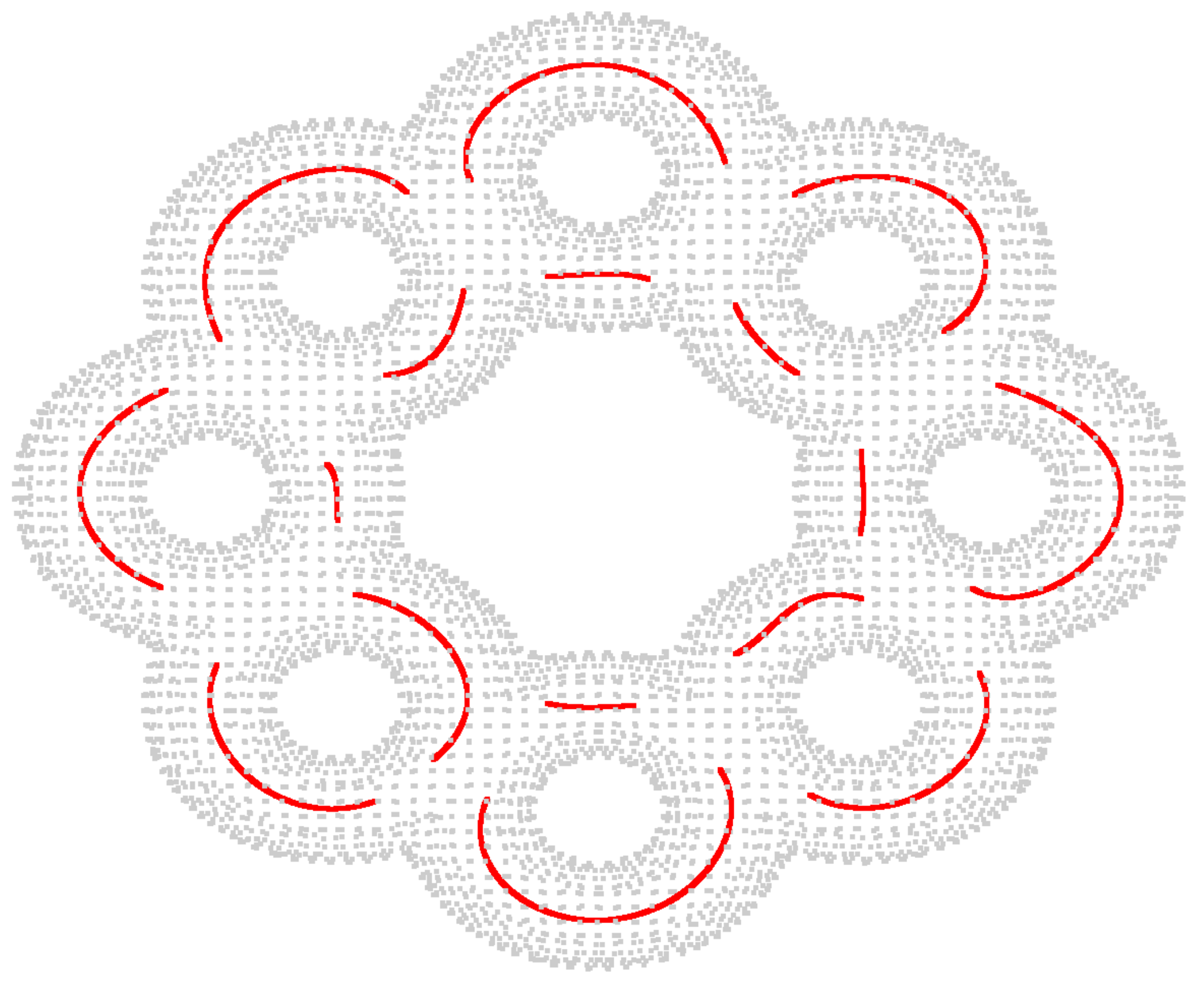}
	\caption{}
\end{subfigure}	
\begin{subfigure}[b]{0.24\textwidth}
\centering
	\includegraphics[width=\textwidth]{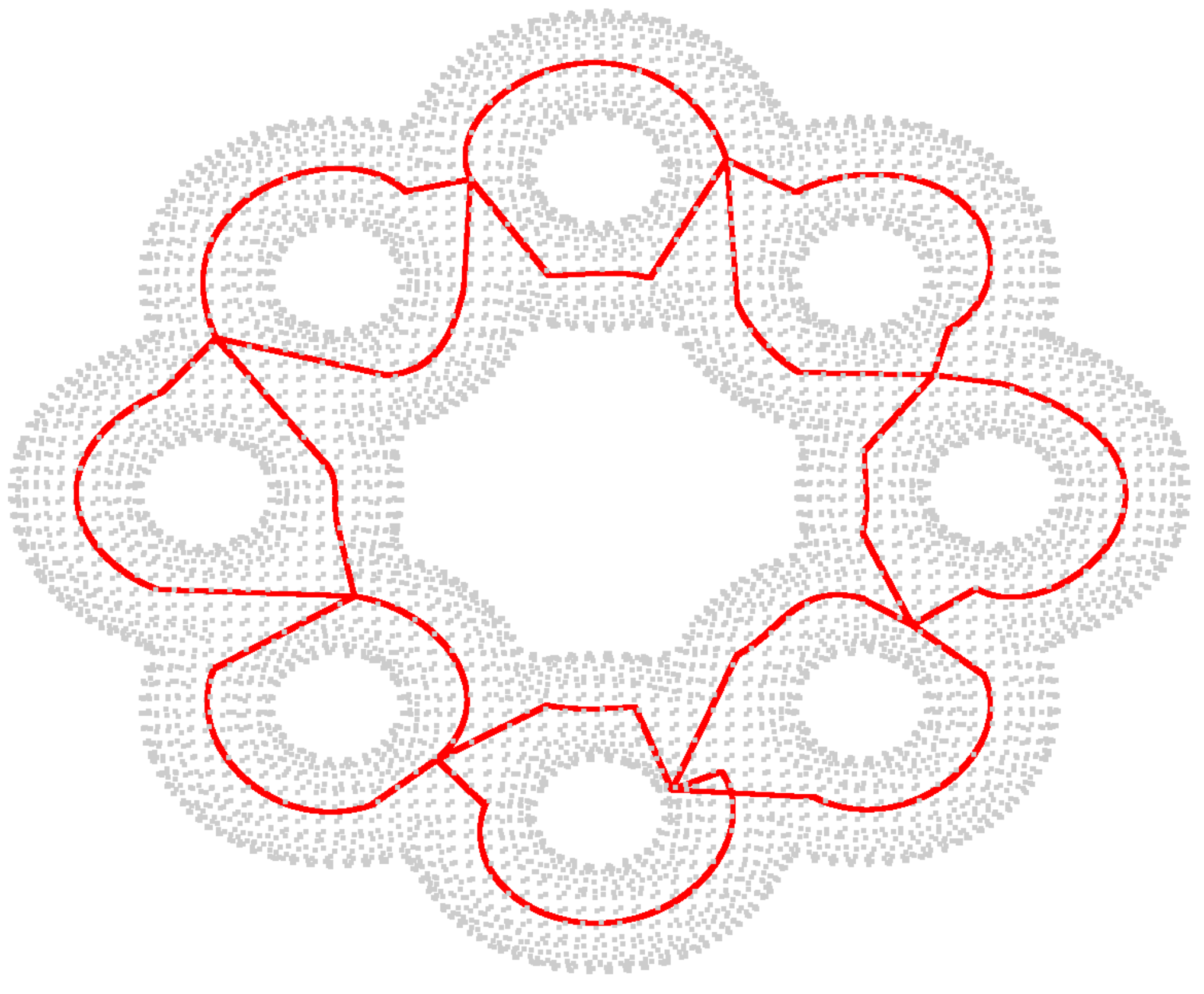}
	\caption{}
\end{subfigure}
\caption{}
\label{fig:torus}
\end{figure}

One of the limitations of our algorithm is that the method we use to link part skeletons, to form a complete skeleton of the shape, can sometimes lead to anomalies like the one shown in Fig. \ref{fig:torus} (b). Keep in mind that we join individual part skeletons together using straight lines. The drawback of this can be clearly seen in Fig. . \ref{fig:torus} (b), where this approach leads to an unnatural looking overall skeleton. The individual part skeletons of the parts detected by our algorithm shown in Fig. \ref{fig:torus} (a). Ideally, the point cloud has to be segmented into eight toruses and each torus would be connected to two its neighbors. However, our algorithm does not succeed in extracting individual torus skeletons due to the presence of large number of junction regions where the translational symmetry of a torus breaks. In this case, the algorithm by Cao et al. succeeds in extracting a better skeletal representation of the shape as shown in Fig. \ref{fig:result_comparison} (c). As pointed out earlier, this is due to the fact that the point cloud at the end of the contraction step closely resembles the skeleton of the shape.

\begin{figure}[htp]
\centering
\begin{subfigure}[b]{0.15\textwidth}
	\includegraphics[width=\textwidth]{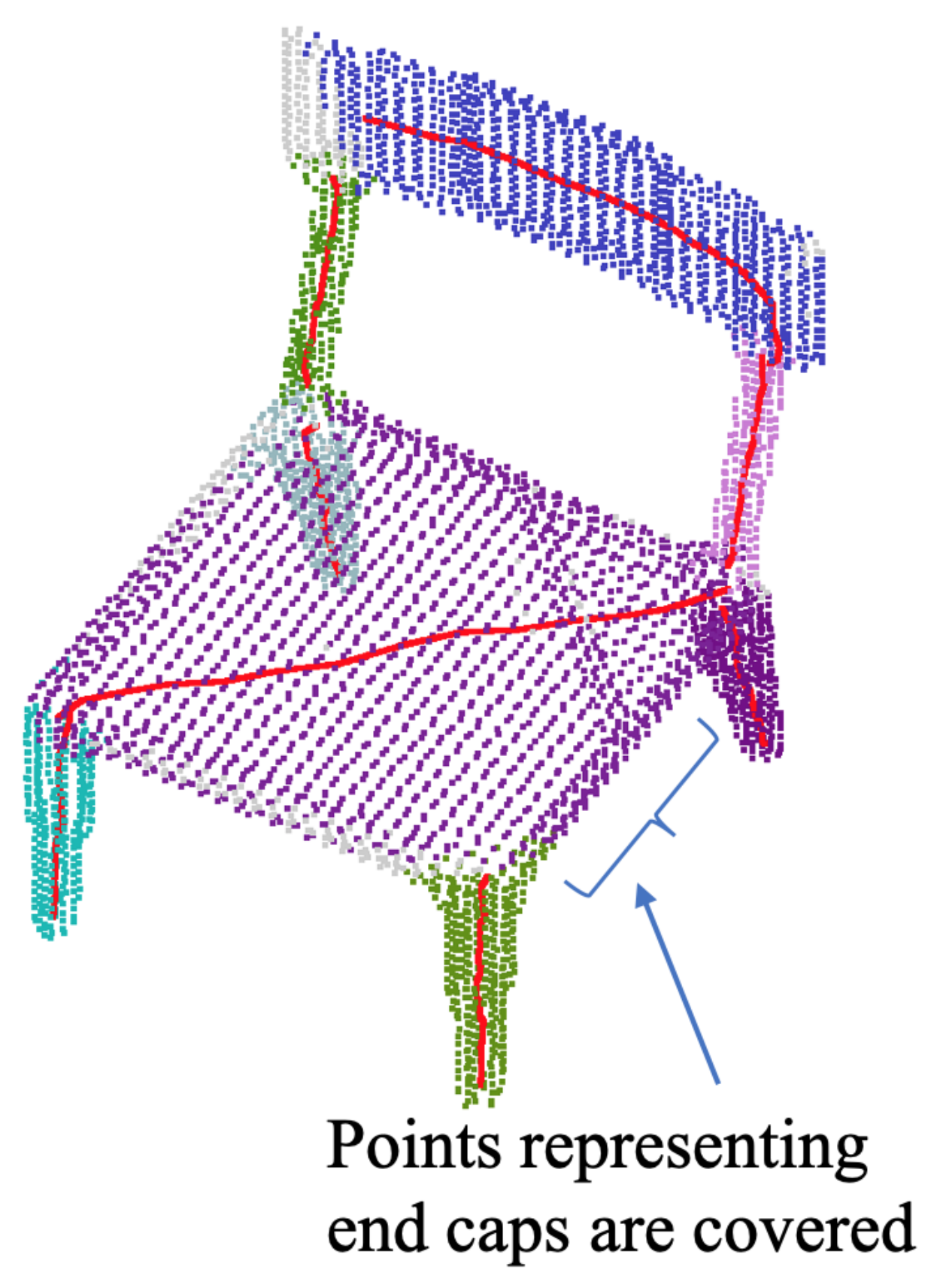}
	\caption{}
\end{subfigure}	
\begin{subfigure}[b]{0.16\textwidth}
\centering
	\includegraphics[width=\textwidth]{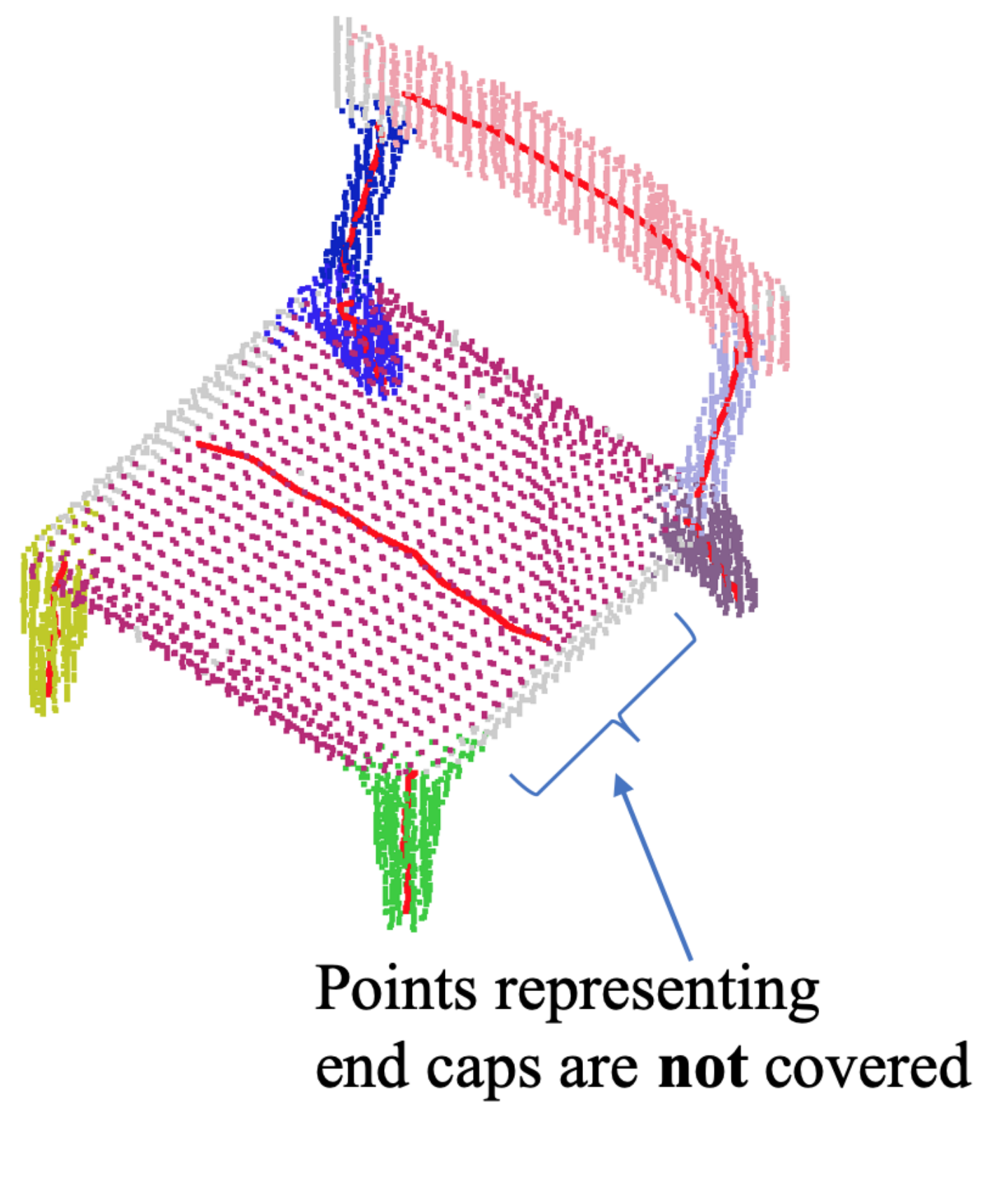}
	\caption{}
\end{subfigure}
\begin{subfigure}[b]{0.15\textwidth}
\centering
	\includegraphics[width=\textwidth]{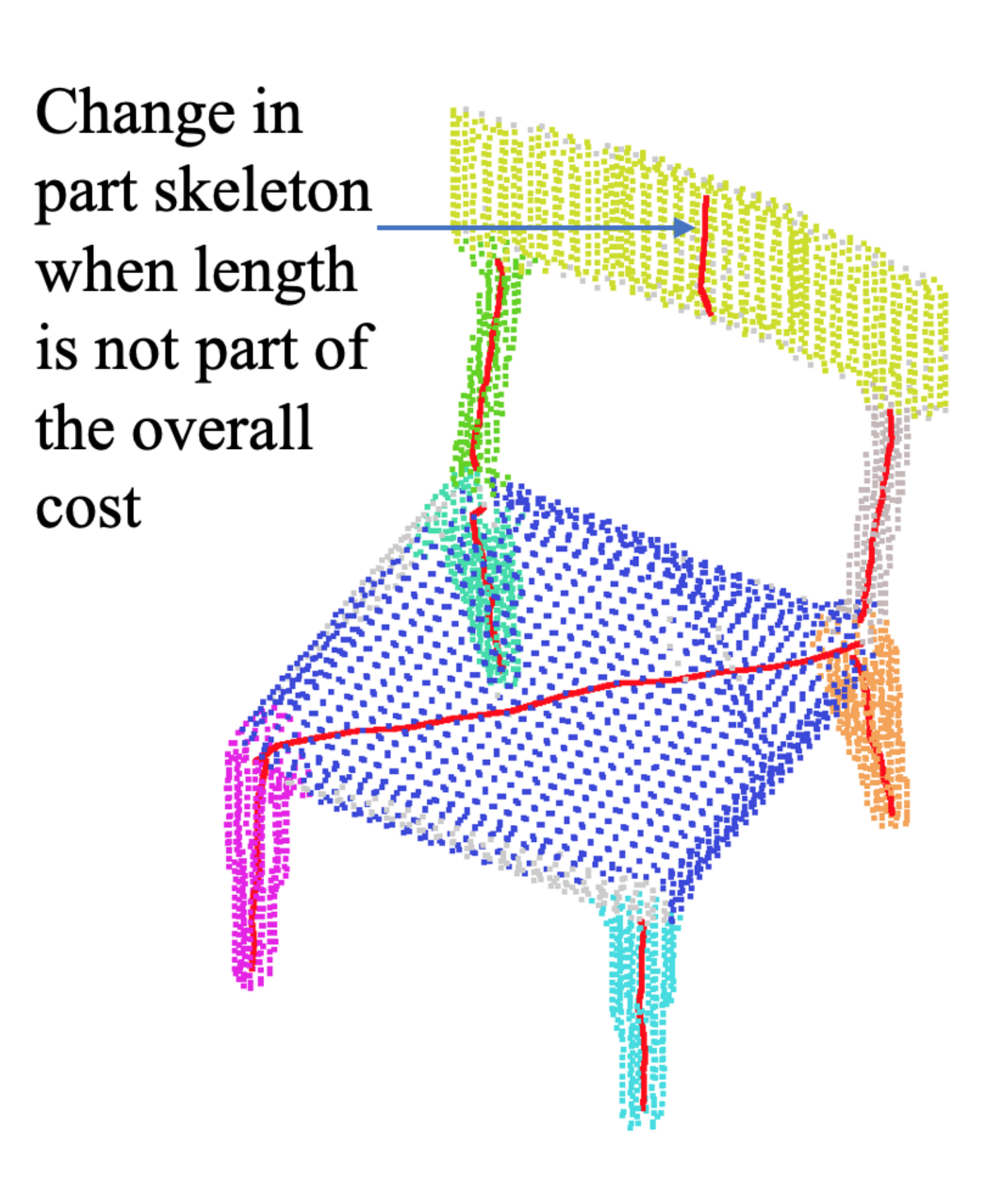}
	\caption{}
\end{subfigure}
\caption{Ambiguity in skeletal representation of parts. (a) The different parts identified, by our algorithm, are shown in different colors along with the skeleton shown as a red curve. (b) Parts identified and part skeletons extracted for a vale of $k_1$ (in equation \ref{eq:optimization} , constrained \ref{cstr:c1}) slighltly lesser than the maximum feasible value. (c) Part skeletons extracted when the length component is not part of the overall cost.}
\label{fig:ambiguity}
\end{figure}

We would like to point out an interesting observation regarding the ambiguity of part skeletons for the same part. The part skeleton extracted for the seat of the point cloud of a chair is shown in Fig. \ref{fig:ambiguity} (a). This result was obtained when the number of clusters was set to 150 and using the maximum feasible value of $k_1 = 0.95$ (in equation \ref{eq:optimization} , constrained \ref{cstr:c1}) for this point cloud. Note that by setting $k_1 = 0.95$, we are asking the algorithm to cover at least 95\% of the points in the cloud. In Fig. \ref{fig:ambiguity} (b), the results obtained by setting $k_1 = 0.92$ is shown. Notice that the parts extracted are almost the same but the skeleton of the part representing the seat of the chair is now different. In fact, this result (i.e., Fig. \ref{fig:ambiguity} (b)) is a qualitative better result than the one in Fig. \ref{fig:ambiguity} (a). Also note that the points at the end cap region of the GC in Fig. \ref{fig:ambiguity} (b) are not included by the algorithm as part of the seat. This shouldn't be surprising because the translational symmetry based definition of a GC will not capture the end caps. However, the GC can capture the end caps if the skeleton is diagonal as is the case with Fig. \ref{fig:ambiguity} (a). In short, very similar set of points representing the same part in a point cloud can be well represented by multiple skeletal curves due to the very general definition of a GC. Another example of such ambiguity can be seen by comparing the part skeletons in Fig. \ref{fig:ambiguity} (a) and (c). Keep in mind that each part was assigned a cost which had four components. One of the components was length of the part skeleton. Longer part skeletons were given preference over shorter ones. Fig. \ref{fig:ambiguity} (c) shows that removing the length component (i.e., not rewarding or penalizing length of parts) would result in a different set of part skeletons extracted. Note that the parts are almost the same, but the part skeletons are different.  To summarize, the same part can be represented by multiple skeletal representations as per the definition of a GC. When such situations arise, there will be ambiguity in choosing the best skeletal representation. Since, our algorithm extracts parts and lists them as shown in Fig. \ref{fig:GUI}, it provides the user with the flexibility of manually choosing the best skeletal representation according to them.

\section{Implementation and Run Time}
The code is implemented in Python and takes about three minutes to run for a point cloud containing five thousand points with the number of clusters set to hundred on a 2.8 GHz Intel Core i7 processor with 16 GB RAM. We use Numba \cite{Lam_2015_numba} to accelerate the cost function evaluation and gradient calculations used for registering two 3D point clouds. Note that the registration function is called multiple times in order to grow a part. Even with a modest estimate of five calls to the registration function per cluster, the total number of calls would be around five hundred. Hence, it is essential to be able to quickly evaluate the cost function and compute gradients for the registration function. Growing parts is the most time consuming step in the algorithm. Since, two parts can be grown completely independently, this process is amenable to parallelization. Therefore, there is ample scope to speed up the computations by parallelizing the part growing stage.

\section{Conclusion}
In this work we introduced a method to extract curve skeletons from unorganized point clouds by decomposing a shape into its parts. The parts are treated as generalized cylinders which are characterized by translational symmetry. Starting from the detection of thin cross-sectional clusters we grew parts by exploiting their fundamental property of translational symmetry. To do so, we introduced a point set registration method which is tailor made for the purpose of growing parts from thin cross-sections. By defining the parts as GCs and then relying primarily on the very definition of a GC to extract parts, we presented a very principled approach to 3D Point cloud decomposition. In the results section we demonstrated how the algorithm is capable of handling a wide variety of shapes. We pointed out the advantages of a part based approach over other state of the art methods. Another advantage of a part based approach is that it makes user interaction intuitive and easy. This was demonstrated with the help of a GUI.
%

\ifCLASSOPTIONcompsoc
  \section*{Acknowledgments}
\else
  \section*{Acknowledgment}
\fi

This research was supported by the National Eye Institute of the National Institutes of Health under award number 1R01EY024666-01. The content of this paper is solely the responsibility of the authors and does not necessarily represent the official views of the National Institutes of Health.

\ifCLASSOPTIONcaptionsoff
  \newpage
\fi

\bibliographystyle{IEEEtran}
\bibliography{bibliography}

%
%
%

\end{document}